\documentclass[12pt]{article}
\usepackage{fullpage}
\usepackage{epsf}
\usepackage{epsfig}
\usepackage{amsthm}
\usepackage{amsfonts}
\usepackage{leftidx}
\usepackage{color}
\usepackage{amsmath}
\usepackage{setspace}
\usepackage{natbib}
\usepackage{hyperref}
\usepackage{algorithmic}
\usepackage[boxed]{algorithm}
\usepackage{graphicx}
\usepackage{booktabs}
\usepackage{multirow}

\begin{document}

\newcommand{\bm}[1]{\mbox{\boldmath $#1$}}
\newcommand{\vect}[1]{\boldsymbol{#1}}
\newcommand{\mb}[1]{#1}
\newcommand{\bE}[0]{\mathbb{E}}
\newcommand{\bV}[0]{\mathbb{V}\mathrm{ar}}
\newcommand{\bP}[0]{\mathbb{P}}
\newcommand{\ve}[0]{\varepsilon}
\newcommand{\mN}[0]{\mathcal{N}}
\newcommand{\iidsim}[0]{\stackrel{\mathrm{iid}}{\sim}}
\newcommand{\NA}[0]{{\tt NA}}
\newcommand{\cB}{\mathcal{B}}
\newcommand{\R}{\mathbb{R}}
\newcommand{\Rp}{\R_+}


\title{\vspace{-1cm}Emulating satellite drag from large simulation experiments}
\author{Furong Sun\thanks{Corresponding author: Department of Statistics, Virginia Tech,
Hutcheson Hall, 250 Drillfield Drive
Blacksburg, VA 24061, USA;
\href{mailto:furongs@vt.edu}{\tt furongs@vt.edu}}
\and Robert B.~Gramacy\thanks{Department of Statistics, Virginia Tech}
\and Benjamin Haaland\thanks{Population Health Sciences, University of Utah}
\and Earl Lawrence\thanks{Computer, Computational and Statistical Sciences, Los Alamos National Laboratory}
\and Andrew Walker\thanks{Space Science and Applications, Los Alamos National Laboratory}
}
\date{}

\maketitle

\vspace{-0.2cm}
\begin{abstract}
Obtaining accurate estimates of satellite drag coefficients in low Earth orbit
is a crucial component in positioning and collision avoidance.  Simulators can
produce accurate estimates, but their computational expense is much too large
for real-time application.  A pilot study by \citet{metha:etal:2014} showed
that Gaussian process (GP) surrogate models could accurately emulate
simulations.  However, cubic runtime for training GPs means that they could
only be applied to a narrow range of input configurations to achieve the
desired level of accuracy.  In this paper we show how extensions to the local
approximate GP ({\tt laGP}) method allow accurate full-scale
emulation.  The new methodological contributions, which involve a multilevel
global/local modeling approach, and a setwise approach to local subset
selection, are shown to perform well in benchmark and synthetic data settings.
We conclude by demonstrating that our method achieves the desired level of
accuracy, besting simpler viable (i.e., computationally tractable) global and
local modeling approaches, when trained on seventy thousand core hours of drag
simulations for two real-world satellites: the Hubble space
telescope (HST) and the gravity recovery and climate experiment (GRACE).

\end{abstract}


\section{Introduction}
\label{sec:intro}

Low Earth orbit (LEO) is becoming increasingly crowded. There is 
a growing fear of collisions and their long term effects on the accessibility
of space \citep{kessler:cour:1978}. One such collision in 2009
\citep{achenbach:2009} has already occurred and could be a preview of things
to come. Satellite operators need precise estimates of satellites' current and
future positions to anticipate and avoid these collisions, and to plan
experiments. The Committee for the Assessment of the U.S. Air Force’s
Astrodynamics Standards recently released a report citing atmospheric drag as
the largest source of uncertainty for LEO objects, due in part to improper
modeling of the interaction between the atmosphere and object. Accurate drag
coefficients, necessary for computing the drag force, can greatly reduce this
uncertainty and produce more accurate orbital predictions. Modern numerical
simulators can produce accurate drag coefficients under varying input
conditions (e.g., temperature(s), velocity, chemical composition of the
atmosphere, object mass, orientation, and geometry). However, these solvers
are too computationally intensive for time-sensitive applications like
navigation and calculation of collision probabilities
\citep{lawrence:etal:2014}.

\citet{metha:etal:2014} suggest a remedy involving Gaussian process (GP)
emulation \citep{sack:welc:mitc:wynn:1989,sant:will:notz:2003}. For idealized
satellite geometries (e.g., a ball) and $N=1000$ simulations at a
space-filling design of input configurations, \citet{metha:etal:2014}'s
emulated drag coefficients were quite accurate: being within 0.3\% of the true
drag as measured by root-mean-squared-percentage-error (RMSPE). For more
realistic geometries, like the Hubble Space Telescope (HST), errors were
upwards of 1\%.  A 1\% relative error is a standard benchmark in the
literature, however, to obtain this result \citet{metha:etal:2014} had to
drastically narrow the orientation inputs, both in terms of design and
out-of-sample predictive locations, to less than 2\% of their desired range.
Expanding out to their full range, while maintaining the same design density
would have required tens of millions of runs.  That's way too big for an
ordinary GP, which buckles under the weight of quadratic space requirements
and cubic-time matrix decompositions. \citet{metha:etal:2014}'s input
sensitivity analysis revealed further shortcomings to a typical GP emulation
strategy: drag surfaces are highly nonstationary.

In this paper we aim to address both issues, design size and nonstationarity,
simultaneously.  A preliminary component is the development of a simulation
framework that would enable a large corpus of drag estimates to be collected
in a massively parallel fashion---on modern supercomputing architectures---in
a reasonable amount of time. More details on estimating satellite drag via
simulation, and a description of the so-called {\em test particle method}
behind our implementation, are provided in Section \ref{sec:review}.  In our
supplementary material we provide an {\sf R} interface that can be used to
perform new runs. Using that setup we were able to compile data files (also
provided in the supplementary material) with the inputs and outputs from more
than seventy thousand core-hours of simulations, comprising of millions of
runs, for two real-world satellites: the HST, and the gravity recovery and
climate experiment (GRACE).  These runs were performed over the span of six
months on nodes managed by the University of Chicago Research Computing
Center.

Using that data, we demonstrate how we are able to tractably build accurate
emulators via local approximate Gaussian processes
\citep{gramacy:apley:2015,laGP}. The {\tt laGP} framework can be seen as a
modernization of so-called ``ad hoc" {\em local kriging} neighborhoods
\citep[][pp.~131--134]{cressie:1993} achieved by deploying {\em active learning}
\citep{seo:etal:2000} to dynamically determine local neighborhoods.  We review {\tt laGP} in Section \ref{sec:laGP}, with
emphasis on how its unique flavor of divide-and-conquer facilitates
fast emulation through approximation; how it offers potential for massive
parallelization through a limited form of statistical independence, while simultaneously offering a degree of nonstationary flexibility.

While the out-of-the-box capability of {\tt laGP} is able to yield more
accurate emulation than cruder, \citet{metha:etal:2014}--style
alternatives, they are unable to meet the 1\% relative error target, even when
trained on millions of runs.  In Sections \ref{sec:mrlaGP} and \ref{sec:joint}
we detail two crucial updates which were developed in order to enhance
capability and thus emulation accuracy.  One is an engineering detail required
to allow anisotropic local fits. The other is a multiresolution
global/local modeling strategy leveraging new results from
\citet{liu:hung:2015} on estimating global GP lengthscales from sub-sampled data. We describe how the application of this
result has the effect of modulating and stabilizing {\tt laGP}'s nonstationary capability.

Finally, a third methodological enhancement is proposed to cope with a demand
unique to the satellite positioning application: local emulation along a
trajectory in the input space.  The modifications introduced above enable {\tt
laGP} to provide pointwise drag surrogates for all potential input
configurations---i.e., for all satellite positions, orientation, velocities,
etc.---quickly, in a vastly parallelized framework that leverages the
statistical independence of its local (approximate)
calculations.\footnote{Conditional on nearby locations and responses,
predictions at distant locations are treated as approximately independent.} On
the one hand, that represents a kind of overkill.  It is nice to know that
{\tt laGP} can furnish so many accurate predictions, but such a large field
will never be fully utilized in practice. On the other hand, pointwise results
(say for today's HST configuration) under-quantify the most immediately relevant uncertainties.  Joint prediction over likely paths in the input
space, or trajectories in low Earth orbit, would be far more useful.  We
extend the {\tt laGP} notion of ``locale'' to sets of predictive locations,
along paths in the input space, and show that these lead to tractable and
accurate joint predictions.  

After detailing the satellite drag application, reviewing GP-based methods
[Section \ref{sec:review}], and then enhancing them to suit [Sections
\ref{sec:mrlaGP}--\ref{sec:joint}],  we embark on a detailed empirical study.
Section \ref{sec:bench} illustrates the methodological enhancements in a
controlled setting, on benchmark and synthetic data sets. Then in Section
\ref{sec:satemu} we return to the satellite drag data, illustrating a cascade
of comparators on millions of simulation runs for the HST and GRACE
satellites. Finally, Section \ref{sec:discuss} concludes with brief
reflection and perspective.

\section{Simulating and emulating satellite drag}
\label{sec:review}

Drag coefficients can be simulated based on the geometry of the object, its
position, and velocity.  Several other factors are somewhat less directly involved,
such as solar conditions, which we will not detail here, but are accounted for
in our simulation apparatus. For a nice survey, see \citet{metha:etal:2014}.
As a more macro consideration, the choice of gas--surface interaction (GSI) model
is important insofar as it impacts the accuracy (or fidelity) of the
simulations and the computation time.  It is worth noting here that we follow
the setup of \citet{metha:etal:2014} in using the so-called
Cercignani--Lampis--Lord (CLL) \citep{cercignani:etal:1971} GSI model, and
this choice impacts the spectrum of input variables we describe below.

There are three sets of inputs.  The first set is described in Table
\ref{t:inputs}.  We refer to these as the ``free parameters'', as they are
the main ones which we vary in our designs.
\begin{table}[ht!]
\centering
\begin{tabular}{llllc}
Symbol & ASCII & Parameter & [units]  & Range \\
\hline
$v_{\mathrm{rel}}$ & {\tt Umag} & velocity & m/s & [5500, 9500] \\
$T_s$ & {\tt Ts} & surface temperature & K & [100, 500] \\ 
$T_a$ & {\tt Ta} & atmospheric temperature & K & [200, 2000] \\
$\theta$ & {\tt theta} & yaw & radians & $[-\pi, \pi]$ \\
$\phi$ & {\tt phi} & pitch & radians & $[-\pi/2, \pi/2]$ \\
$\alpha_n$ & {\tt alphan} & normal energy AC & unitless & [0, 1] \\
$\sigma_t$ & {\tt sigmat} & tangential momentum AC & unitless & [0, 1]
\end{tabular}
\caption{The free parameters: inputs to the CLL GSI model for satellite drag
coefficients, typical ASCII representation, units and ranges. AC stands for
``accommodation coefficient''.}
\label{t:inputs}
\end{table}
On the first row of the table is velocity, which is mostly self-explanatory, 
however, note that this is a direction-less measure of speed in LEO.  The next two are temperatures, with surface temperature being 
measured on the satellite's exterior and atmospheric temperature being derived 
from raw position information (latitude, longitude and altitude), which couples 
it to our second set of inputs described momentarily.  The angles in the middle 
of the table describe orientation, and thereby the effect of this input is 
intimately linked to satellite geometry, comprising our third set of inputs.  
The final two rows in the table are the accommodation coefficients that are 
particular to the CLL GSI.
 
At LEO altitudes the atmosphere is primarily comprised of atomic oxygen (O),
molecular oxygen (O$_2$), atomic nitrogen (N), molecular nitrogen (N$_2$),
helium (He), and hydrogen (H) \citep{picone:etal:2002}.\footnote{Argon (Ar)
and anomalous oxygen (AO) are also present, but their relatively low mole
fractions mean their effect on drag is negligible, so these are dropped from
the analysis.} The mixture of these so-called chemical ``species'' varies with
position, and there are calculators such as
\begin{center}
\url{https://ccmc.gsfc.nasa.gov/modelweb/models/nrlmsise00.php}
\end{center}
which deliver mixture weights provided position and time coordinates. While
these six weights technically qualify as inputs to the drag simulator, it is
more typical to perform a ``blocked'' experiment separately for each pure
species.  That is, if the simulator is invoked with weights $e_k$,
where $e_5 = (0,0,0,0,1,0)$ denotes pure He say, producing drag
coefficient $C_{D_k}$ for $k=1,\dots, 6$, then a total drag coefficient in any
mixture can be calculated as
\begin{equation}
C_D = \frac{\sum_{k=1}^6 C_{D_k} \chi_k \cdot m_k }{\sum_{k=1}^6  \chi_k \cdot m_k}. \label{eq:mix}
\end{equation}
Above, $\chi_k$ and $m_k$ are the mole fraction (i.e., the mixture weight) and
particle mass for the $k^\mathrm{th}$ species, respectively.  If only a single
drag simulation is desired, for a single chemical species composition, then
clearly six simulations are more work than one.  However, the nature of the
highly parallelized simulator implementation described in Section
\ref{sec:tpm} strongly favors a blocked approach with fixed species weights.
Blocking is also favorable from a data modularization and modeling
perspective, which is discussed in more detail below.

The final set of inputs comprise the satellite geometries.  These are
specified as so-called ``mesh files'', which are ASCII representations of 
the volumes that describe the shape of the satellites.
\begin{figure}[ht!]
\centering
\includegraphics[scale=0.35]{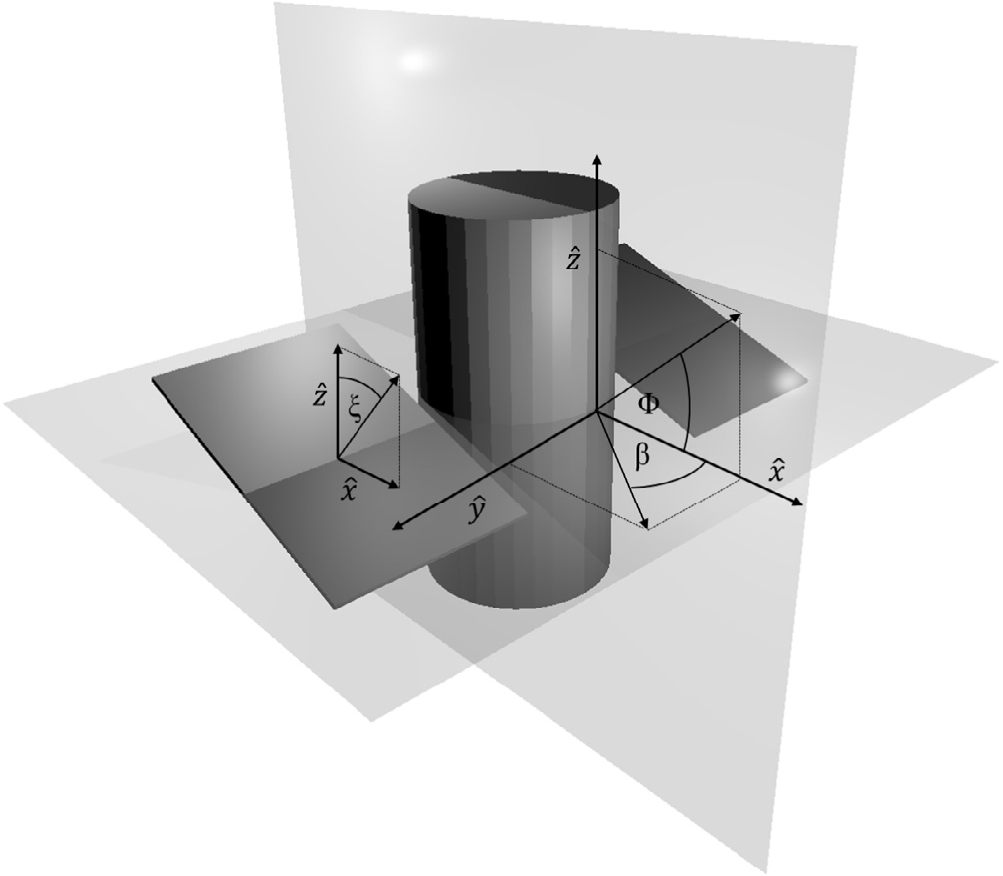} \hspace{0.5cm}
\includegraphics[scale=0.4]{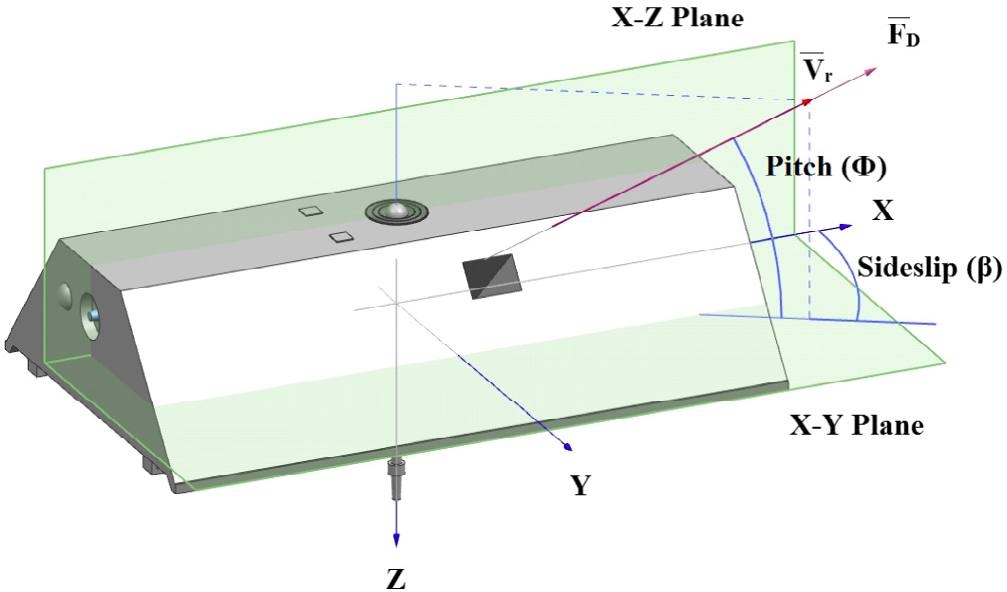}
\caption{Mesh visuals for the HST {\em (left)} and 
the GRACE {\em (right)} satellite.}
\label{f:mesh}
\end{figure}
Figure \ref{f:mesh} shows visualizations of mesh files for two satellites, HST
on the {\em left} and GRACE to the {\em right}.  Satellite configuration in LEO is
modulated by the orientation angles $\theta$ and $\phi$ in Table
\ref{t:inputs}, which have the effect of rotating the meshes.  Although each
vector on the mesh is technically an input coordinate, meaning that one could
take the parameter space to be every degree-of-freedom in the satellite design
imaginable, we do not consider that setup in our analysis.  It is typical to
take the meshes as fixed except in the case of movable features of the
satellite such as the panels attached to the HST.  We have been provided with
ten sets of meshes for HST, from panel angle zero to 90 degrees, in ten
degree increments, which effectively creates an eighth ``free'' input
parameter augmenting the seven in Table \ref{t:inputs}.

We used so-called test particle Monte Carlo (TPMC) method, via a {\sf C}
program developed at Los Alamos National Laboratory, to simulate the
environment encountered by satellites for drag coefficient calculations in
LEO.  TPMC is a computational method used primarily for studying gas dynamics 
in the free molecular regime where particles have mean-free paths which are 
much longer than the characteristic lengthscale of the flow. This means 
that intermolecular collisions can be ignored and TPMC becomes a ray tracing 
technique that treats each particle independently.  In TPMC simulations, the only 
collisions that matter are those that occur with a surface, such as the surface 
of a spacecraft.  For more details and a comparison to alternative methods, see 
Appendix \ref{sec:tpm} and \cite{mehta:etal:2014b}.  TPMC takes in a mesh file, 
settings for the free parameters (Table \ref{t:inputs}), and weights on the chemical 
species, which we usually provide as unit vectors.  TPMC supports several GSIs, 
but we utilize the CLL throughout.  In Appendix \ref{sec:tpm} we describe an 
enhanced implementation leveraging {\tt OpenMP} parallelization, an {\sf R} \citep{cranR} 
wrapper we call {\tt tpm}, and features in {\sf R} for cluster parallelization via the 
built-in {\tt parallel} library.  Timing information on a modern multicore desktop is 
also provided for reference. Source code for this implementation, as well as mesh 
files for a dozen or so satellites including HST, GRACE, and the International 
Space Station, and data files containing runs from \cite{metha:etal:2014} and those 
involved in our suite of experiments in Section \ref{sec:satemu}, are furnished as 
supplementary material.

\subsection{GP review}
\label{sec:gp}

A GP is technically a prior over functions \citep{stein:1999}, with finite-dimensional 
distributions defined by a mean $\mu(x)$ and positive definite
covariance $\Sigma(x, x')$, for $p$-dimensional input(s) $x$ and $x'$. For
regression applications with $N\times p$ design matrix $X_N$ and
$N$-vector of responses $Y_N$, combined as data $D_N = (X_N, Y_N)$, that prior
implies the multivariate normal sampling model $Y_N \sim \mN_N(\mu_N,
\Sigma_N)$. Usually,  a small number of parameters $\theta$ determine $\mu_N$
and $\Sigma_N$. For example, linear regression is a special case where $\theta
= (\beta, \tau^2)$, $\mu_N = X_N \beta$, and $\Sigma_N = \tau^2 I_N$.

Whereas the linear case puts most of the ``modeling'' in the mean, the term GP
emulation is typically used when the emphasis is covariance. When modeling
computer experiments \citep[e.g.,][]{sant:will:notz:2003} often $\mu_N = 0$,
which is a simplifying assumption we make throughout.  Let $K_\theta(x, x')$
be a correlation function so that $Y_N \sim \mN_N(0, \tau^2 K_N)$, where $K_N$
is the positive definite matrix comprised of $K_\theta(x_i, x_j)$ evaluations
on pairs of rows of $X_N$. Here we are changing the notation slightly,
reserving $\theta$ for $K_\theta$ and introducing $\tau^2$ as a separate scale
parameter. Choices of $K_\theta(\cdot,\cdot)$ determine properties of the
input--output relationship including stationarity, smoothness,
differentiability, and the decay of spatial correlation.

A simple first choice is the so-called {\em isotropic Gaussian}: 
$K_\theta(x, x') = \exp\{-\sum_{k=1}^d (x_k - x'_k)^2/\theta\}$, where correlation decays
exponentially fast in the squared distance between $x_k$ and $x_k'$ at rate
$\theta$, called the characteristic {\em lengthscale}.  A more common
generalization is a so-called {\em anisotropic} or {\em separable} Gaussian
which uses a separate lengthscale for each input coordinate: $K_\theta(x, x')
= \exp\{-\sum_{k=1}^d (x_k - x'_k)^2/\theta_k\}$, with vectorized
$\theta=(\theta_1,...,\theta_k)$.
With these choices the sample paths are very
smooth (infinitely differentiable) and the resulting predictor is an
interpolator, which is appropriate for many deterministic computer
experiments.  For smoothing and numerical stability, a {\em nugget} can be
added to $K_{\theta, \eta}(x, x') = K_\theta(x, x') + \eta
\mathbb{I}_{\{x=x'\}}$.\footnote{Throughout we fix $\eta$ to the
smallest value that leads to stable matrix decompositions in our empirical
examples---detailed later.} 
When a large degree of smoothness is not appropriate, the Mat\'ern family of
correlation functions is a common choice \citep{stein:1999}, however the
software libraries we augment here only support the Gaussian
correlation family at this time. Technical aspects of
our presentation below are generic to choices of $K(\cdot, \cdot)$, excepting
that it be differentiable in all parameters.  There are, however, some
important computational considerations to do with inference for the unknown
parameters which will require some care in the local neighborhood(s) context
of Section \ref{sec:laGP}.

GP emulation is popular because inference for $(\theta,\eta)$ is
theoretically easy, and prediction is highly accurate and,
conditionally on $(\theta,\eta)$, analytic.  Under the reference prior
$\pi(\tau^2) \propto 1/\tau^2$, \citet{berger:deiliveira:sanso:2001} provides
a marginal likelihood for the remaining unknowns as\footnote{Eq.~(\ref{eq:gpk})
drops $\eta$ to streamline notation.
Everything applies to $K_{\theta, \eta}(\cdot, \cdot)$ as well.}
\begin{equation}
p(Y_N|K_\theta(\cdot, \cdot)) = \Gamma[N/2] (2\pi)^{-N/2}|K_N|^{-1/2} \times
\left(\psi_N/2\right)^{\!-\frac{N}{2}},
\label{eq:gpk}
\;\;\;\;\; \mbox{where} \;\;\;
\psi_N = Y_N^\top K_N^{-1} Y_N.
\end{equation}
Derivatives are available analytically and lead to more numerically stable (if
not always computationally efficient) Newton-like schemes for maximizing
to obtain Maximum Likelihood Estimation (MLE) lengthscale estimates, $\hat{\theta}$, which is our
preferred method of inference throughout, unless otherwise noted.
The predictive distribution $p(y(x) | D_N, K_\theta(\cdot, \cdot))$, is
Student-$t$ with degrees of freedom $N$:
\begin{align} 
  \mbox{mean:} && \mu(x|D_N, K_\theta(\cdot, \cdot)) &= k_N^\top(x)  K_N^{-1}Y_N,
\label{eq:predgp} \\ 
\mbox{scale:} && 
 \sigma^2(x|D_N, K_\theta(\cdot, \cdot)) &= \textstyle{\frac{1}{N}}   
\psi_N [K_\theta(x, x) - k_N^\top(x)K_N^{-1} k_N(x)],
\label{eq:preds2}
\end{align}
where $k_N(x)$ is the $N$-vector whose $i^{\mbox{\tiny th}}$ component is
$K_\theta(x,x_i)$. The variance of $Y(x)$ is $V_N(x) \equiv
\sigma^2(x|D_N,K_\theta(\cdot, \cdot))\times N/(N - 2)$. Its form will be 
important later when we discuss local approximation (Section \ref{sec:laGP}).
Qualitatively,  $V_N(x)$ is small/zero for $x$ being in the design $X_N$ and
increasing in a quadratic fashion when moving $x$ away from elements of $X_N$. 

\subsection{Proof-of-concept GP emulation of satellite drag}
\label{sec:poc}
The reason we are careful to say ``if
not always computationally efficient'' above is because the trouble with all
this is $K_N^{-1}$ and $|K_N|$, appearing in several instances in
Eqs.~(\ref{eq:gpk}--\ref{eq:preds2}), usually requiring $\mathcal{O}(N^3)$
computation and $\mathcal{O}(N^2)$ storage for dense matrices. That limits
data size to $N \approx 5000$ in reasonable time using optimized linear
algebra libraries such as the Intel MKL. This limitation led
\citet{metha:etal:2014} to consider $N=1000$, separately for each of the six
pure chemical species, in their Bayesian Markov Chain Monte Carlo (MCMC) setup---posterior sampling
requiring orders of magnitude more matrix inversions compared to
derivative-based MLE optimization, which implied further limitations on $N$.
An $N=1000$-sized space-filling design in the full input space, described in
Table \ref{t:inputs}, leads to an underwhelming predictor.  In our own
empirical work in Section \ref{sec:satemu}, we show that relative
out-of-sample predictive error is upwards of 15\%, far off the typical 1\%
benchmark.

This led \citet{metha:etal:2014} towards a more local analysis, focusing
on a limited range of the orientation angles in Table \ref{t:inputs}. 
\begin{table}[ht!]
\centering
\begin{tabular}{lllccr}
Symbol & ASCII & Parameter & Ideal Range & Reduced Range & Percentage \\
\hline
$\theta$ & {\tt theta} & yaw & $[-\pi, \pi]$ & [-0.052313, 0.052342] & 1.7\% \\
$\phi$ & {\tt phi} & pitch & $[-\pi/2, \pi/2]$ & [1.059e-05, 5.232e-02] & 1.7\% 
\end{tabular}
\caption{Reduced ranges for the orientation inputs first described in Table \ref{t:inputs}.}
\label{t:reduced}
\end{table}
Table \ref{t:reduced} shows the reduced ranges for these two variables.  The
combined omission of 98.3\% for each of two variables results in an input
space sized to less than 0.3\% of its original volume---even with the other
variables continuing to span their entire range. For GRACE, these new limits
are reasonable considering that this satellite is attitude stabilized. That
range spans actual observed angles over a several-year period.  However, other
angles may be encountered in the future.  For HST, this range is extremely
limiting, as the satellite frequently rotates when performing different
imaging experiments. Appendix \ref{sec:tpm} provides a coded example showing
that a space-filling design in this reduced domain (data files are provided as
supplementary material) leads to an out-of-sample relative accuracy of 0.73\%
for GRACE in pure He, beating the 1\% benchmark.  The results are similar,
with nearly identical code, for the other pure chemical species and for HST.

Although these results are promising, there are some obstacles to scaling
things up to the full domain, particularly the issue of GPs with large $N$.
One option is a patchwork of little-$n=1000$ analyses, but that would require
upwards of $N=4$ million runs for each chemical species and would result in a
global predictive surface that had hard breaks between the regions.  That
could prove problematic when trying to predict jointly for an orbital trajectory through the input
space spanning the disjoint regions, i.e., spanning more than 1.7\% of either
of the orientation angles.  However, the spirit of local analysis has merits.

\subsection{Local approximate Gaussian processes}
\label{sec:laGP}

The limitation of GP modeling with moderate/large $N$ is not an exactly
new problem in the literature.  However as data sets have become ever larger
over time, research into remedies, via approximation, has become ever more
frenzied. Inducing sparsity in the covariance structure is a recurring theme.
For a review, see \cite{gramacy:apley:2015} which describes the foundation of
the methodology we prefer for the satellite drag problem as its emphasis is on
local sub-problems.  Specifically, it is faithful to the spirit of
\citet{metha:etal:2014}'s data-subsetting approach but without hard
breaks. Both can also be seen as leveraging sparsity, but in fact only work
with small dense matrices. Both also involve statistically and algorithmically
independent calculations, and thus facilitate massive parallelization. They
differ, however, in how they define the local data subsets.  Whereas
\citet{metha:etal:2014} would divvy up the space in terms of the
orientation inputs, \citeauthor{gramacy:apley:2015} offer a more fluid notion
of locale.

The idea centers around deriving {\em approximate} predictive equations at
particular generic location(s), $x$, via a subset of the data $D_n(x)
\subseteq D_N$, where the sub-design $X_n(x)$ is (primarily) comprised of
$X_N$ close to $x$. The reason is that, with the
typical choices of $K_\theta(x, x')$, where correlation with elements $x' \in
X_N$ decays quickly for $x'$ far from $x$, remote $x'$ have vanishingly small
influence on prediction.  Ignoring them in order to work with much smaller,
$n\times n$, matrices brings big computational savings
with little impact on accuracy. Although an exhaustive search for the best $n$
out of $N$ things is combinatorially huge, \citeauthor{gramacy:apley:2015}
show how a greedy search can provide designs $X_n(x)$ where predictors based
on $D_n(x)$ furnish local predictors which are at least as accurate as simpler
alternatives (such as nearest neighbor (NN)) yet require no extra computational
time.

Their greedy search starts with a small nearest neighbor set $D_{n_0}(x)$, and
then, successively chooses $x_{j+1} \in X_N \setminus X_j(x)$ to
augment $X_j(x)$ and form $D_{j+1}(x)$ according to one of several simple
objective criteria. The criterion they prefer, approximately minimizing an
estimate of mean-square prediction error, involves choosing $x_{j+1}$ to
maximize the {\em reduction} in variance at $x$:
\begin{align}
 v_j(x&; \theta)  - v_{j+1}(x; \theta), \quad \mbox{(dropping $\theta$
below)} \label{eq:dxy} \\ 
&= k_j^\top(x) G_j(x_{j+1}) v_j(x_{j+1}) k_j(x) +
2k_j^\top(x) g_j(x_{j+1}) K(x_{j+1},x) + K(x_{j+1},x)^2 / v_j(x_{j+1}),
\nonumber
\end{align}
where $G_j(x_{j+1}) \equiv g_j(x_{j+1}) g_j^\top(x_{j+1})$, and $g_j(x_{j+1}) = -K_j^{-1}
k_j(x_{j+1})/v_j(x_{j+1})$. 
To recognize a similar {\em global} design criterion called {\em active
learning Cohn} \citep{cohn:1996}, \citeauthor{gramacy:apley:2015} called
this criterion ALC. They demonstrate that building local designs in this way
with small $n$, say $n=50$, takes fractions of a second even with $N$ in the
tens of thousands, say $N=4\times 10^4$, and the predictions are highly accurate.
Inverting a $(4 \times 10^4)
\times (4 \times 10^4)$ matrix is not possible on most modern desktops because of memory
swapping issues.  For a point of reference, inverting a $(4 \times 10^3)
\times (4 \times 10^3)$
matrix on a modern desktop takes about five seconds.  

The $\mathcal{O}(n^3)$ calculations above are independent for each $x$, so
predicting over a dense grid of locations $x \in \mathcal{X}$ is trivially
parallelizable. Shared memory parallelization, exploiting now-standard
multi-core desktop architectures, proceeds in {\sf C} via one additional line
of code: \verb!pragma omp parallel for! placed above the serial {\tt for} over
$x \in \mathcal{X}$.  That statistical independence can also be leveraged to
obtain a nonstationary effect by independently fitting local lengthscales to
the local designs at each $x$.  Global predictors so-derived tend to be more
accurate than local ones sharing a common lengthscale, and better than global
full-GP ones when those can feasibly be calculated.  A comprehensive suite of
benchmarking results on {\tt laGP} variations can be found in the literature
\citep[e.g.,][]{gramacy:apley:2015,gramacy:niemi:weiss:2014,gramacy:haaland:2015,gramacy:jss:2016}.

We note two caveats here, which are important for our applied work. One is
that although \citeauthor{gramacy:apley:2015} developed their theory generic
to the correlation structure, in fact, their initial implementation in the {\tt
laGP} package \cite{laGP,gramacy:jss:2016} only supported an isotropic version
owing to the peculiarities of {\sf R}'s internal optimization subroutines.
Those routines use static {\sf C} variables which are not thread-safe, meaning
that separable covariance MLE calculations could not (easily) occur in
parallel. Using somewhat elaborate {\tt OpenMP} pragmas we were able to
largely circumvent this issue with an update to {\tt laGP}. As we illustrate
in Sections \ref{sec:bench}--\ref{sec:satemu}, these enhancements are
essential for good performance in many benchmark problems and for our
satellite drag application, in particular.  We view this as a substantial
technological innovation if not a traditionally methodological one.  The other
caveat is that building up local designs $X_n(x)$ requires iterating over all
$N-j$ remaining candidates in $X_N \setminus X_j(x)$ in search of each
$x_{j+1}$. This can be quite expensive when $N$ is big.
\citeauthor{gramacy:apley:2015} suggest limiting the search to a set of $N'$
nearest neighbors, where $n \ll N' \ll N$.  However this is somewhat
unsatisfying, and is particularly disadvantageous in the context of our
proposed trajectory-wise extensions to local design search in Section
\ref{sec:joint}.

\section{Multiresolution global/local GP emulation}
\label{sec:mrlaGP}

As we illustrate in our drag emulation empirical work in Section
\ref{sec:satemu}, {\tt laGP} is able to accommodate much larger data sets and
thereby drive down relative out-of-sample prediction error substantially
compared to the 10-15\% global GPs can accomplish based on N in the small
thousands.  The technological ability to circumvent the thread-safety issue in
order to efficiently fit local anisotropic covariance structures (in parallel)
leads to substantial improvements over the existing locally isotropic
capability. However, even with N as big as 2 million runs, its predictions
are unable to meet the 1\% benchmark error rate, being at best upwards of
2.5\%. Generating even larger data sets is an option, potentially requiring
a doubling of the simulation effort which already stands at seventy thousand
core hours.  Here we outline a simple hierarchical approach to global/local
emulation that offers big improvements on accuracy without larger design,
ultimately substantially beating the 1\% benchmark.

The idea involves recognizing that a sparsity-inducing correlation structure
biases fits towards local effects. This is well known in the literature, and
remedies abound.  For example, \citet{kaufman:etal:2012} compensate by
utilizing a low degree Legendre polynomial basis in the mean structure of the
GP.  Unfortunately this idea is not easily extended to {\tt laGP} since joint
(and fully Bayesian) inference for mean and residual spatial structure would
break the independent, and thus easily parallelized, nature of the
calculations (and MCMC would prohibitively limit data sizes).  In any case, as
we show in Appendix \ref{sec:csc} on much more modestly-sized examples, that
approach is uncompetitive on accuracy grounds (ignoring computation) relative
to what we propose here, even though there are many similarities, at least in
spirit.  The basic idea is to first fit a global GP, rather than a
basis-expanded mean, and use that fit to set up a more appropriate (and
primed) local (sparsity-inducing) prediction problem. We show in several
variations [Appendix \ref{sec:csc}, Sections \ref{sec:bench} and
\ref{sec:satemu}] that this leads to highly accurate, and computationally
tractable predictors in a large $N$ setting.  


\subsection{Bootstrapped block Latin hypercube subsamples}
\label{sec:blhs}

The idea is to first obtain global lengthscale estimates, $\hat{\theta}$,
rather than a global predictor, even though our ultimate goal is prediction.
\citet{liu:hung:2015} demonstrate that a global lengthscale for the separable
GP can be efficiently estimated via MLE calculations on manageably-sized data
subsets. The simple option of random subsetting works well in practice, but
does not offer theoretical guarantees. Intuitively, simple random subsampling
fails to ensure that the spectrum of pairwise distances in the subsample
reflects that of the original data set. \citeauthor{liu:hung:2015} show that a
so-called block-bootstrap Latin hypercube subsampling (BLHS) strategy can
yield global lengthscales which are consistent with the full data
MLE under increasing domain asymptotics. Basically, the BLHS guarantees a good mix of short and long
pairwise distances. The method is similar in flavor to so-called {\em
full scale approximation} \citep{sang:huang:2012,zhang:etal:2015}.  It also
has aspects in common with composite likelihood approaches
\citep{varin:etal:2011,gu:berger:2016}.  Yet BLHS is simpler both conceptually
and in implementation, in part because it offers far less---only consistent
lengthscales.  The ingredients are: (1) a block Latin hypercube (LH) subsample routine; (2) a GP
fitting library; (3) a {\tt for} loop.  Items (2) and (3) are easy.  Our {\tt
blhs} in the {\tt laGP} package, following \citeauthor{liu:hung:2015}'s
description provided below, is just 35 lines of {\sf R} code.

A single BLHS in $d$ dimensions may be obtained by dividing
each dimension of the input space equally into $m$ intervals, yielding $m^d$
mutually exclusive hypercubes.  The average number of observations in each
hypercube is $Nm^{-d}$ if there are $N$ samples in the original design. From
each of these hypercubes, $m$ {\em blocks} are randomly selected following the
LH paradigm, i.e., so that only one interval is chosen from each of the $m$
segments. The average number of observations in the subsample, combining the
$m$ randomly selected blocks, is $Nm^{-d+1}$. Ensuring a subsample size $\geq
1$ requires having $m\leq N^{\frac{1}{d-1}}$, thereby linking the parameter
$m$ to computational effort.  Smaller $m$ is preferred so long as GP inference
on data of that size remains tractable.  Since the blocks follow a LH
structure, the resulting sub-design inherits the usual LHS properties, e.g.,
retaining univariate stratification modulo features present in the original,
large $N$, design. The computational cost to estimate lengthscale via
BLHS is $\mathcal{O}({n^{\star}}^3)$, where $n^{\star}=Nm^{-d+1}$.

\begin{figure}[ht!]
\centering
\includegraphics[scale=0.4]{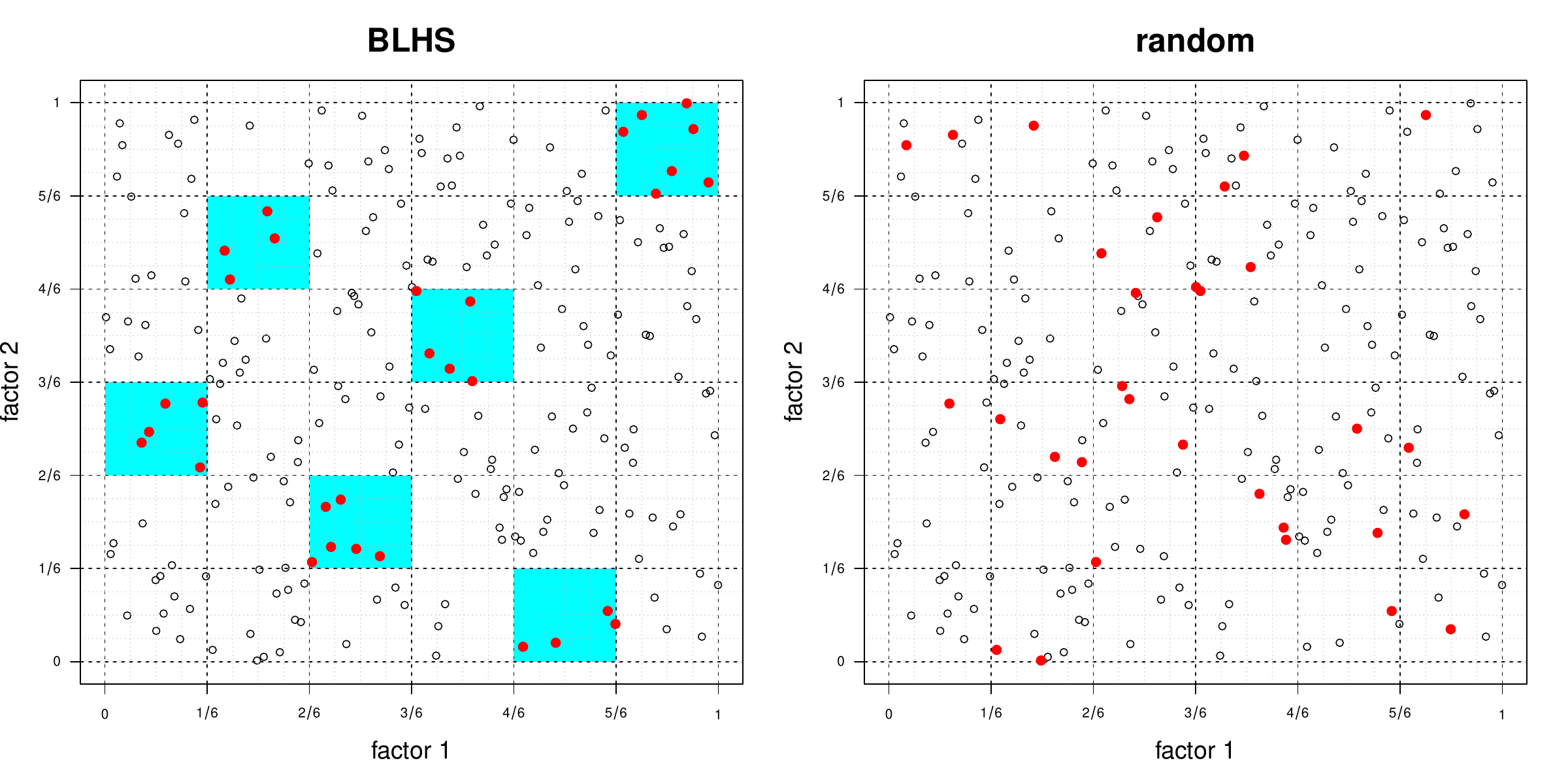}
\vspace{-0.4cm}
\caption{A visual comparison of the BLHS based subsample ({\em
left}) and a random subsample ({\em right}), both derived from an original
sample from a LHS. The original $N$-sized design locations are shown as open
circles, cyan-filled areas are the randomly chosen blocks, and the red-filled dots
indicate the random subsample.}
\label{f:blhs}
\end{figure}

As an example, consider the {\em left} panel of Figure \ref{f:blhs} where we
have $N=216$, $d=2$, and have chosen $m=6$. All together there are $m^d=36$
hypercubes, which are outlined by black dashed lines. The cyan-filled areas
correspond to the selected blocks, and the red dots are the resulting
subsamples. It is clear that the selected samples have desirable space-filling
properties. Observe that---compared say to the random subsample exemplified by
the {\em right} panel in the figure---the BLHS guarantees short and long
pairwise distances.

\begin{figure}[ht!]
\centering
\includegraphics[scale=0.32,trim=0 10 0 15,clip=TRUE]{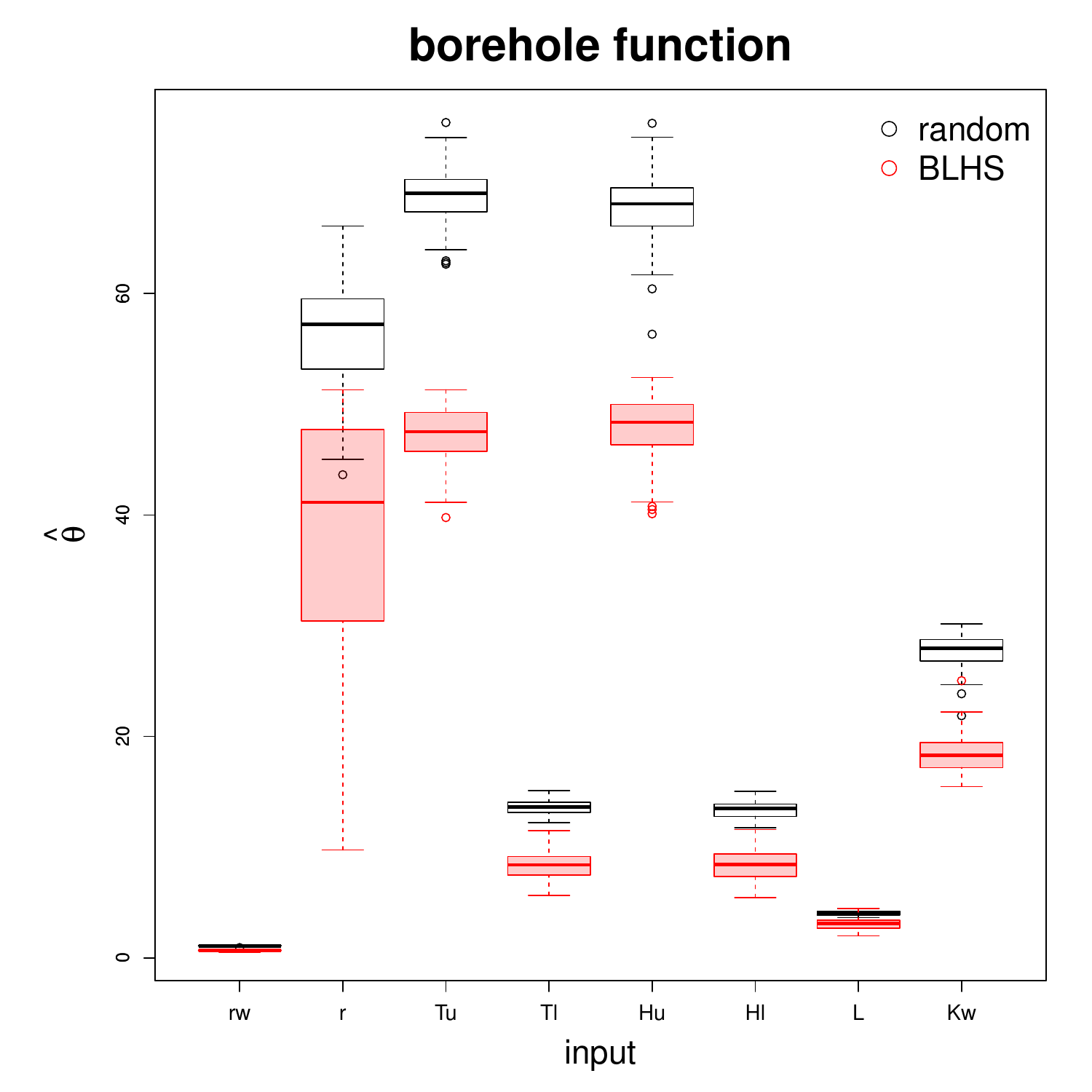}
\includegraphics[scale=0.32,trim=30 10 0 15,clip=TRUE]{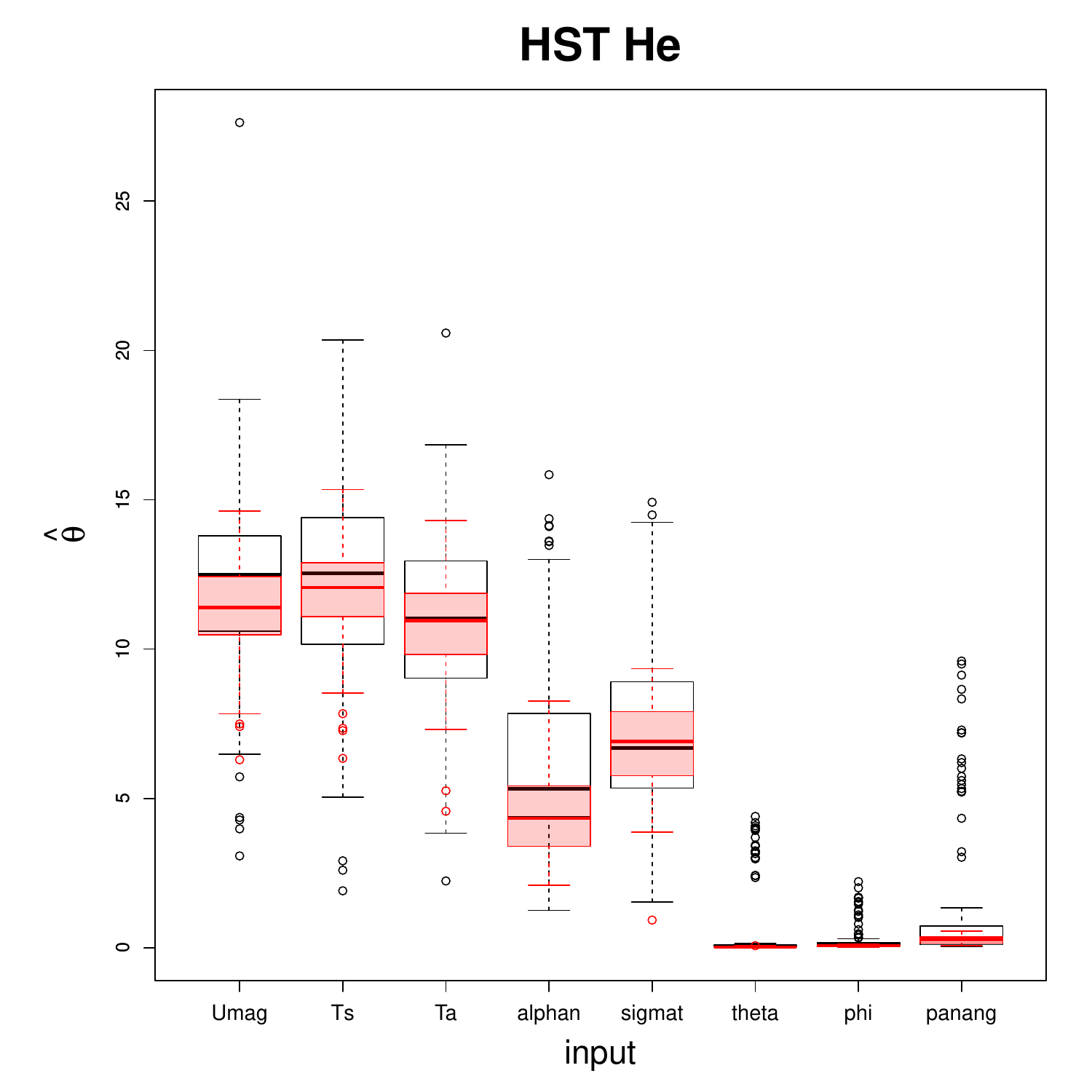}
\caption{Comparing MLE lengthscale parameters obtained via BLHS and random subsampling.  The {\em left} panel corresponds to the
borehole example (see Section \ref{sec:bench}), using $N=10^5$ and
$n=Nm^{-d+1}=781$ with $m=2$ and $d=8$; the {\em right} panel is HST in pure
He with $N=10^6$ and $n=Nm^{-d+1}=457$ with $m=3$ and $d=8$.}
\label{f:blhs_theta}
\end{figure}

Subsample in hand, one can efficiently obtain estimated lengthscale parameters
via typical MLE calculations.  Figure \ref{f:blhs_theta} anticipates some of
our later empirical work by illustrating the bootstrap distribution of
estimated lengthscales for the classic borehole problem ({\em left}) and HST
in pure He ({\em right}) (For more details on the experimental setup, see
Section \ref{sec:borehole} and \ref{sec:epp}, respectively.).
Details such as data sizes $N$ are provided in the figure caption, and $m$ was
chosen as small as possible, yet yielding a subsample with less than $1000$
runs. To ensure a fair comparison, random subsample sizes were chosen to match
their BLHS counterparts. Observe that, in both cases, the lengthscales
estimated under BLHS are consistently smaller than the ones under random
subsampling, with differences being particularly stark for the borehole
example. Intuitively, this is due to the larger degree of short pairwise
distances in the BLHSs.  Theoretically, we know that
the BLHS estimates are asymptotically consistent, and therefore the random
subsample ones must be biased.  Also observe that estimates under random
subsampling for HST have a long right-hand tail.  This suggests that when
extracting point estimates from the random samples, medians are preferred
over means.

\subsection{From global lengthscales to local prediction}
\label{sec:gllp}

Having a computationally tractable method for estimating lengthscales is
important, but not directly useful in many settings where GP surrogates are
typically deployed.\footnote{This may be one reason they are called
hyperparameters rather than full-fledged parameters.}  \citet{liu:hung:2015}
developed BLHS in order to perform variable selection on a parametric
modeling component, which is not really relevant to our emulation problem.
Although lengthscale estimation may have avoided expensive $\mathcal{O}(N^3)$
computation, prediction conditional on those parameters still requires large
matrix decompositions (see, e.g., Eq.~(\ref{eq:predgp})).  

This is where {\tt laGP} comes in. One option would be to condition {\tt laGP}
on these estimated global lengthscales $\hat{\theta}^{(g)}$, rather than an
otherwise arbitrary default value $\theta^{(0)}$. However this is equivalent to scaling the
input coordinates as $x_{ij}/\sqrt{\theta_j^{(g)}}$, for both training and testing 
locations, and initializing with a lengthscale of $1$ for all coordinates. 
We prefer this latter option from an implementation perspective, considering
how {\tt laGP} derives default priors, parameter ranges, and other
MLE-calculating specifications.  Then {\tt laGP} calculations can proceed as
usual.  Inferring local lengthscales, with the search initialized at $1$,
produces a kind of multiresolution effect:  the local design $X_n(x) |
\hat{\theta}^{(g)}$ is determined by the global lengthscale, and subsequent
local estimation of the $\hat{\theta}(x) | D_n(x)$, is determined by
the observations at that local design:  $D_n(x) = \{X_n(x), y(X_n(x))\}$. 

Pre-scaling with suitably estimated global
lengthscales enhances numerical stability and efficiency, by priming the local
optimizer, and tailors local adjustments (to both $D_n(x)$ and
$\hat{\theta}(x)$) to a globally inferred reference $\hat{\theta}^{(g)}$
rather than one-size-fits-all default $\theta^{(0)}$.  Even in the special case of
NN local design, the $X_n(x)$ in the re-scaled space ``feel'' the effect of
the global lengthscale. Back on the original scale of the inputs, the
distances upon which ``nearest'' is judged are no longer Euclidean.  By
down-weighting inputs with longer lengthscales, the effect is akin to utilizing
a Mahalanobis distance \citep{bastos:ohagan:2009} but without the prohibitive
$\mathcal{O}(N^3)$ computational cost that that would entail.


\begin{figure}[ht!]
\centering
\includegraphics[scale=0.35]{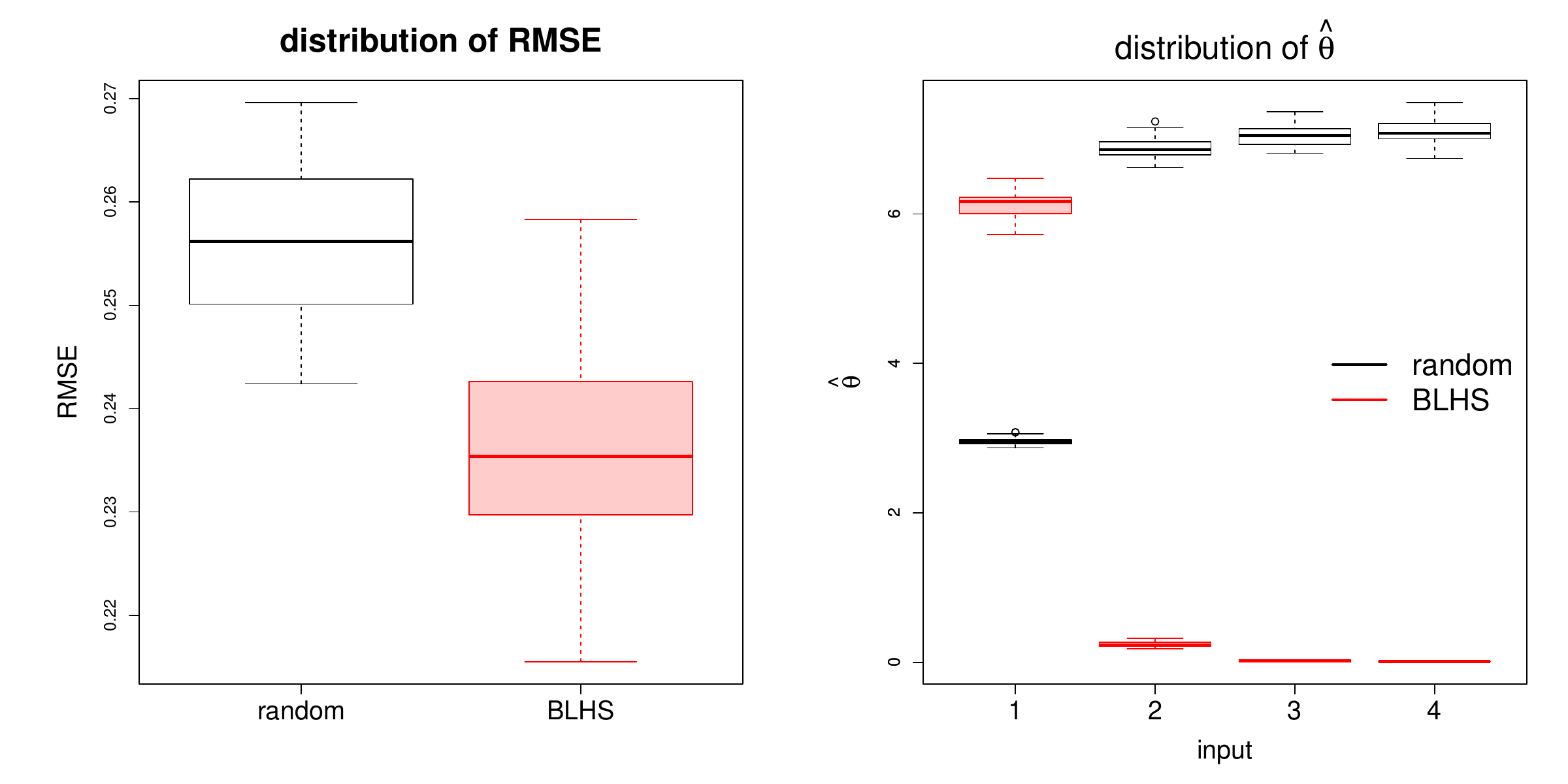}
\vspace{-0.4cm}
\caption{Comparing RMSE ({\em left}) and MLE lengthscales ({\em right}) on
simulated Michalewicz data obtained via BLHS and random subsampling,
respectively, using $N=5\times 10^4$ and $n=Nm^{-d+1}=781$ with $m=4$ and $d=4$.}
\label{f:michal}
\end{figure}

The {\em left} panel of Figure \ref{f:michal} shows a root-mean-squared-error
(RMSE) distribution, and the {\em right} panel shows MLE lengthscales for
$N=5\times 10^4$ LHS, both on the Michalewicz function [Section
\ref{sec:michalewicz}], with $p=4$ and $M=10$.  The RMSEs reported were
obtained by predicting on a testing set of size $1000$, and employing the
global-local hybrid {\tt laGP} described above. Section \ref{sec:michalewicz} offers
more details, and an expanded comparator set for this example. Observe that
the distribution of RMSEs is clearly more favorable for BLHS compared to
random subsampling.  Looking at the MLEs, three of the four inputs exhibit the
same feature as in the borehole data: the random subset ones are biased high.
The first input is the other way around, however. One might speculate that
gross over-estimates for the latter three inputs may have contributed to
overcompensation in the first. 

One of the advantages of the Michalewicz function, over say the
borehole, is that one can easily generate variations based on the input
dimension, $p$. For example, we can empirically study MLE of lengthscale
parameters under $p \in \{2, 4, 6, 8\}$ with a sensible data size counterpart 
$N \in \{10^4, 5\times 10^4, 10^5, 10^6\}$ to acknowledge how volume increases
quickly with dimension. Tables
\ref{t:michal:rmse}--\ref{t:michal:that} summarize the results from just such
an experiment.
\begin{table}[ht!]
  \centering
 \begin{tabular}{c l r r r}
  \toprule
  \multirow{2}{*}{$p$} & 
  \multirow{2}{*}{method} & 
  \multicolumn{3}{c}{quantile}\\ \cmidrule(lr){3-5} 
  && { 5 \%} & { 50 \%} & { 95 \%} \\ \cmidrule(lr){1-5} 
  2 & BLHS & 3.18e-05 & 3.36e-05 & 3.83e-05 \\ 
  2 & random & 3.30e-05 & 3.48e-05 & 3.72e-05 \\ 
  \hline
  4 & BLHS & 0.225 & 0.235 & 0.250 \\ 
  4 & random & 0.246 & 0.256 & 0.266 \\ 
  \hline
  6 & BLHS & 0.477 & 0.494 & 0.512 \\ 
  6 & random & 0.524 & 0.540 & 0.562 \\ 
  \hline
  8 & BLHS & 0.641 & 0.655 & 0.675 \\ 
  8 & random & 0.652 & 0.673 & 0.708 \\ 
  \bottomrule
 \end{tabular}
 \caption{Distribution of RMSEs. Each column, from left to right, shows the dimensionality, 
 subsampling method, and quantile information of RMSEs (5\%, 50\%, and 95\%, respectively). Each row 
 is composed of two sub-rows, each of which shows the quantile distribution using the specific subsampling method 
 under certain dimensionality.}
 \label{t:michal:rmse}
\end{table}
Table \ref{t:michal:rmse} continues where the left panel of Figure
\ref{f:michal} left off, detailing the distribution of RMSEs for varying
dimension $p$ and sample size $N$.   Observe that BLHS consistently
outperforms random subsampling in this exercise, excepting the upper 95\% quantile for $p=2$.
\begin{table}[ht!]
\small
 \centering
 \begin{tabular}{l l r r r r r r r r}
  \toprule
  \multirow{2}{*}{method} & 
  \multirow{2}{*}{quantile} & 
  \multicolumn{8}{c}{lengthscale in each coordinate}\\ \cmidrule(lr){3-10}
  && { 1} & { 2} & { 3} & { 4} & { 5} & { 6} & { 7} & { 8} \\ \cmidrule(lr){1-10} 
  BLHS & 10 \% & 0.0087 & 0.0024 &  &  &  &  &  &  \\ 
  BLHS & 50 \% & 0.0088 & 0.0024 &  &  &  &  &  &  \\ 
  BLHS & 90 \% & 0.0089 & 0.0024 &  &  &  &  &  &  \\ 
  \hline
  random & 10 \% & 0.0107 & 0.0029 &  &  &  &  &  &  \\ 
  random & 50 \% & 0.0108 & 0.0029 &  &  &  &  &  &  \\ 
  random & 90 \% & 0.0108 & 0.0029 &  &  &  &  &  &  \\ 
  \hline
  \hline
  BLHS & 10 \% & 5.8116 & 0.1995 & 0.0207 & 0.0109 &  &  &  &  \\ 
  BLHS & 50 \% & 6.1631 & 0.2299 & 0.0212 & 0.0127 &  &  &  &  \\ 
  BLHS & 90 \% & 6.3919 & 0.2957 & 0.0219 & 0.0142 &  &  &  &  \\ 
  \hline
  random & 10 \% & 2.8905 & 6.7090 & 6.8810 & 6.8940 &  &  &  &  \\ 
  random & 50 \% & 2.9512 & 6.8596 & 7.0491 & 7.0767 &  &  &  &  \\ 
  random & 90 \% & 2.9987 & 7.1045 & 7.2477 & 7.3142 &  &  &  &  \\ 
  \hline
  \hline
  BLHS & 10 \% & 2.7158 & 1.1104 & 4.3722 & 1.4843 & 2.8059 & 4.4307 &  &  \\ 
  BLHS & 50 \% & 2.9419 & 1.3741 & 4.5539 & 1.5865 & 3.0854 & 4.6223 &  &  \\ 
  BLHS & 90 \% & 3.2309 & 1.5153 & 4.6883 & 1.7911 & 3.5958 & 4.8184 &  &  \\ 
  \hline
  random & 10 \% & 0.0457 & 0.0280 & 0.1874 & 0.2970 & 0.3352 & 0.3366 &  &  \\ 
  random & 50 \% & 0.0555 & 0.0342 & 0.2467 & 0.3603 & 0.3972 & 0.4050 &  &  \\ 
  random & 90 \% & 0.0730 & 0.0419 & 0.2907 & 0.4159 & 0.4290 & 0.4526 &  &  \\ 
  \hline
  \hline
  BLHS & 10 \% & 4.6204 & 2.4794 & 5.5265 & 2.8159 & 4.8700 & 5.5874 & 5.8449 & 6.1936 \\ 
  BLHS & 50 \% & 4.8334 & 2.7317 & 5.8004 & 3.1132 & 5.1314 & 5.8291 & 6.0473 & 6.2677 \\ 
  BLHS & 90 \% & 5.0122 & 2.9202 & 5.9439 & 3.3995 & 5.2812 & 6.0269 & 6.1813 & 6.4472 \\ 
  \hline 
  random & 10 \% & 0.0581 & 0.2794 & 0.3801 & 0.4269 & 0.4267 & 0.4641 & 0.4456 & 0.4742 \\ 
  random & 50 \% & 0.0805 & 0.3436 & 0.4429 & 0.4824 & 0.5100 & 0.5230 & 0.5166 & 0.5472 \\ 
  random & 90 \% & 0.0930 & 0.3928 & 0.4823 & 0.5616 & 0.5879 & 0.5996 & 0.5935 & 0.6627 \\ 
  \bottomrule
 \end{tabular}
 \caption{Distribution of $\hat{\theta}$.  Each column, from left to right, shows the subsampling method, 
 quantile information, and $\hat{\theta}$ under different dimensionality. Each row is composed of two big 
 sub-rows, each of which shows the quantile distribution (10\%, 50\%, and 90\%) using the specific subsampling 
 method under certain dimensionality.}
\label{t:michal:that}
\end{table}

Table \ref{t:michal:that} focuses on lengthscales estimated from that
experiment, this time summarizing boxplots numerically (like in the right
panel of Figure \ref{f:michal}) for varying input dimension $p$. Notice
how BLHS estimates differ from those from random
subsampling.  When $p=2$, the BLHS estimates are consistently smaller at each
quantile.  However the pattern is exactly the opposite when 
$p=8$. The pattern for $p=4$ was already discussed above.

\section{Joint path sampling}
\label{sec:joint}

Obtaining predictions over a dense global input ``grid'' $\mathcal{X}$ may be
overkill in many settings. For example, we are rarely interested in drag
coefficients for satellites like the HST positioned anywhere in orbit or in
arbitrary orientation. Rather, it is far more useful to be able to understand
drag at potential nearby locations and configurations in space and time.  A
satellite's trajectory in the input space is limited to $x's$ near its current
configuration, $x_0$ say, since aspects like velocity and orientation can not
be altered instantaneously, and others like temperature and chemical
composition evolve slowly and smoothly in time and space. Our LANL colleagues
desire {\em joint} predictions along paths $\mathcal{W}(x_0)
\subset \mathcal{X}$ in the input space emanating from $x_0$ values. 
Yet {\tt laGP} is an inherently pointwise procedure.  Although it derives its
speed from parallelization, via statistical and computational independence,
that comes at the expense of punting on a joint predictive capability. Here we develop a new
criterion for sub-design local to a set, alongside methods for finding solutions efficiently.

\subsection{An aggregate criterion}
\label{sec:alcw}

Consider a fixed, discrete and finite set of input locations $W \subset \mathcal{X}$, perhaps
specified by the practitioner, or derived from a satellite's current trajectory/flight plan in the motivating drag context.
Let $k_j(w)$ denote 
the $j \times 1$ correlation vector for each $w \in W$.  As described above, 
we target $W \equiv \mathcal{W}(x_0)$ as satellite trajectories, but the
development here is for a generic set $W$ with potential for wider
application.  
The extension of our reduction in variance (ALC) criterion 
$v_j(x) - v_{j+1}(x)$, from Eq.~(\ref{eq:dxy}), is
\begin{align}
&v_j(W)- v_{j+1}(W) = \frac{1}{|W|}\sum_{w \in W} \left\{v_j(w)-v_{j+1}(w)\right\}, \label{eq:vdxy} \\
&=  \frac{1}{|W|}\sum_{w \in W} \left\{ k_j^\top(w)G_j(x_{j+1})v_j(x_{j+1})k_j(w) + 2k_j^\top(w)g_j(x_{j+1})K(x_{j+1}, w) + \frac{K(x_{j+1}, w)^2 }{v_j(x_{j+1})} \right\}. \nonumber
\end{align}
Observe that this quantity is a scalar, measuring the average reduction in
predictive variance over $W$, which we wish to maximize over the
choice of new $x_{j+1}$.  With our trajectory application in mind, we refer to
Eq.~(\ref{eq:vdxy}) interchangeably as a ``joint'' and/or ``path'' ALC criterion.

As an illustration of sequential design under the criterion in
Eq.~(\ref{eq:vdxy}), consider a two-dimensional $\mathcal{X} \in [-2,2]^2$ and $W$
comprised of a path of locations $w(x) = (x-0.75, x^3 + 0.51)$ with $100$
$x$-values ranging uniformly in $[-0.85, 0.45]$.  The response, and other
details, are described in \citet[][Section 3.4]{gramacy:apley:2015}, but are
not particularly germane here.  
\begin{figure}[ht!]
\centering
\includegraphics[scale=0.4,trim=0 10 20 50,clip=TRUE]{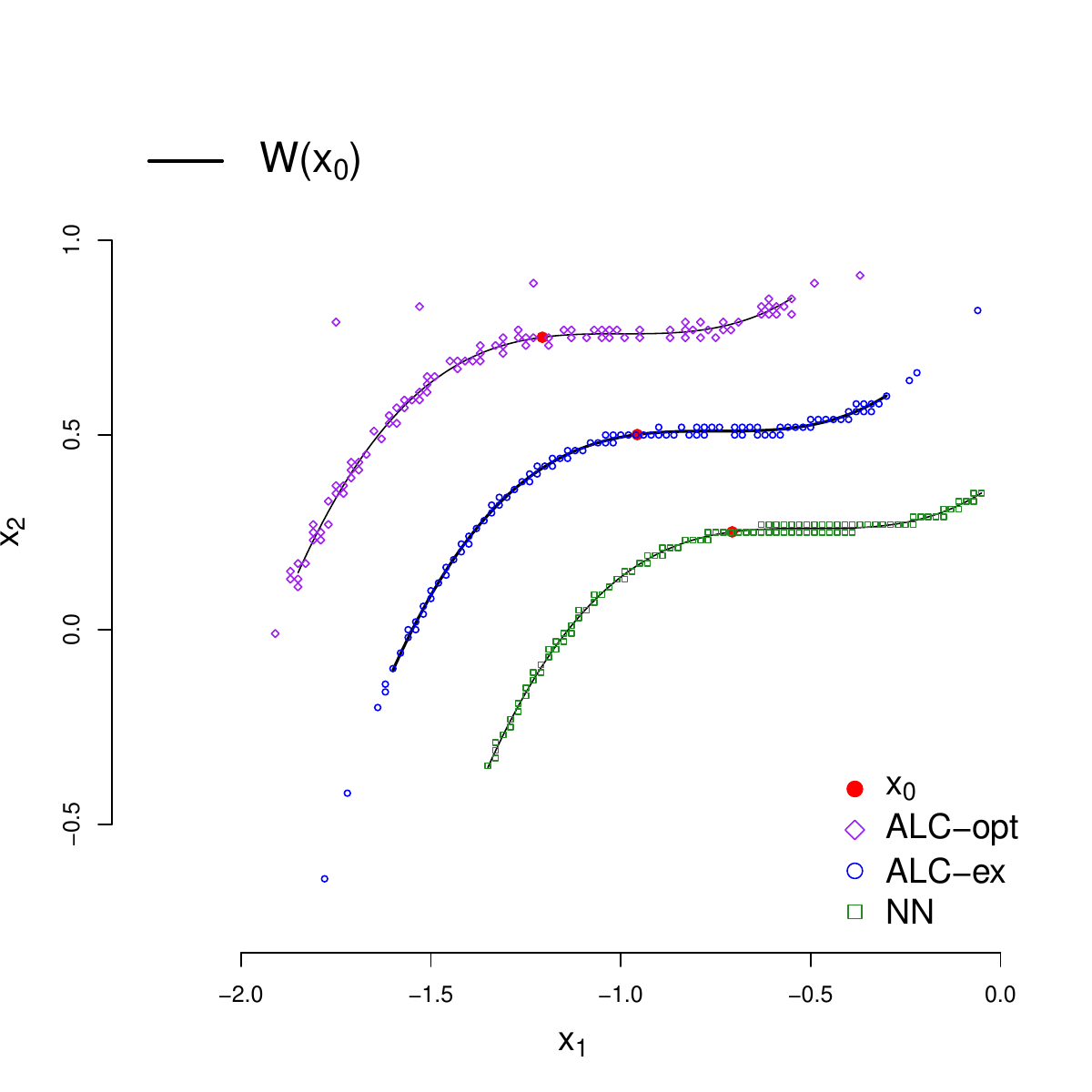}\hspace{0.1cm}
\includegraphics[scale=0.4,trim=0 10 0 50,clip=TRUE]{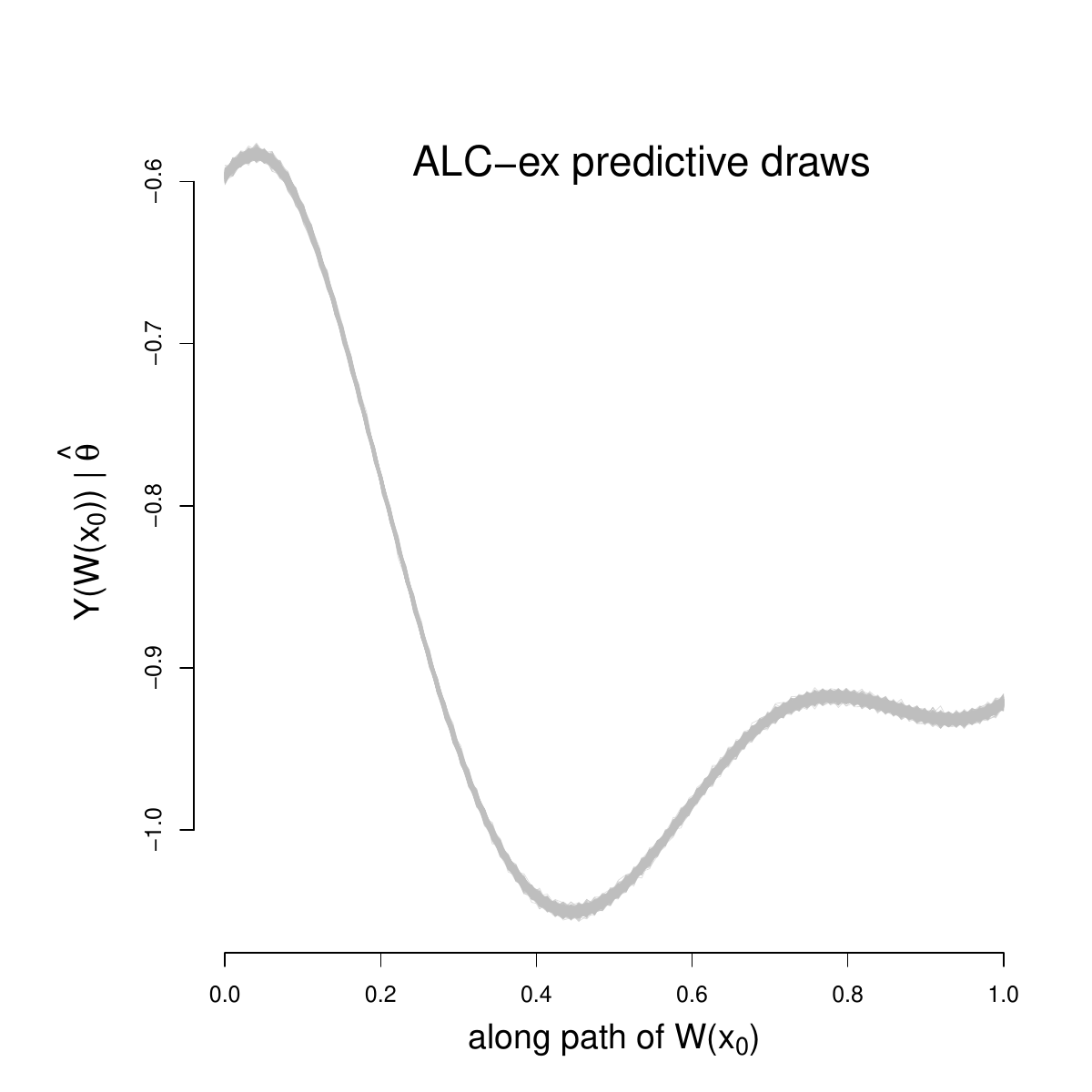}\vspace{-0.4cm}
\caption{
The {\em left} panel shows local designs under three criteria (shifted for
visibility) for identical paths $\mathcal{W}(x_0) \in [-2, 2]^2$,
respectively; the {\em right} panel shows predictive samples along
$\mathcal{W}(x_0)$ corresponding to one of the three designs (the other two
are nearly identical).  The point labeled $x_0$ is the midpoint along the
path.}
\label{f:path_design_opt}
\end{figure}
Figure \ref{f:path_design_opt} shows local designs of size $n=100$ from an
$N \approx 4 \times 10^4$-sized grid, and predictive samples along that path. The {\em
left} panel shows two ``path" ALC-based local designs, i.e., following
Eq.~(\ref{eq:vdxy}), and one NN-based design, with plots shifted for visibility.
The NN sub-design is comprised of elements of the data grid which are closest
in Euclidean distance
to any element of $W$. The sub-design labeled ALC-ex was found in a manner
similar to ordinary pointwise ALC search where every location (limited to the
$N'=10^4$ locations closest to $W$)
is entertained, in turn, at each iteration of the sequential design search, at
substantial computational expense. Observe the clear distinction with the much
thriftier NN-based sub-design:  it has fewer neighbors and more satellite
points. The ALC-opt set will be discussed momentarily.

The {\em right} panel in Figure \ref{f:path_design_opt} shows $1000$
sample paths from the local design along the path under the ``joint" ALC
criterion, illustrating its value over an independent pointwise alternative.
Pointwise sampling could not generate such sample paths, lacking a full
predictive covariance structure. Only one set of samples is shown, with the ones
for the other two methods being strikingly similar.  However, as we show
shortly, the three approaches do not always predict equally well over a
collection of many diverse predictive paths.

An exhaustive search for $X_n(W) \subset X_N$, where $X_n(W)$ denotes
the local design for $W$, following the new path criterion in
Eq.~(\ref{eq:vdxy}) (ALC-ex in the figure) is more computationally expensive
than the pointwise analog $X_n(x)$ (Eq.~\ref{eq:dxy}) for two reasons.  One
reason is that the former makes $m$ ($=100$ in our example) calculations where
the latter makes one. The other is that, relative to a single point $x$,  a
path $W$ traversing a swath of the input space demands inspecting a larger
portion of $X_N$.  That is, a larger searching set $N'$ is needed. Using
the nearest $N'=10^3$ in our example above is too
small, excluding about 10\% of the points selected via ALC(s). Combined,
those multiple orders of magnitude increase in computational expense represent
a heavy burden that we shall quantify shortly.

\subsection{Derivative-based ALC search}
\label{sec:dalc}

Towards lighter-weight sub-design search we considered swapping out exhaustive
enumeration (a discrete search) with a continuous analog utilizing
derivatives.  This is similar in spirit to the ray-based search of
\citet{gramacy:haaland:2015} for pointwise local designs at each $x$.  Rays,
however, would not be suitable for exploring over joint characteristics for
sets $W$.  Although our ultimate usage will be for paths $\mathcal{W}(x_0)$,
we note that these derivatives may also be used for pointwise search, however
the computational savings may not be as impressive as the ray-based
alternative in that setting.

Below we provide the expression for the components of the gradient of
the extended ALC expression in Eq.~(\ref{eq:vdxy}), assuming a separable
covariance structure.  For $\ell=1, \dots, p$
\begin{align}
\frac{\partial}{\partial (x_{j+1})_\ell} &
\left\{v_j(W) - v_{j+1}(W)\right\} \label{eq:dalc} \\
=&\frac{1}{|W|}\sum_{w \in W} \left\{-2k^\top_j(w)K^{-1}_J\left[ 
\frac{\partial k_j (x_{j+1}) }{\partial (x_{j+1})_\ell}  + \frac{ k_j (x_{j+1}) a}{v_j(x_{j+1})} \right]b-abc + 
2 c \frac{\partial K(x_{j+1}, w)}{\partial (x_{j+1})_\ell}\right\}, \nonumber
\end{align}
where the macros $a$, $b$ and $c$ (all scalars) are
\begin{align*}
a &= 2 \frac{\partial k_j^\top (x_{j+1}) }{\partial (x_{j+1})_\ell}K_j^{-1} k_j (x_{j+1}), & 
b &=g_j^\top (x_{j+1})k_j(w) - K(x_{j+1},w)/v_j(x_{j+1}), \\
&\quad \quad \mbox{and}& c & =g_j^\top (x_{j+1})k_j(w) + K(x_{j+1},w)/v_j(x_{j+1}).  
\end{align*}
A detailed derivation is provided in Appendix \ref{sec:dd}. 
Simplifications arise under an isotropic lengthscale, via a single derivative
calculation rather than the $p$ components of a gradient, $\ell=1, \dots, p$.
Any correlation family may be used so long as derivatives may be calculated
relative to the spatial locations, $x$, e.g., for $
\frac{\partial}{\partial (x_{j+1})_\ell} k_j^\top (x_{j+1})$, which is straightforward
for the typical power exponential (Gaussian) and the Mat\'ern families.

Derivative in hand, we turn to numerical optimization using {\tt L-BFGS-B}
\citep{byrd:etal:1995}, a quasi-Newton method implemented in the {\tt optim}
library for {\sf R}. Owing to the linear nature of variance reduction in
Gaussian models as the sample size $n$ increases (particularly with nearby
locations in GP prediction), the magnitude of the {\em reduction} in variance
criterion like Eq.~(\ref{eq:vdxy}) becomes numerically tiny as $n$ grows, easily
matching typical convergence tolerances like $10^{-8}$, even when $n$ as small as
$20$, which can result in premature convergence of the numerical optimizer.
Instead, supplying the logarithm of the criterion as an objective to {\tt
optim}---along with the derivative of the $\log$ which may be calculated as a
ratio of the derivative and the original criterion---leads to better behavior.
The numerator and denominator in ratio comprising the log derivative may
involve small numbers which can result in ``extra'' iterations of the solver
close to convergence.  We have found that these can be avoided by supplying a larger
\verb!pgtol! argument\footnote{The default is {\tt pgtol=0}, however, we
find that {\tt pgtol=0.1} works just as well and can halve the number of
iterations, leading to a much faster search without deleterious effects.}
which controls an aspect of the stopping rule determined by ineffective line
searches in the \verb!L-BFGS-B! algorithm.

Our use of {\tt optim} and the derivative replaces an exhaustive discrete
search with a continuous one, providing a solution $x^{\ast}_{j+1}$ which is
``off'' the candidate set $X_N$.  So the final step involves ``snapping"
$x^{\ast}_{j+1}$ onto $x_{j+1} \in X_N \setminus X_j(W)$, by minimizing
Euclidean distance. It is worth noting that the result is not guaranteed to be
a global minimum of the ALC surface.  A multi-start scheme could be employed,
however, we find that this does not offer a good cost-benefit trade-off in
terms of computational effort.  Rather, we developed an initialization scheme
for the solver which recognizes the sequential nature of our sub-design
search, over the iterations $j=n_0, \dots, n$, via a round-robin on NN similar
to \citet{gramacy:haaland:2015}. Specifically, we maintain a stack of starting
locations from $X_N$ which are sorted via (minimum) distance to $W$.  To start
a search for the next $x_{j+1}$, the next initializer is popped off.
If the $x_{j+1}$ thus found matches any (remaining) stack element, it too is popped off to
prevent it from initializing future searches.  Although
this scheme does not guarantee global optima, for each individual search
or for the final local design $X_n(W)$, it has the advantage of being
deterministic. We find that it leads to local designs that are just as good as
the exhaustive search on average and, most importantly, are better than
NN.

Before turning to empirical results, let's revisit our illustration in
Figure \ref{f:path_design_opt}. The design denoted as ALC-opt in the {\em
left} panel corresponds to the derivative-based search described above.
Observe that this local design is qualitatively similar to the exhaustive one,
having more satellite points and fewer neighbors than the NN version.  In this
instance, ALC-opt's satellite points differ from ALC-ex's, but that is not always the case.

\subsection{Empirical results}
\label{sec:er}

Here we compare out-of-sample performance for {\tt laGP} predictors based on
the following local design schemes: ALC-opt, ALC-ex, NN, the original ALC
criterion (Eq.~(\ref{eq:dxy})) applied pointwise (ALC-pw), and NN-pw. We utilize two
metrics: Mahalanobis distance, $\sqrt{(y - \mu)^\top \Sigma^{-1} (y -
\mu)}$, where $\mu$ is a predictive mean vector with components from
Eq.~(\ref{eq:predgp}), and $\Sigma$ is the predictive covariance matrix whose
diagonal is in Eq.~(\ref{eq:preds2}); and RMSE, which follows the same formula but
with $\Sigma^{-1} =
\mathrm{Diag}(1/N)$.  Although RMSE is more conventional, Mahalanobis distance
is emerging as the gold standard for data arising from deterministic
computer simulation where predictive accuracy and covariance structure are
equally important \citep{bastos:ohagan:2009}.  In both cases, smaller is better.

\begin{figure}[ht!]
\centering
\includegraphics[scale=0.38,trim=0 10 0 50,clip=TRUE]{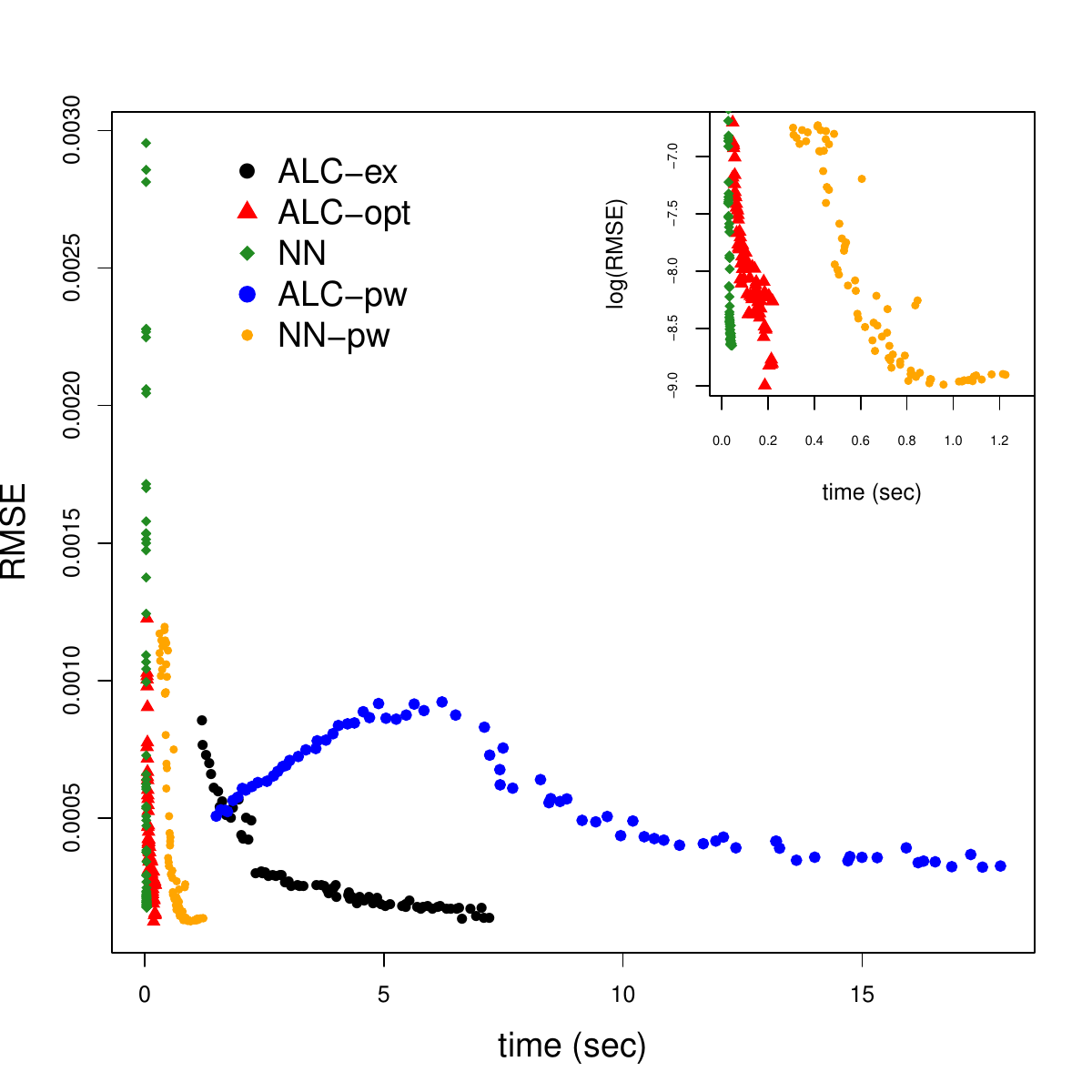} \hspace{0.1cm}
\includegraphics[scale=0.38,trim=0 10 0 50,clip=TRUE]{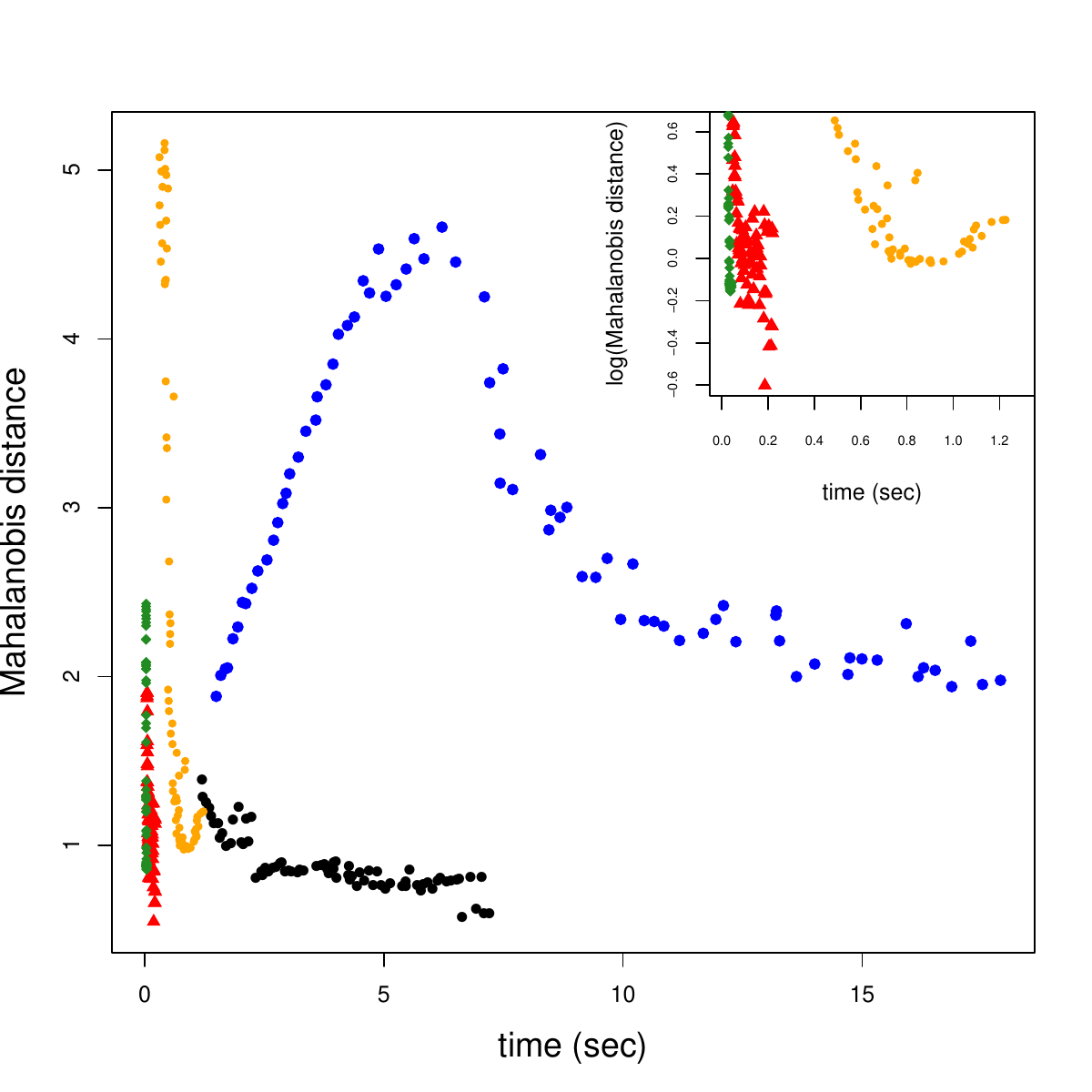} \vspace{-0.26cm}
\caption{Comparing RMSEs {\em (left)} and Mahalanobis distances {\em (right)}
versus time; smaller is better.  Comparators: joint NN, pointwise
NN (NN-pw), joint ALC (`ex'haustive and
derivative-based `opt'imization), and pointwise ALC (ALC-pw).
Zoomed-in glimpses of the lower left of each plot are provided in log space to
better visualize separation.}
\label{f:metric_time}
\end{figure}	

Our first experiment involves the predictive line from Figure
\ref{f:path_design_opt} with local designs of increasing size, $n=25,\dots,
100$, and thus increasing computational effort. Figure \ref{f:metric_time}
shows RMSEs ({\em left}) and Mahalanobis distances ({\em right}) versus
wall-clock time. Excepting ALC-pw and NN-pw, observe how the comparators'
accuracy tends to improve with added time.   Also notice that NN and ALC-opt
are much faster than the other comparators: it takes relatively little
additional computational effort to grow the design and improve out-of-sample
performance. Achieving the same gains via exhaustive search, however, brings
seriously diminishing returns. Otherwise, i.e., ignoring time, the ALC-ex and
ALC-opt searches give about the same out-of-sample performance for fixed local
design size. The strikingly poor performance of ALC-pw and NN-pw can be
attributed to a number of factors. The most important factor may be the (statistical)
independence of calculations involving inherent estimation risk which is
mitigated by the joint designs.  These two pointwise design schemes perform
exceedingly poorly via Mahalanobis distances since a zero off-diagonal
covariance is a gross underestimate of the true underlying structure. 

\begin{figure}[ht!]
\centering
\includegraphics[scale=0.48]{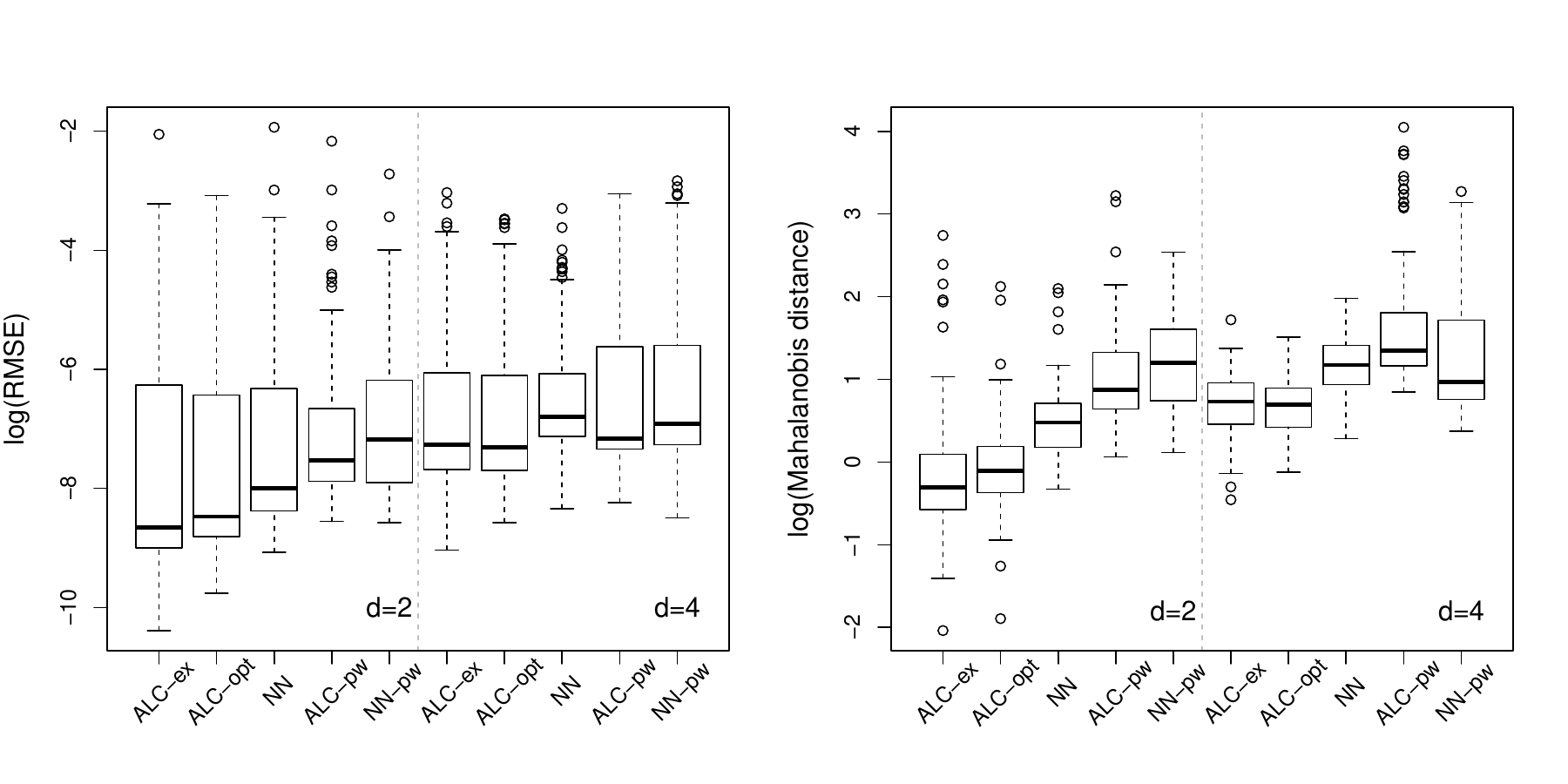}
\vspace{-0.4cm}
\caption{Comparing RMSEs and Mahalanobis distances among ALC-ex,
ALC-opt, NN, ALC-pw, and NN-pw. The {\em left} panel shows boxplots of log RMSEs
based on $100$ random paths; the {\em right} panel shows boxplots of log
Mahalanobis distances.}
\label{f:metric_compare}
\end{figure}	

Our second experiment holds the local design size fixed (at $n=100$) and
varies the two-dimensional lines comprising the set(s), $\mathcal{W}(x_0)$,
in order to more completely span the input space and the space of sets, with
the satellite trajectory analogy in mind.  Appendix \ref{sec:rpp} describes
how the sets were randomly generated, and shows what they look like (Figure
\ref{f:rpp}).  We consider the distribution of RMSEs and Mahalanobis distances
obtained from $100$ predictions on $100$ such lines separately in two and four
input dimensions.  In the latter four-dimensional (4d) case, the two input dimensions on which
the line $\mathcal{W}(x_0)$ was varied were chosen at random, and the output
is derived as the product of two independent two-dimensional (2d) ones.  Figure
\ref{f:metric_compare} summarizes the results in terms of log RMSE and
log Mahalanobis distance. A vertical dashed line separates 2d and 4d experiments.
Observe that these results are largely in line with the previous experiment:
the joint ALC methods (ALC-ex and ALC-opt) beat the pointwise methods (ALC-pw
and NN-pw) to a significant extent.  The pattern is more distinct at the lower
end of the ranges, with outliers at the top end arising from predictive paths
with a substantial component outside the range of the training data, requiring
extrapolation. 

\section{Benchmark and synthetic results}
\label{sec:bench}

Here we focus on pointwise prediction on two benchmark data sets from the
Virtual Library of Simulation Experiments \citep{simulationlib}, anticipating
similar experiments on satellite drag in Section \ref{sec:satemu}.  Eighteen
comparator variations are trained on $N=10^5$ and $N=10^6$ runs, paired with
testing sets of size $10^4$. Sixty such training and testing pairs are
generated jointly via LHS, to create a thirty-replicate Monte Carlo experiment
comprising a total of 36 variations. The comparators combine a choice of local
sub-design (one of ``nn'', ``alc'' and ``alc2'', with the latter being a
second stage of ALC design as recommended by \citet{gramacy:apley:2015}),
paired with a choice of correlation structure (either isotropic or
separable---``sep''), and a choice of global pre-processing of lengthscale as
described in Section \ref{sec:blhs} (either none, random subset---``s'', or a
BLHS subset---``sb'').

Although many aspects of the two experiments are similar, variations in the
input dimension necessitated some customizations. Our first experiment, on the
borehole function (more details in Section \ref{sec:borehole}), is in eight dimensions.
To have manageable BLHS sizes we chose $m=2$ for the
$N=10^5$ version, leading to a global subset of size $n=781$; for
$N=10^6$ we used $m=3$ yielding $n=457$.  Choosing $m=2$ for this latter
case would have produced $n=7812$, too big for fast GP inference.
Similarly, for the Michalewicz function (see Section \ref{sec:michalewicz}),
in four dimensions we chose $m=12$ ($n=578$) for $N=10^6$, and $m=6$
($n=462$) for $N=10^5$.

\begin{figure}[ht!]
\centering
\includegraphics[scale=0.5]{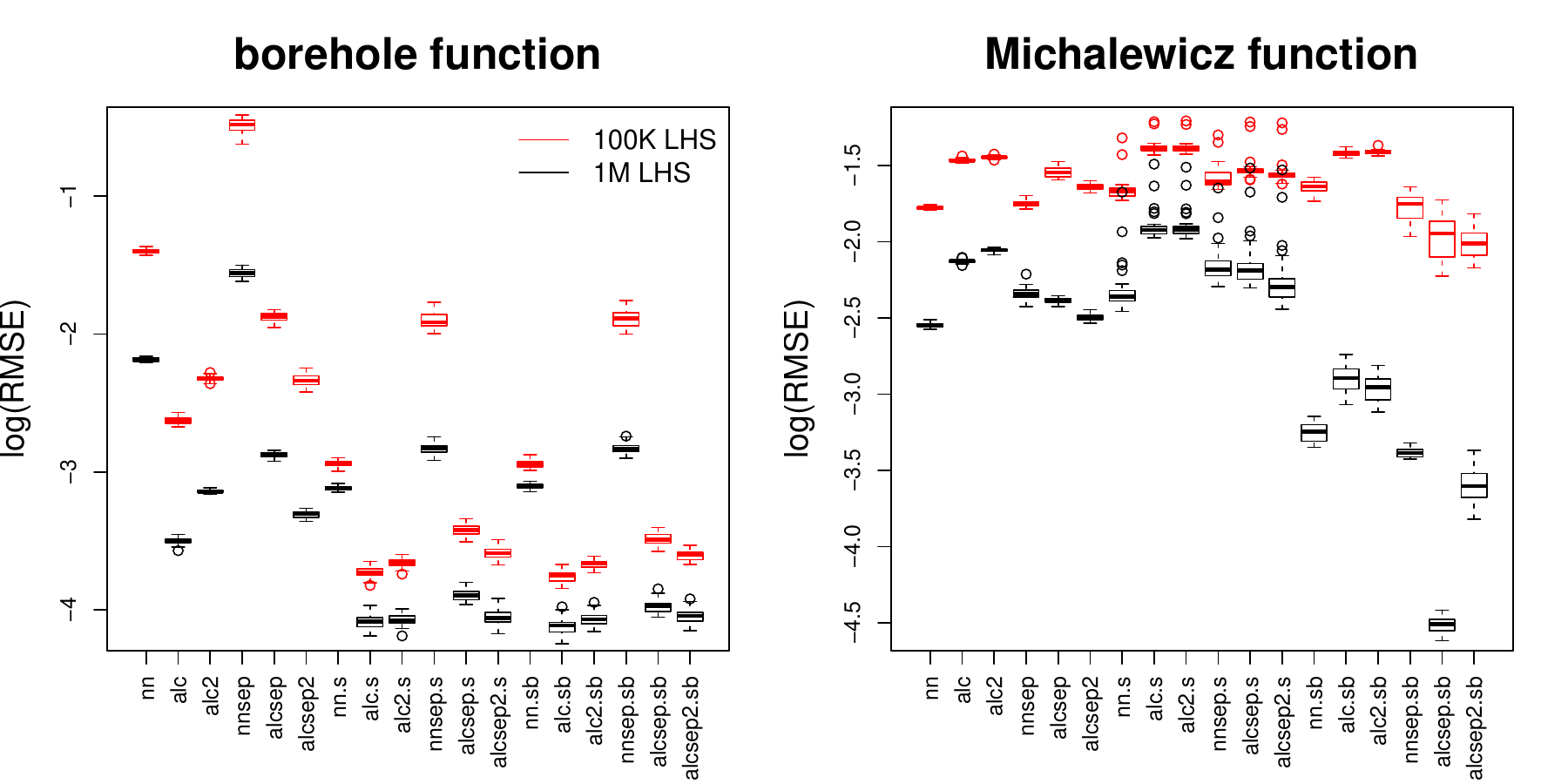}
\vspace{-0.25cm}
\caption{Comparing {\tt laGP} variations on  
borehole ({\em left}) and Michalewicz ({\em right}) functions.}
\label{f:synthetic_compare}
\end{figure}

Figure \ref{f:synthetic_compare} contains boxplots summarizing out-of-sample
RMSE for both experiments, with discussion following shortly. The comparators
along the $x$-axis are blocked in groups of six from left to right, first
without global scaling, and then random or BLHS subsets.  Black boxplots
correspond to the larger experiment ($N=10^6$), which are uniformly lower than
the red ones: more training data means more accurate predictions.

\subsection{Borehole experiment}
\label{sec:borehole}

The borehole function is a classical example from the computer experiments
literature \citep{morris:mitchell:ylvisaker:1993} with inputs in eight dimensions.  The
response $y$ is given by
$$
y=\frac{2\pi T_u\left[H_u-H_l \right]}{
\log\left(\frac{r}{r_w}\right)\left[1+\frac{2LT_u}{\log\left(r/r_w\right)r^2_wK_w}+\frac{T_u}{T_l} \right]}
$$  
where the inputs are constrained to lie in a rectangular domain:
\begin{align*}
r_w\in [0.05, 0.15],\qquad  r\in[100, 5000],\qquad  T_u\in [63070, 115600],\qquad  T_l\in[63.1, 116] \\
H_u\in [990, 1110],\qquad  H_l\in[700, 820],\qquad  L\in[1120, 1680],\qquad  K_w\in[9855, 12045].
\end{align*}
Before any modeling, we scale the inputs to lie in the unit cube.  

The distribution of $\log(\mathrm{RMSE})$ obtained for our comparators on this
borehole data is shown in the {\em left} panel of Figure
\ref{f:synthetic_compare}. There are several noteworthy results.  First,
global pre-scaling (whether by random subsets or BLHS, shown by the latter
$12$ boxplot pairs) outperform their counterparts without scaling (first six).
Second, BLHS-based subsets (latter six) slightly outperform random subsets
(middle six).  This result is consistent across all local models, but the
discrepancy is marginal. Third, observe that NN-based local modeling is poor,
with separable correlation kernels being surprisingly poor in each
pre-processing regime. The overall best comparator is ``alc.sb'' which
corresponds to BLHS-based pre-scaling and a single application of ALC-based
isotropic modeling. Apparently, global estimates of lengthscale are sufficient
to capture local scaling.
Finally, observe that a second stage of ALC design (comparators with a ``2'')
is not consistently better than the first stage alone.  

A referee requested further experimentation in order to compare with other
libraries, primarily in a smaller-$N$ setting.  Those results and discussion
may be found in Appendix \ref{sec:rgasp}.

\subsection{Michalewicz experiment}
\label{sec:michalewicz}

\begin{figure}[ht!]
\centering
\includegraphics[scale=0.40,trim=0 15 15 55]{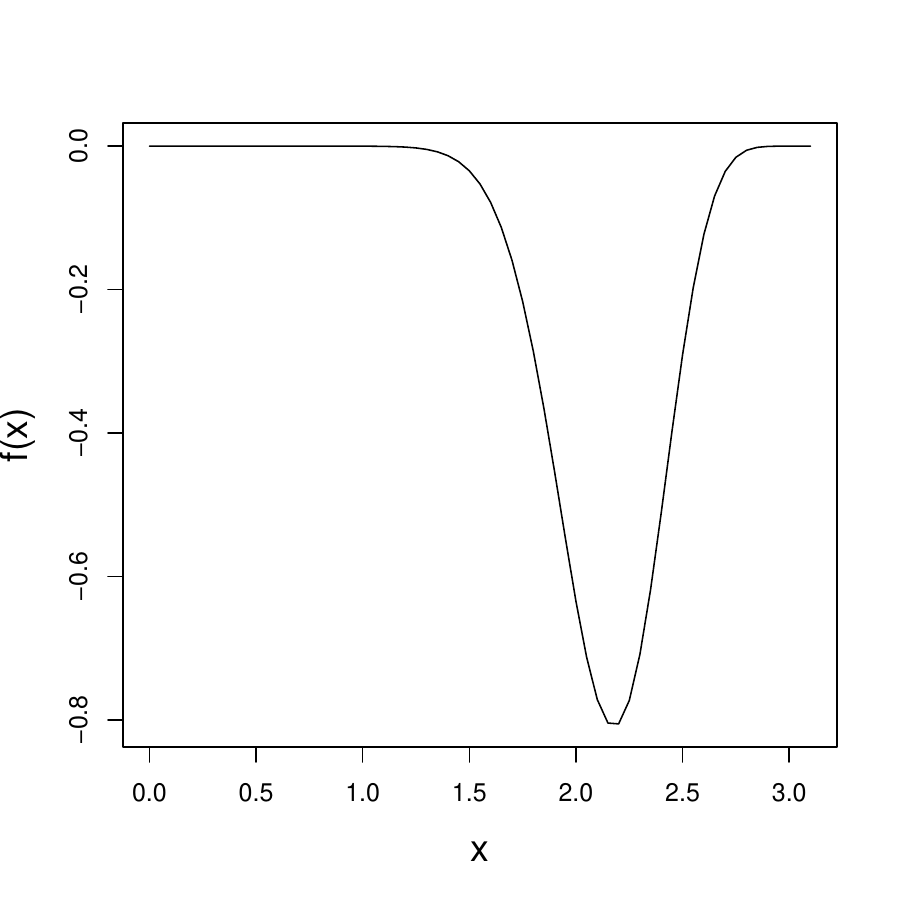} \hspace{1cm}
\includegraphics[scale=0.35,trim=75 60 40 55]{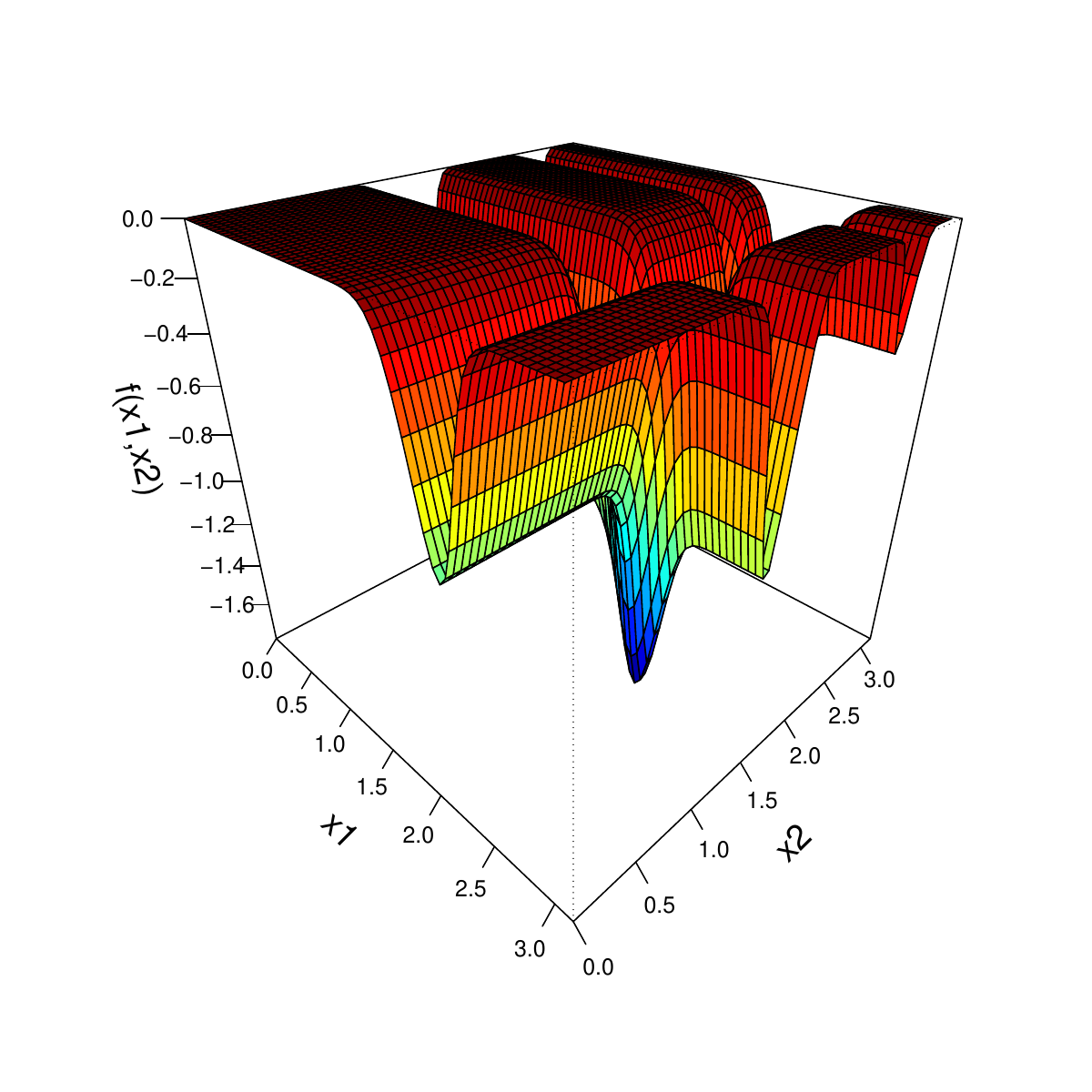}
\caption{The Michalewicz function. The {\em left} panel shows the surface with $(p=1, M=4)$, whereas
the {\em right} panel corresponds to $(p=2, M=10)$.}
\label{f:michale_visual}
\end{figure}

The Michalewicz function is widely used in the optimization literature
\citep{molga:smutnicki:2005}.  One unique feature is that it can be defined
in arbitrary input dimension, $p$:
$$
f(x)=-\sum_{i=1}^{p}\sin(x_i)\sin^{2M}\left(\frac{ix_i^2}{\pi}\right), 
\quad \mbox{ where } x \in \mbox{[0, $\pi$]}.
$$
Larger values of the parameter $M > 0$ yield steeper transitions in the output
space.  Figure \ref{f:michale_visual} shows two examples where $(p=1,M=4)$ and
$(p=2,M=10)$. There are $p!$ local minima, so the response surface is quite
complex under high dimensionality, especially when $M$ is also large.  In the
experiment we report here, we used $p=4$ and $M=10$.

The {\em right} panel of Figure \ref{f:synthetic_compare} shows the
distribution of $\log(RMSE)$ obtained in that setting. Although the results bear
some similarity to the borehole experiment, there are some nuances worth
attention. First, a general global pre-scaling is no longer as powerful as
before.  For example, consider the RMSEs for comparators whose names ended
with ``s'', which are based on random subsets.  These underperform their
counterparts without scaling.  By contrast, scaling based on BLHS
substantially outperforms random subsets, as well as those which do not deploy
pre-scaling. Apparently, for a response surface with $4!=24$ minima, only a
carefully curated random subset is able to properly capture global
lengthscales. Also, observe that NN-based comparators fare relatively better
on this example, even outperforming ALC-based ones in some settings.  The most
outstanding comparator, ``alcsep.sb'', combines several ingredients:
BLHS-based pre-scaling and a single application of ALC-based separable
modeling. Finally, the comparators trained on $N=10^6$ substantially
outperform the ones trained on the smaller data, especially for the last six
comparators, suggesting that a relatively larger training data set is needed when
modeling a complex surface.

\section{Accurate satellite drag emulation}
\label{sec:satemu}

Here we return to our motivating satellite drag examples, first via point
prediction (Section \ref{sec:epp}) and then via trajectories (Section
\ref{sec:satpath}).  In both cases the scheme involves combining predictions
derived on pure chemical species via Eq.~(\ref{eq:mix}) with compositions
obtained for real locations in LEO.

\subsection{Exhaustive point prediction}
\label{sec:epp}

Following \citet{metha:etal:2014}, we begin by building surrogates for the six
pure chemical species separately, considering a span of global--local GP
alternatives matching the synthetic data experiments of Section
\ref{sec:bench}. As described in Section \ref{sec:intro}, the preferred
benchmark is out-of-sample RMSPE, i.e., via percentages, for benchmarking
purposes. Since the TPMC data generating mechanism is expensive, we deploy
$10$-fold cross-validation (CV) rather than pure randomizations over training
and testing partitions. For each of the six species, separately, we generated
random designs of size $N=2 \times 10^6$ for HST 
and $N=10^6$ for GRACE, so
that collecting all species resulted in 18 million TPMC runs to obtain the
full suite of outputs. For the He species of HST, which our collaborators at
LANL indicated would be the hardest to predict, we first generated a
size one-million sample, which was deemed to be inadequate in early tests,
and was ultimately supplemented with a further one million runs, combining for
$N=2 \times 10^6$ in order to explore the potential for larger designs.  For
GRACE, that initial one million was sufficient.

GRACE's LHS was over the seven-dimensional input space, described in Section
\ref{sec:review}. For HST, which has an extra (eighth) panel angle input, a
discrete ``parameter'' indexing the ten sets of mesh files in $\{0^\circ,
10^\circ, \dots, 90^\circ\}$, the full design is comprised of ten separate
equally sized, smaller LHSs.  We remind the reader here that the computational
effort of collecting TPMC runs on these designs, including the ensemble
testing ones mentioned below, was enormous, requiring 70 thousand core hours.
GRACE TPMC runs were more expensive than HST, with the former taking roughly about
two minutes and the latter taking near ten seconds.\footnote{GRACE's mesh
has more triangles defining its geometry, compared to HST.  Since TPMC must
search for collisions between particles and each mesh triangle, its simulation
is slower.} Finally, for HST, our BLHS schemes used $M=3$ generating $n=411$
sized sub-designs when $N=10^6$, and $M=3$ 
generating $n=823$ when $N=2 \times 10^6$, while for 
GRACE, we used $M=3$ leading to $n=1234$ for $N=10^6$.

\begin{figure}[ht!]
\includegraphics[scale=0.413, trim=0 1 30 15, clip=TRUE]{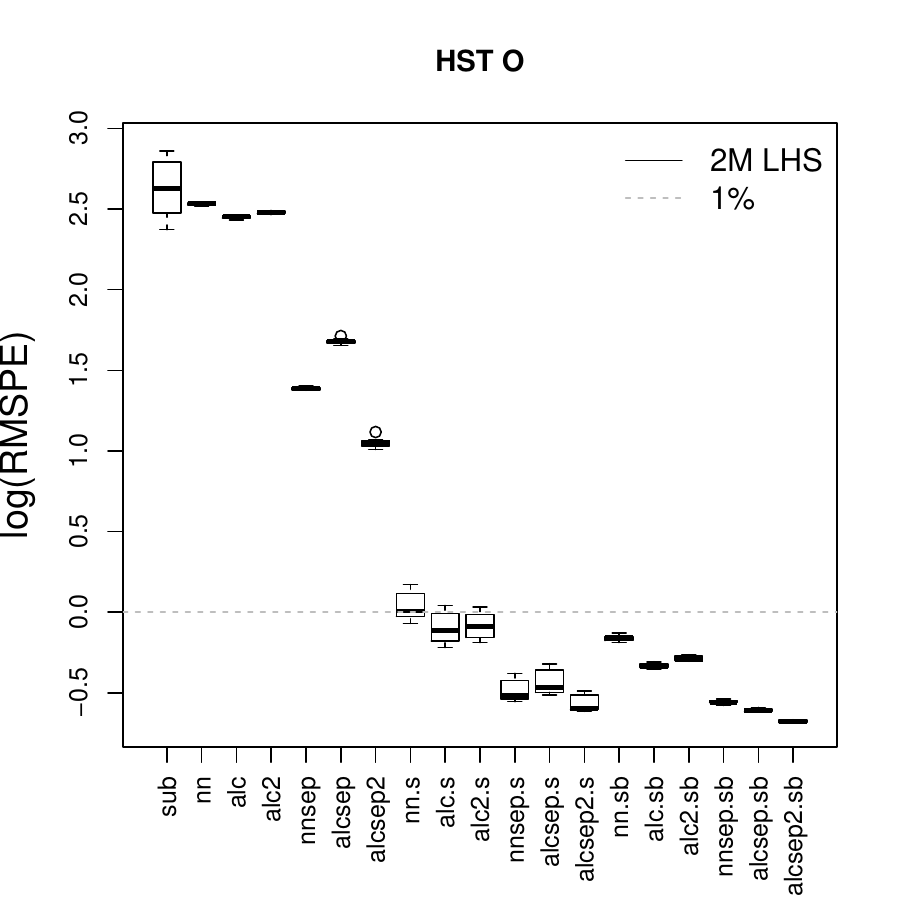}
\includegraphics[scale=0.413, trim=50 1 30 15, clip=TRUE]{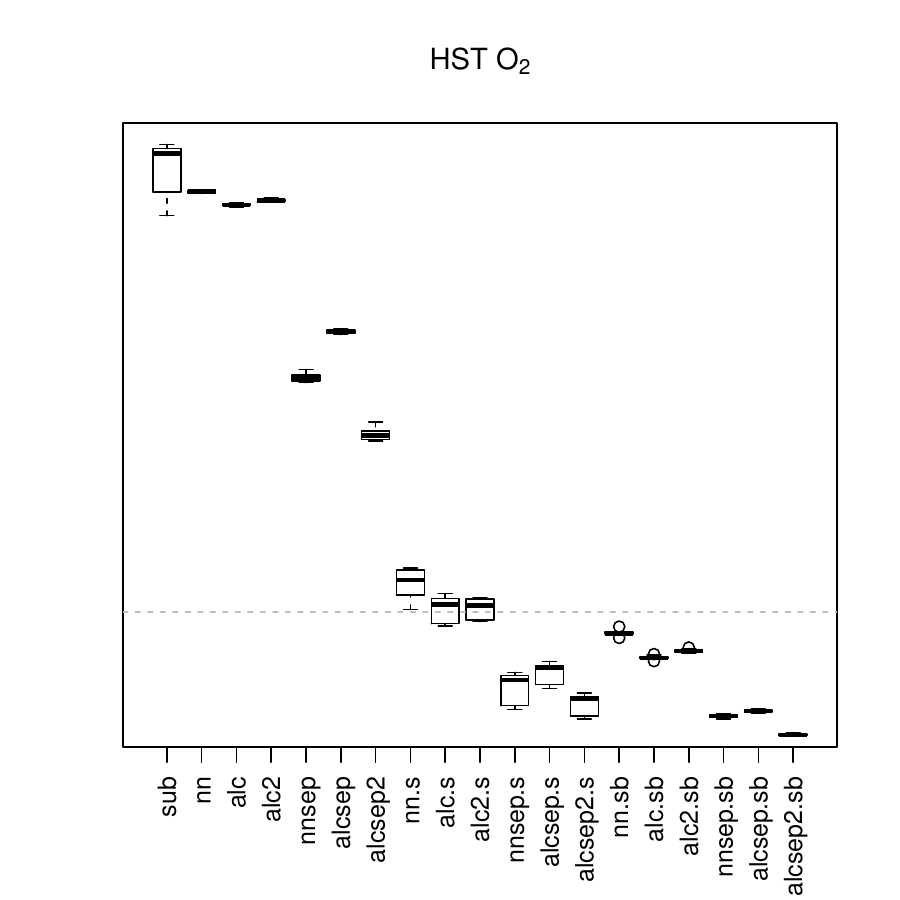}
\includegraphics[scale=0.413, trim=50 1 30 15, clip=TRUE]{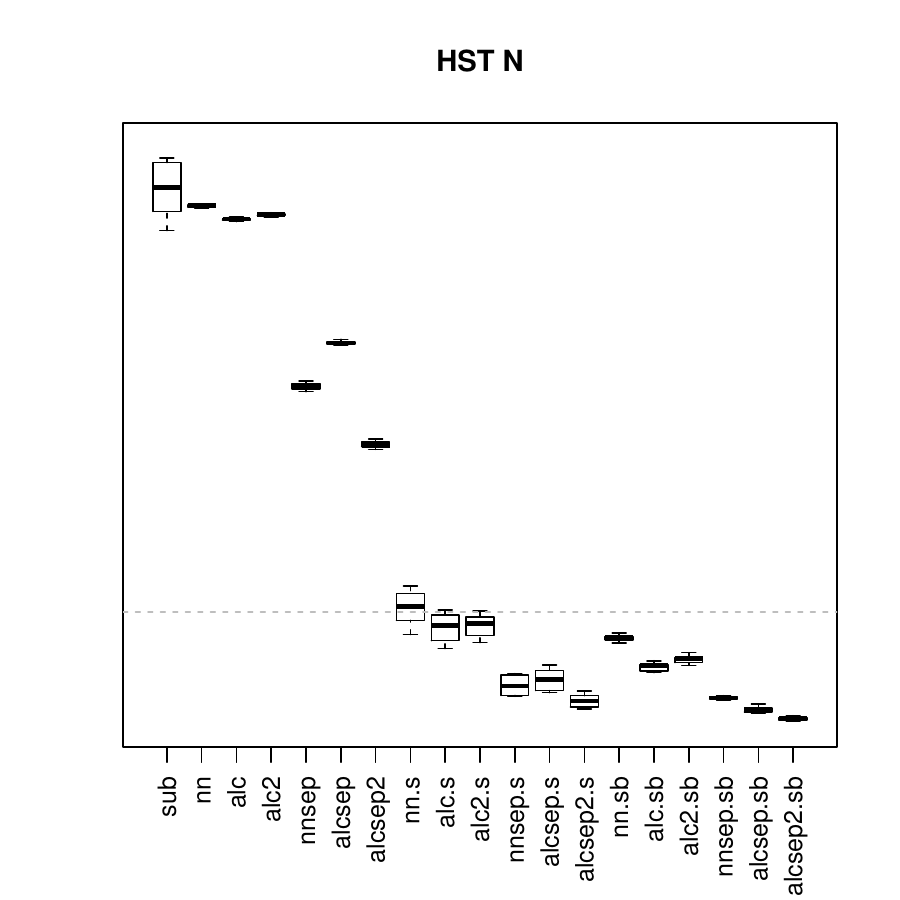}\\

\includegraphics[scale=0.413, trim=0 1 30 20, clip=TRUE]{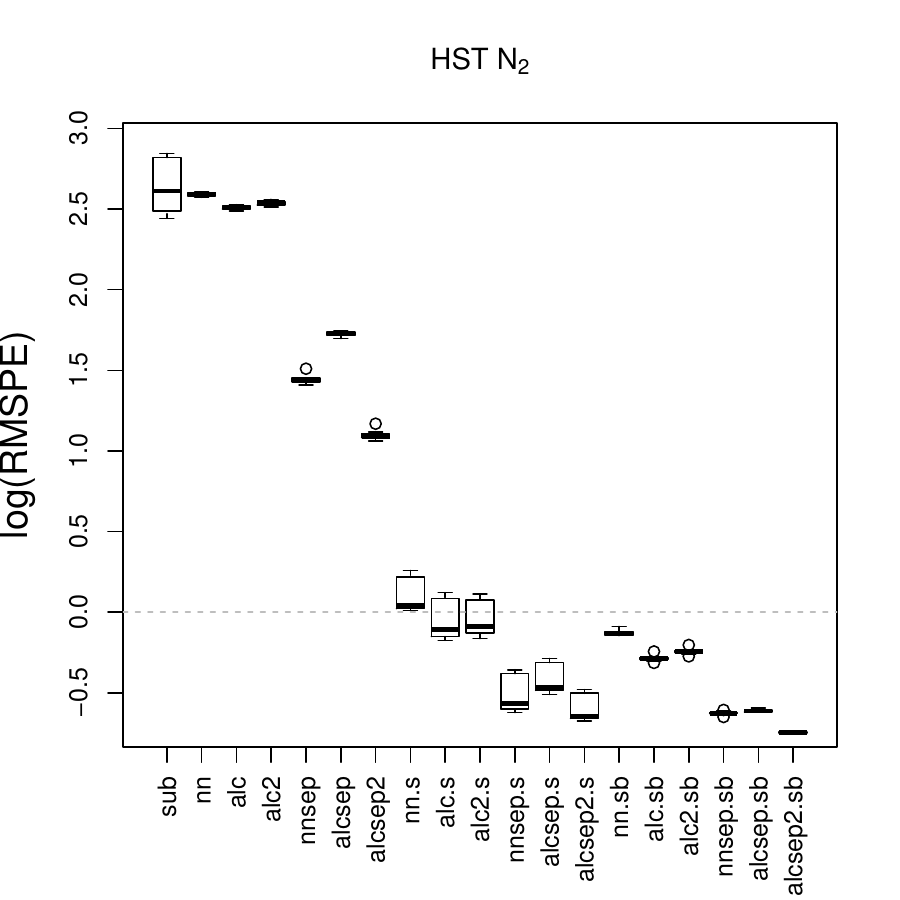} 
\includegraphics[scale=0.413, trim=50 1 30 20, clip=TRUE]{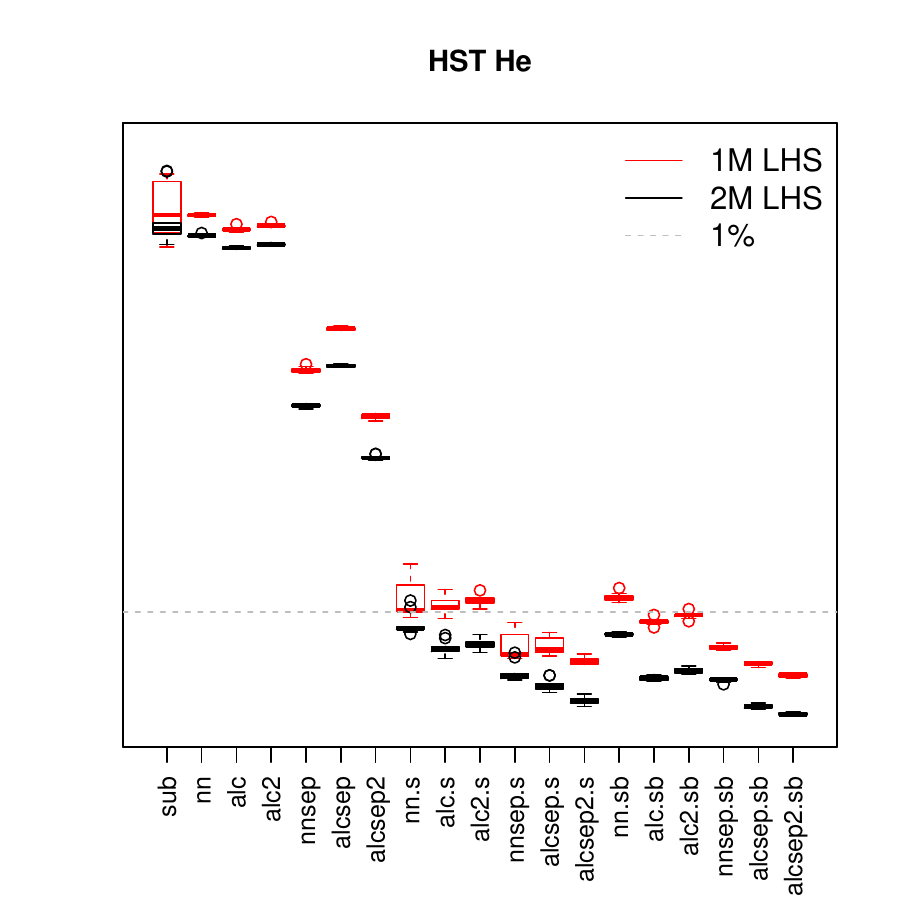}
\includegraphics[scale=0.413, trim=50 1 30 20, clip=TRUE]{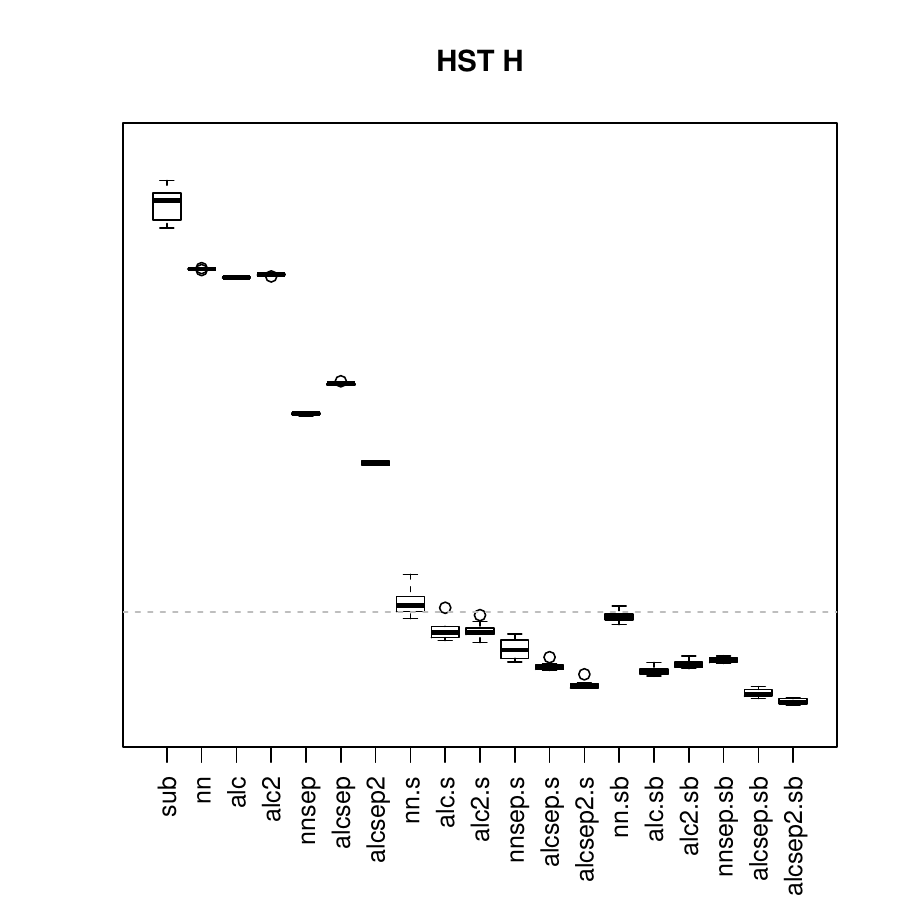}
\caption{{\tt laGP} results on the HST geometry via RMSPE calculated over 10
CV folds.  Each panel is a separate experiment on a pure
chemical species.}
\label{f:hst_all_cv_compare}
\end{figure}	

Figure \ref{f:hst_all_cv_compare} shows the distribution of RMSPEs obtained
from our 10-fold CV on the HST data.  There is a separate panel in the figure
for each pure species, and the y-axis is in log space to ease inspection.
First let's focus on the results for He, shown in the {\em bottom-middle}
panel of the figure. Observe that hybrid global--local modeling is essential
to approach the 1\% benchmark.  Global or local (subset) modeling on their own
are insufficient.  With $N=2 \times 10^6$ runs, random subsetting is sufficient
for all local methods to consistently beat the 1\% benchmark, but BLHS
subsetting is better. In the smaller $N=10^6$ experiment, it is
essential to either use a locally separable correlation structure, or BLHS
pre-scaling, and using both is better still. The best comparator here is
``alcsep2.sb'', i.e., using a second local design stage as recommended by
\citet{gramacy:apley:2015}, in contrast to our earlier synthetic results which
suggested deleterious effects.

The other panels in the figure, for the other five species, show a similar
pattern.   Besides being similar to He, the most important observation here is
that the BLHS pre-scaled comparators (last six boxplots in each panel) show
far lower variability than their random subsample analog (penultimate six).
That the former also has lower mean than the latter is also noteworthy, but is
not as striking visually.

\begin{figure}[ht!]
\includegraphics[scale=0.413, trim=0 1 30 15, clip=TRUE]{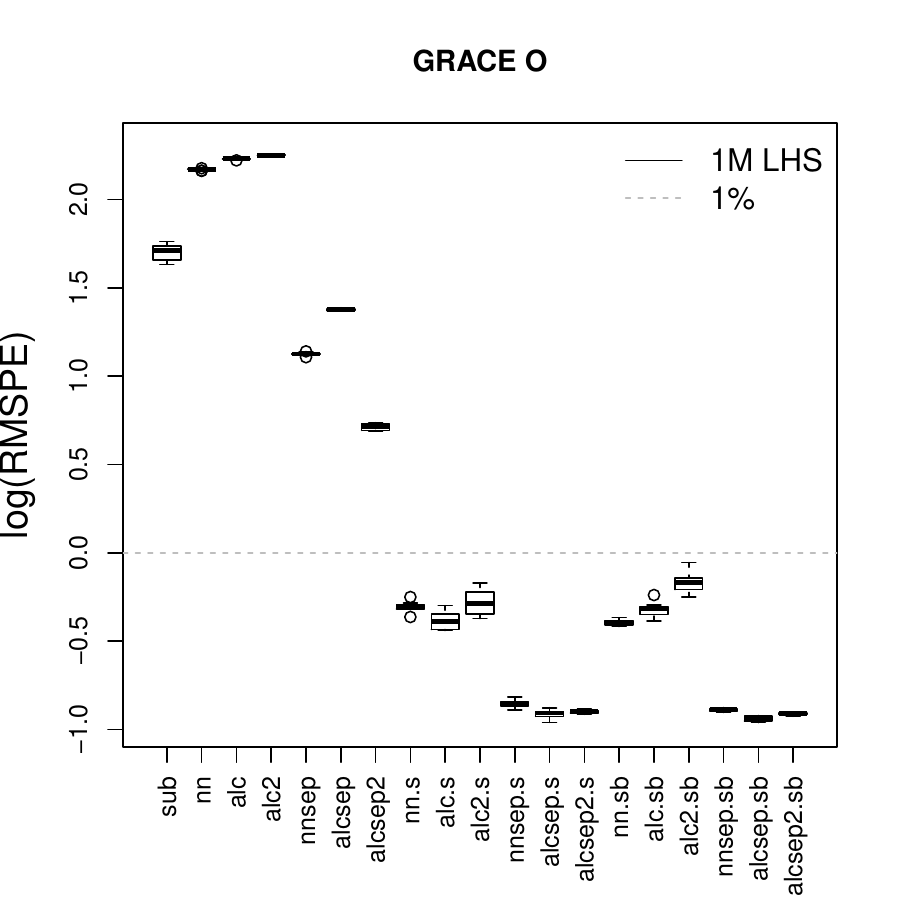}
\includegraphics[scale=0.413, trim=50 1 30 15, clip=TRUE]{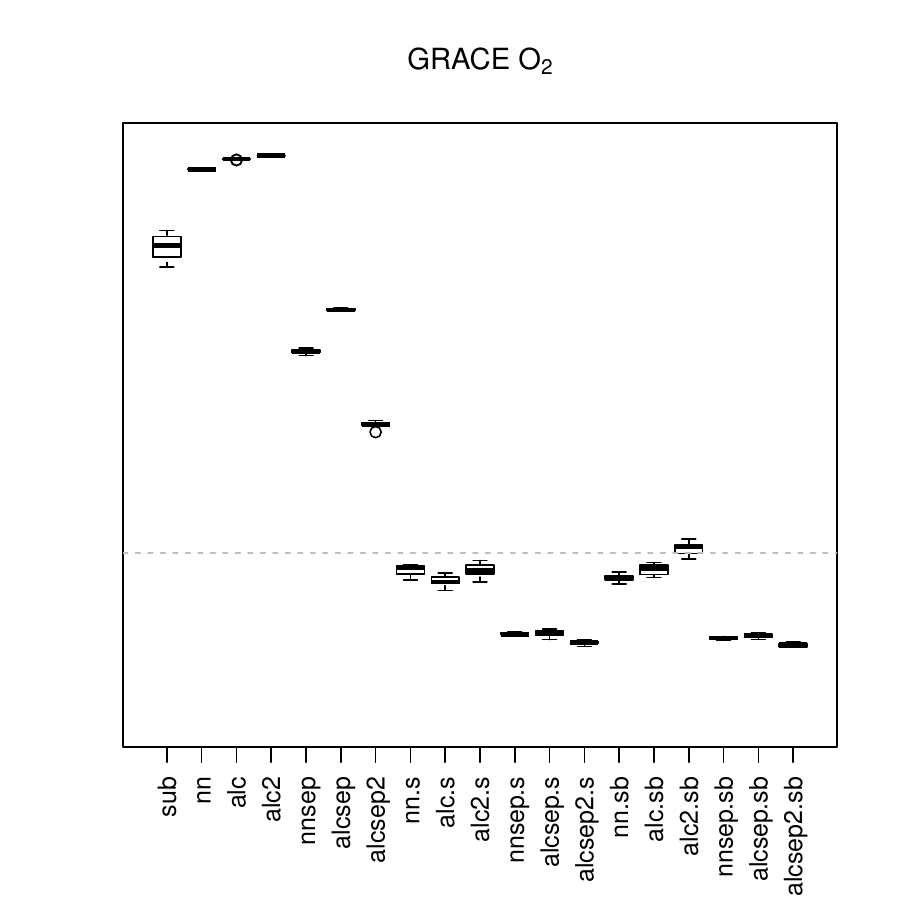}
\includegraphics[scale=0.413, trim=50 1 30 15, clip=TRUE]{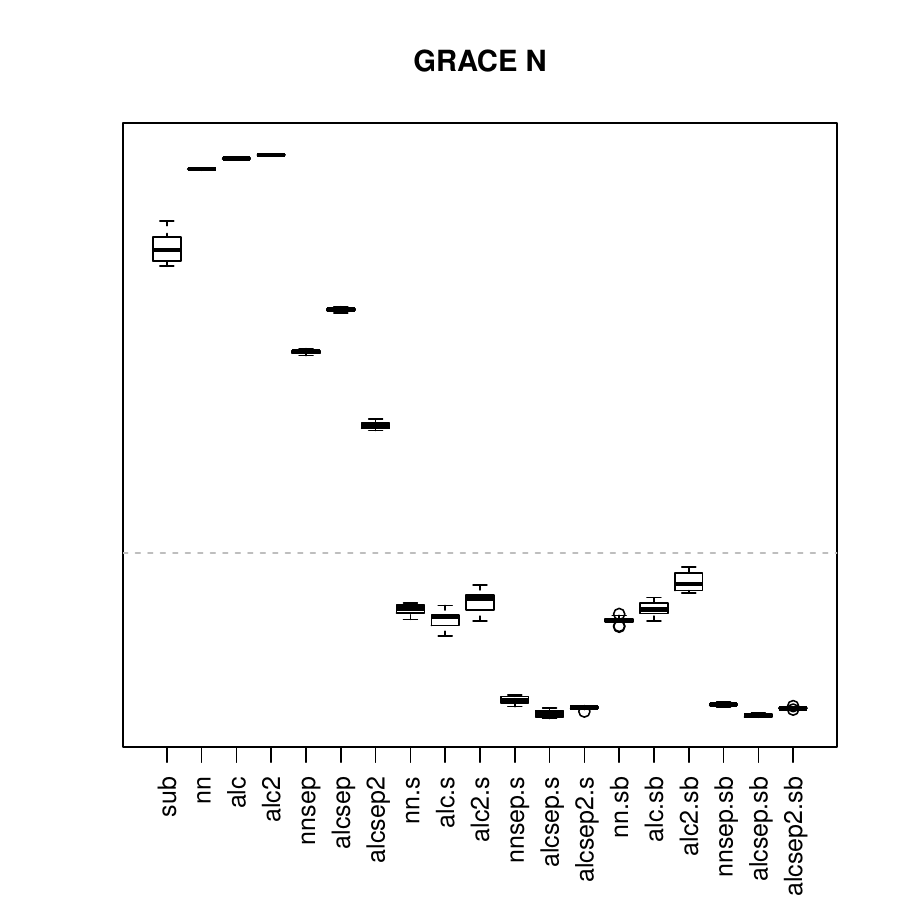}\\

\includegraphics[scale=0.413, trim=0 1 30 20, clip=TRUE]{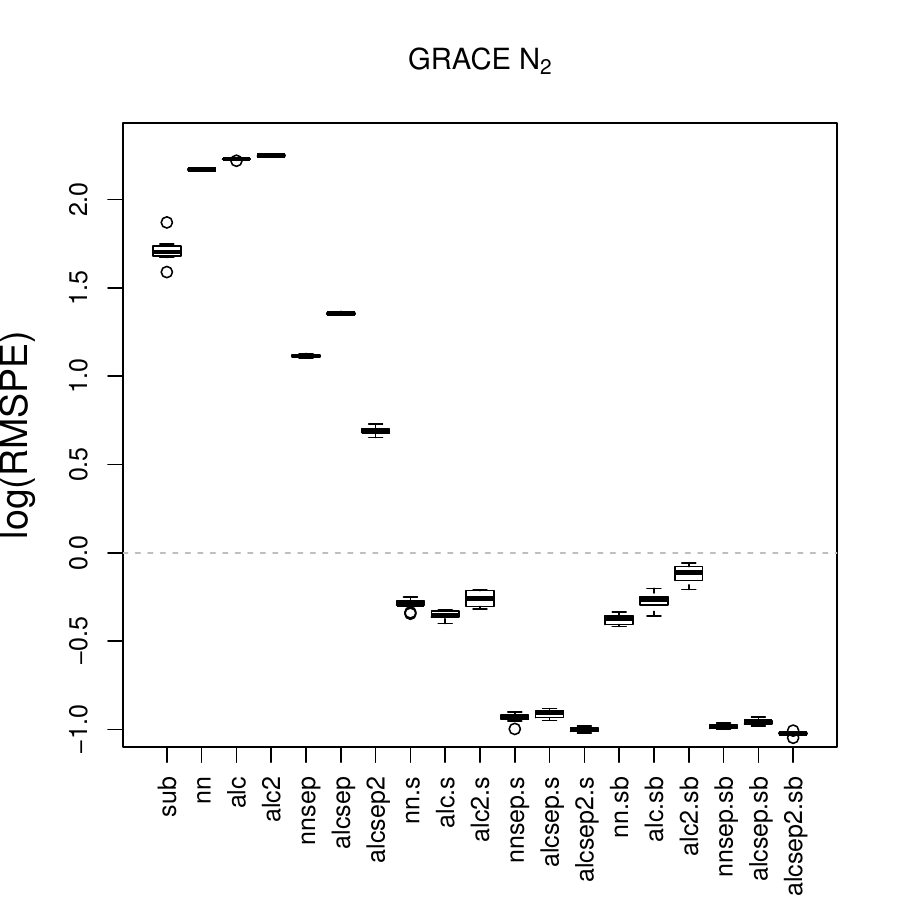} 
\includegraphics[scale=0.413, trim=50 1 30 20, clip=TRUE]{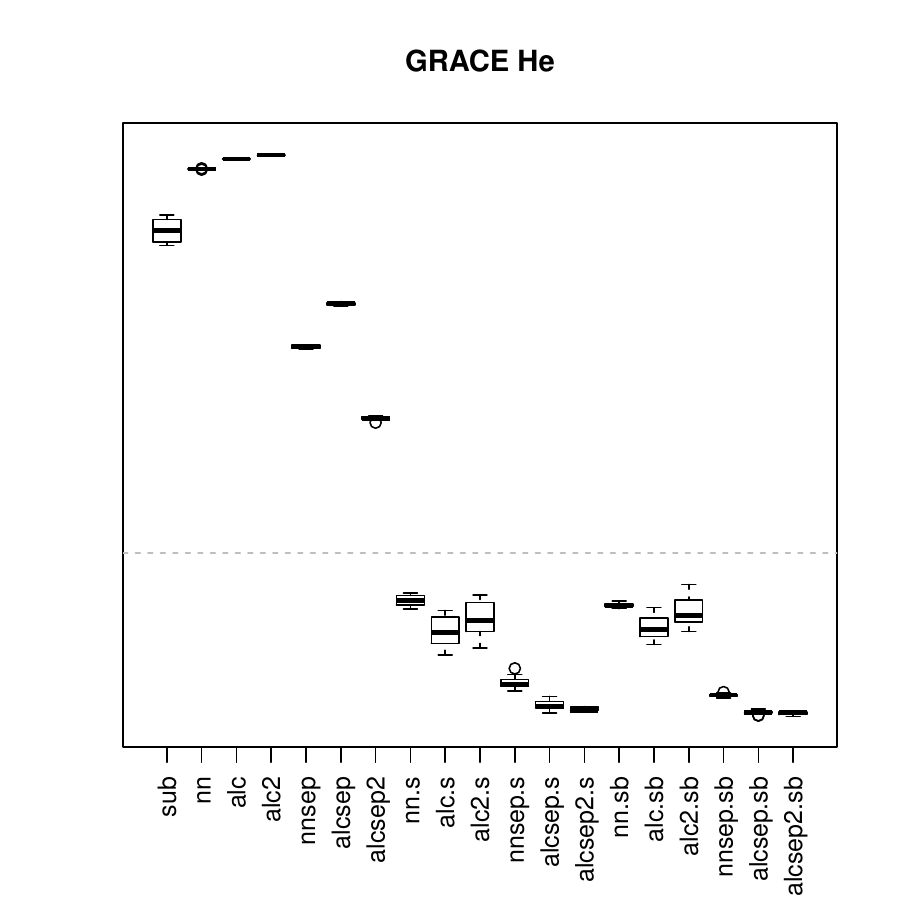}
\includegraphics[scale=0.413, trim=50 1 30 20, clip=TRUE]{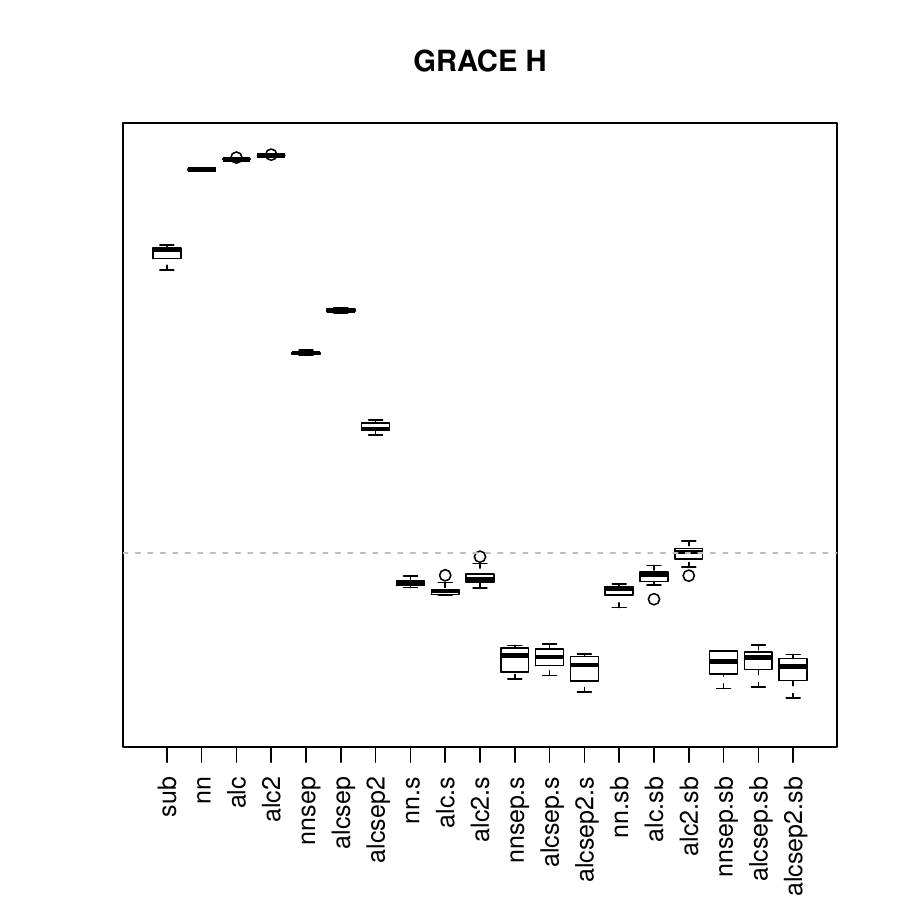}
\caption{{\tt laGP} results on the GRACE geometry via 10-fold CV; otherwise results are similar to
Figure \ref{f:hst_all_cv_compare}.}
\label{f:grace_all_cv_compare}
\end{figure}

Figure \ref{f:grace_all_cv_compare} shows a similar suite of results for the
GRACE satellite.  Many of the high level observations from the HST experiment
port over to this one.  Perhaps the most noteworthy difference is that, despite 
the extra TPMC effort required for GRACE runs relative to HST ones, the
ultimate emulation problem is actually easier.  The RMSPEs are quite a bit
lower, on the whole, compared to those in Figure \ref{f:hst_all_cv_compare},
even with a training data set of half the size. Besides that, note that
global--local modeling is essential to beat the 1\% benchmark, separable
local modeling is better than isotropic, and BLHS subsets offer superior
accuracy.  In terms of variance, BLHS and random sub-setting performs about
the same in this case.  

GRACE may be easier to emulate because it has a simpler, more convex geometry
compared to HST.  In general, the more concave an object, the more multiple
reflections it will have.  Besides making the mean surface more complicated,
those reflections may increase the (albeit very low) noise of the TPMC
simulations, making the signal harder to extract.

\begin{figure}[ht!]
\centering
\includegraphics[scale=0.45]{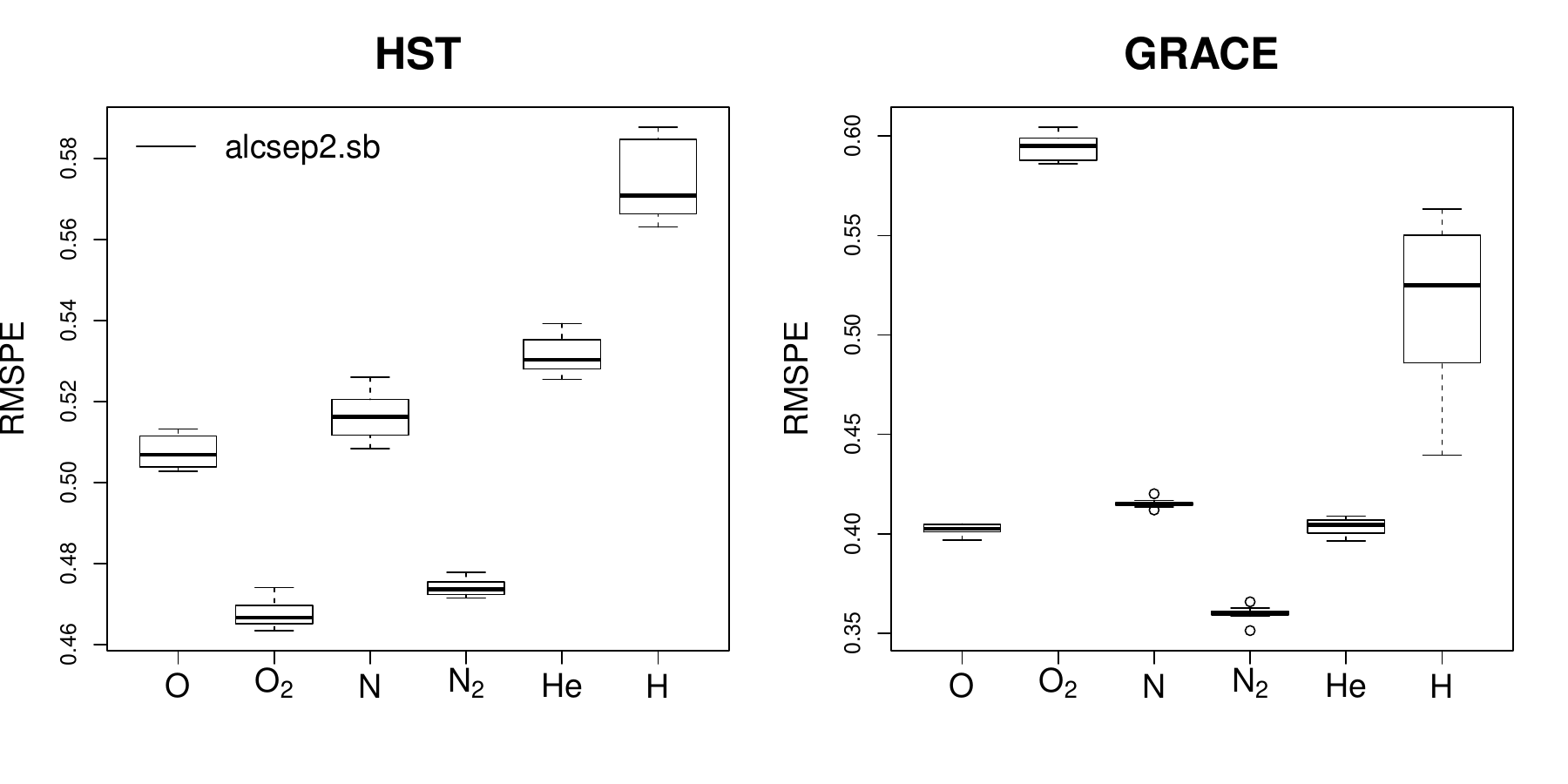}
\vspace{-0.5cm}
\caption{Best RMSPEs for each chemical species, zooming in on Figures \ref{f:hst_all_cv_compare}--\ref{f:grace_all_cv_compare}.}
\label{f:best_species}
\end{figure}	

For a finer-scale comparison we collect the best competitor (``alcsep2.sb'')
from both pure-species experiments for side-by-side visualization in Figure
\ref{f:best_species}, in effect zooming in on parts of Figures
\ref{f:hst_all_cv_compare}--\ref{f:grace_all_cv_compare}.  In the case of He
for HST, these results are from the $N=2 \times 10^6$ experiment, like the others.
Observe that, in contrast to our LANL collaborators' prior hunch, the He
species is not the hardest to predict.  For HST the hardest is H. In fact,
difficulty in emulation of each species appears to be monotonically increasing
with molecular mass from the heaviest (O$_2$) to the lightest (H).  An
explanation may be that lighter species have higher average speed for a given
atmospheric temperature, and/or larger variability in directional velocity.
Hydrogen atoms, for example, will be more likely to undergo multiple
collisions which the surface, complicating the input--response relationship.
The results are similar for GRACE, with O$_2$ being an exception. Perhaps the
GRACE geometry encourages reflections of O$_2$ particles that make O$_2$ have
a larger number of multiple reflections from the surface than other species.
Our collaborators at LANL plan to investigate further.

We close this experiment with some remarks on computational demands.
Entertaining prediction on millions of testing locations in a CV format for
twenty comparators is not a small task.  For the prevailing $N=2\times 10^6$
experiments, a single species CV takes about 4500 core hours.  Although
substantial, this pales in comparison to the cost of TPMC simulation.  The
best method ``alcsep2.sb'' requires about 5-10\% of that effort, depending on
how pre-processing costs are amortized---i.e., at most 450 hours---which means
that emulation is roughly two orders of magnitude faster than TPMC.  Emulation
runtime is also more predictable, taking less than a second per run for every
single input.  TPMC runs may average near ten seconds (HST) or minutes (GRACE),
however the variability is quite high.  Sometimes the Monte Carlo converges
quickly, and sometimes it does not.  Occasionally it does not converge at all.
About $2$ in $10^5$ TPMC runs fail, and must be (randomly) restarted,
which substantially increases the time required to obtain that response.

Inspired by \citet{metha:etal:2014}, we form a final ensemble predictor from
the best pure-species predictor for HST and GRACE, respectively, following
Eq.~(\ref{eq:mix}). To evaluate that predictor we construct a testing set via
LHS in exactly the same fashion as described above (using $N=2\times 10^6$ for
HST and $N=10^6$ for GRACE), except using a mixture of species rather
than six separate pure ones.  The mole fractions in the mixture were obtained
by entering the following specification into NASA's web-based calculator: date
1/1/2000, 0 hours, at 0 deg Latitude and Longitude, and 550 km altitude.  
These are:
\begin{center}
\begin{tabular}{ccccccc}
O & O$_2$ & N & N$_2$ & He & H \\
0.835756795 & 0.000040988 & 0.014095898 & 0.005918278 & 0.137959854 & 0.006228188 
\end{tabular}
\end{center}
The out-of-sample testing RMSPE we obtained is 0.3860 under our best
comparator, ``alcsep2.sb'', which was based on an average BLHS subset size of
$n=1371$ using $M=3$. For HST, based on an average BLHS subset size of $n=914$
using $M=3$, the RMSPE is 0.4877 under ``alcsep2.sb''.  Observe that these
numbers are better than the best ones in Figure \ref{f:best_species} as the
training sets were 10\% larger (since they did not require CV).

\subsection{Quantifying uncertainty in drags over trajectories}
\label{sec:satpath}

Here we consider a more realistic application of satellite drag emulation, via
a demonstration of the value of joint modeling over a trajectory (Section
\ref{sec:joint}), as opposed to a simpler pointwise approach.  Four experiments
are entertained. For each of HST and GRACE geometries, settings for the
``free'' input parameters (Table \ref{t:inputs}) are derived from a
representative orbit in 2017 under periods of ``quiet'' (5.0) and ``active''
(200.0) daily geomagnetic indices.
\begin{table}[ht!]
\small
\centering
\begin{tabular}{l|rrrrrr}
{\bfseries Satellite} & {\bfseries SMA (km) } & {\bfseries Eccen.} & {\bfseries Inc (deg)} & 
{\bfseries AP (deg)} & {\bfseries RAAN (deg)} & {\bfseries TA (deg)} \\ 
  \hline
HST & $\ 6918$ & $\ 0.00027$ & $\quad 28.5$ & $\quad 350$ & $\quad 318$ & $\quad 125$ \\ 
  GRACE & $\ 6713$ & $\ 0.0015$ & $\quad 89$ & $\quad 95$ & $\quad 314$ & $\quad 275$ 
\end{tabular}
\caption{Keplerian Orbit via {\tt SMA}: Semi-Major Axis; {\tt
Eccen.}: Eccentricity; {\tt Inc}: Inclination; {\tt AP}: Argument of Perigee;
{\tt RAAN}: Right-Ascension of the Ascending Node; {\tt TA}: True Anomaly.}
\label{t:UQ_oe}
\end{table}
Table \ref{t:UQ_oe} shows the Keplerian orbital elements for each satellite,
which were entered into the satellite propagator program called
IMPACT-PROP\footnote{IMPACT stands for Integrated Modeling of Perturbations in
Atmosphere for Conjunction Tracking; IMPACT-PROP was developed at LANL
by Michael Shoemaker and Andrew Walker.} to determine the sequence of input
settings for one full day (86,400 seconds), January 1$^\mathrm{st}$, 2017,
starting at 00:00:00 UTC and sampled at 10-second intervals. The result
comprises of several orbits around Earth as depicted by a set of 8641 input
configurations for each satellite, and each geomagnetic regime.

\begin{figure}[ht!]
\centering
\includegraphics[scale=0.305,trim=0 10 5 35,clip=TRUE]{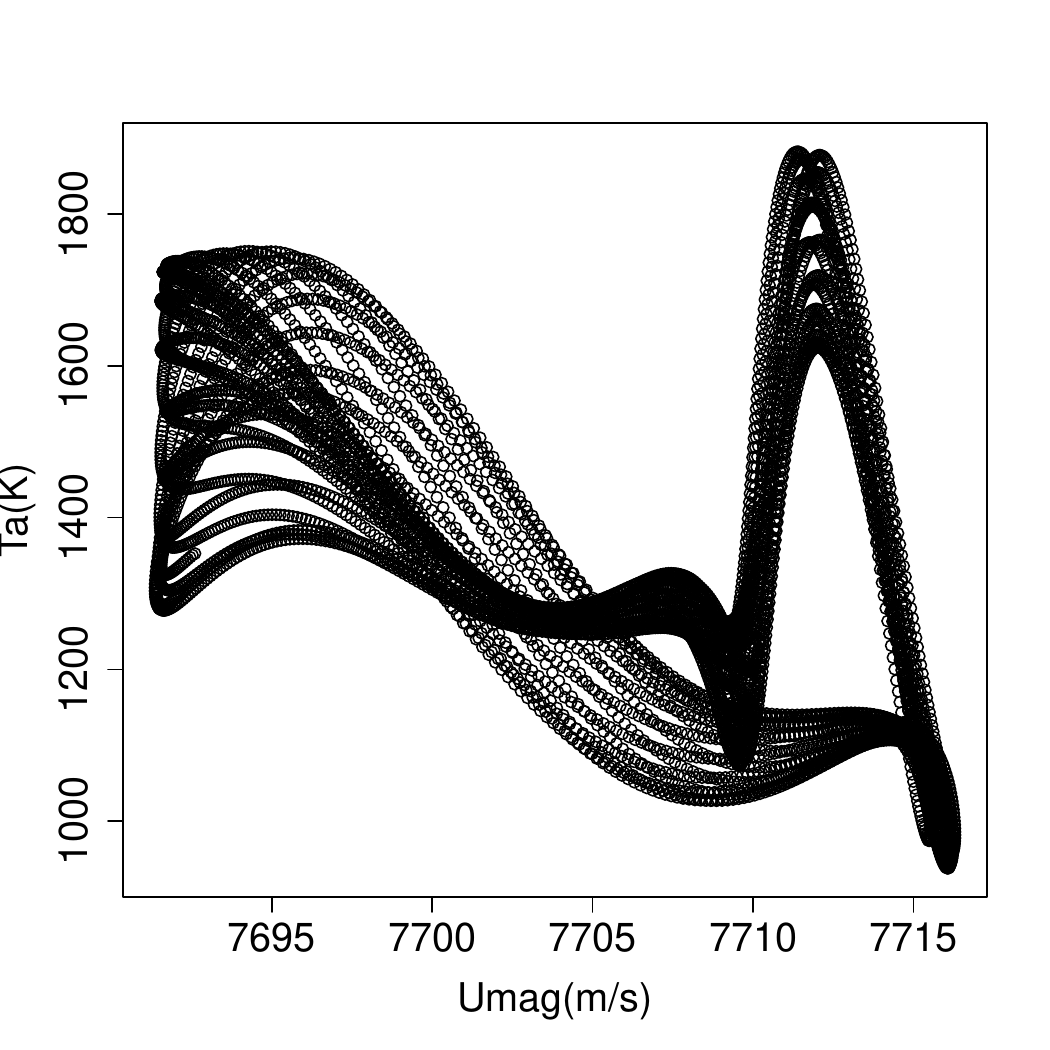}
\includegraphics[scale=0.305,trim=2 10 5 35,clip=TRUE]{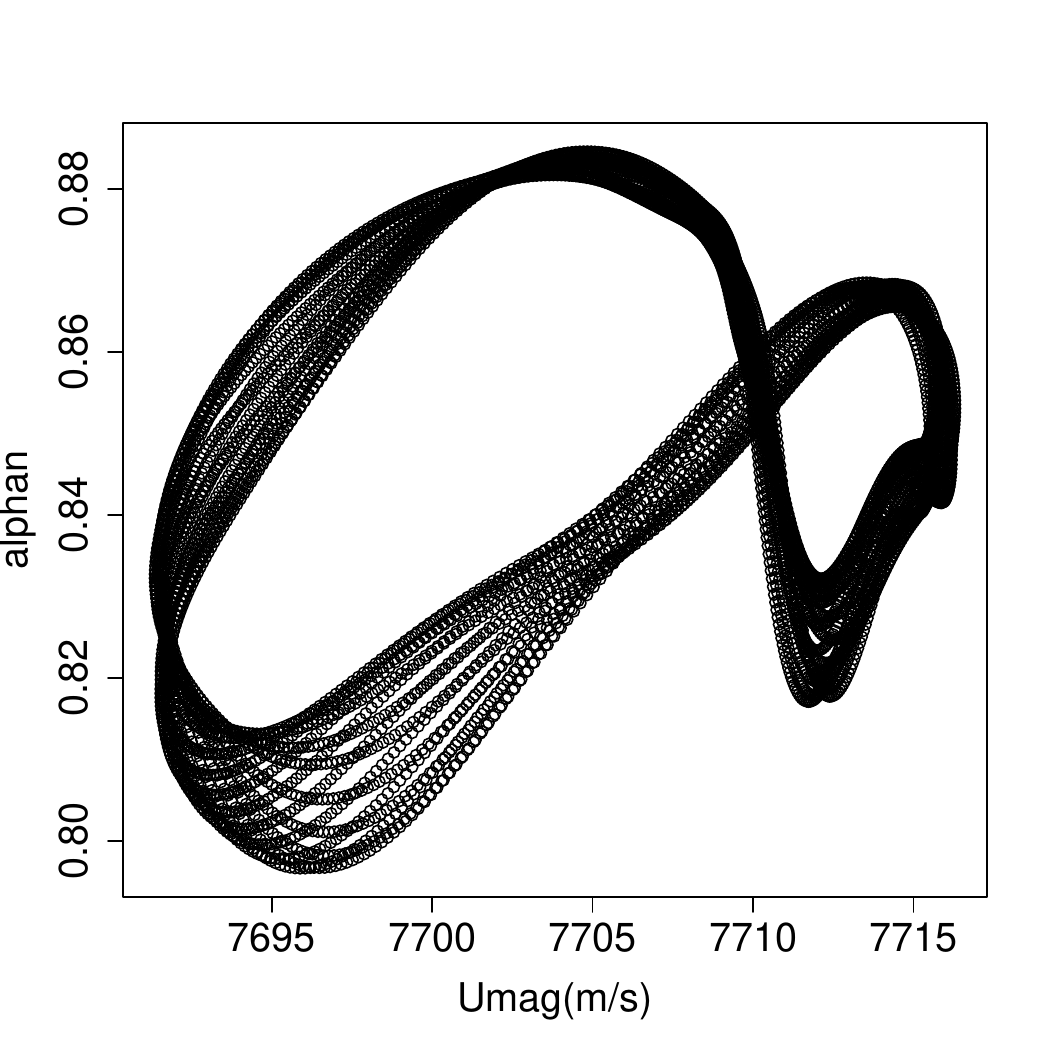}
\includegraphics[scale=0.305,trim=2 10 5 35,clip=TRUE]{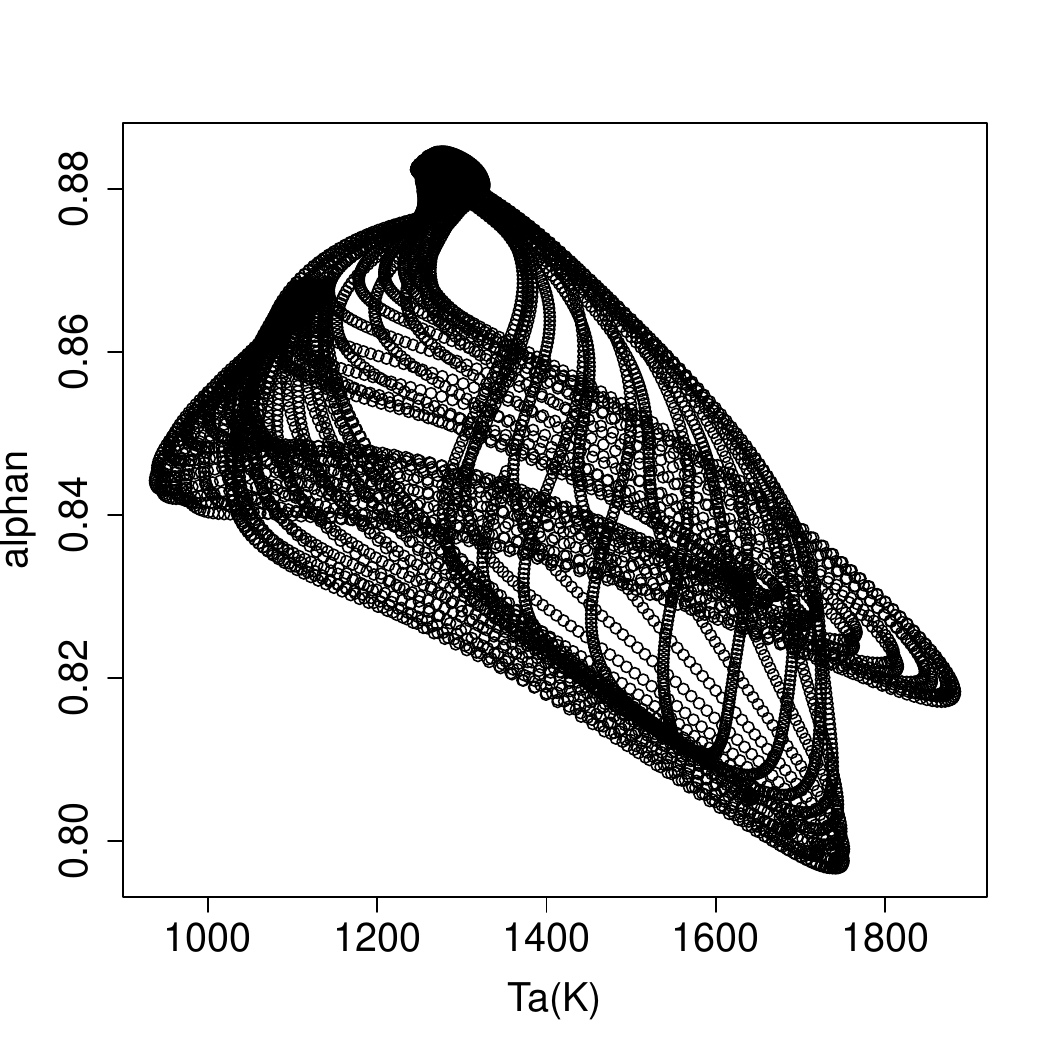}
\vspace{-0.25cm}
\caption{Input coordinates for GRACE-active: {\tt Umag}, {\tt Ta}, and {\tt alphan}.}
\label{f:UQ_par}
\end{figure}

Not all parameters from Table \ref{t:inputs} are varied by IMPACT-PROP. Figure \ref{f:UQ_par} 
shows the nature of the orbits in the parameter space obtained for GRACE-active.  
Besides noting the interesting nature of the resulting trajectories, with slightly 
shifted circuits for each ``pass'' around the Earth, observe (from the axis labels) 
the rather limited range of values these parameters take on compared to the full 
range of study from Table \ref{t:inputs}.  The other parameters, i.e., besides these 
three, were fixed near their midway values. The propagator also provides the chemical
composition in LEO, which is similarly changing with the inputs, however this is 
less easily visualized since it is six-dimensional.

We fed those inputs and chemical compositions into the TPMC simulator to
obtain a testing set of drag coefficients, with total computation times averaging
about 34 hours for HST and 600 hours for GRACE, across both
regimes.\footnote{Since each input was paired with a unique chemical
composition, these runs could not be parallelized by block as described in
Appendix \ref{sec:tpm}, leading to particularly slow wall-clock time.} More
precise timings are provided in Table \ref{t:UQ_time}.  Those inputs, their
corresponding chemical compositions and drag coefficients,\footnote{These data
are also available in our supplementary material.} were used as the basis of
the following out-of-sample predictive experiment, mimicking the setup for our
synthetic exploration in Section \ref{sec:er}. Specifically, we broke the
10-second time steps into 86 sets of one hundred, contiguously, and
considered each to comprise of a single trajectory through position and
parameter space. Predictions under each pure chemical species (i.e., six
predictions) were developed for each trajectory, separately via variations on
joint and point-wise schemes, and then the species were combined via
Eq.~(\ref{eq:mix}) using chemical compositions provided for each of the
associated positions in the data.
\begin{figure}[ht!]
\centering
\includegraphics[scale=0.24,trim=0 0 30 15,clip=TRUE]{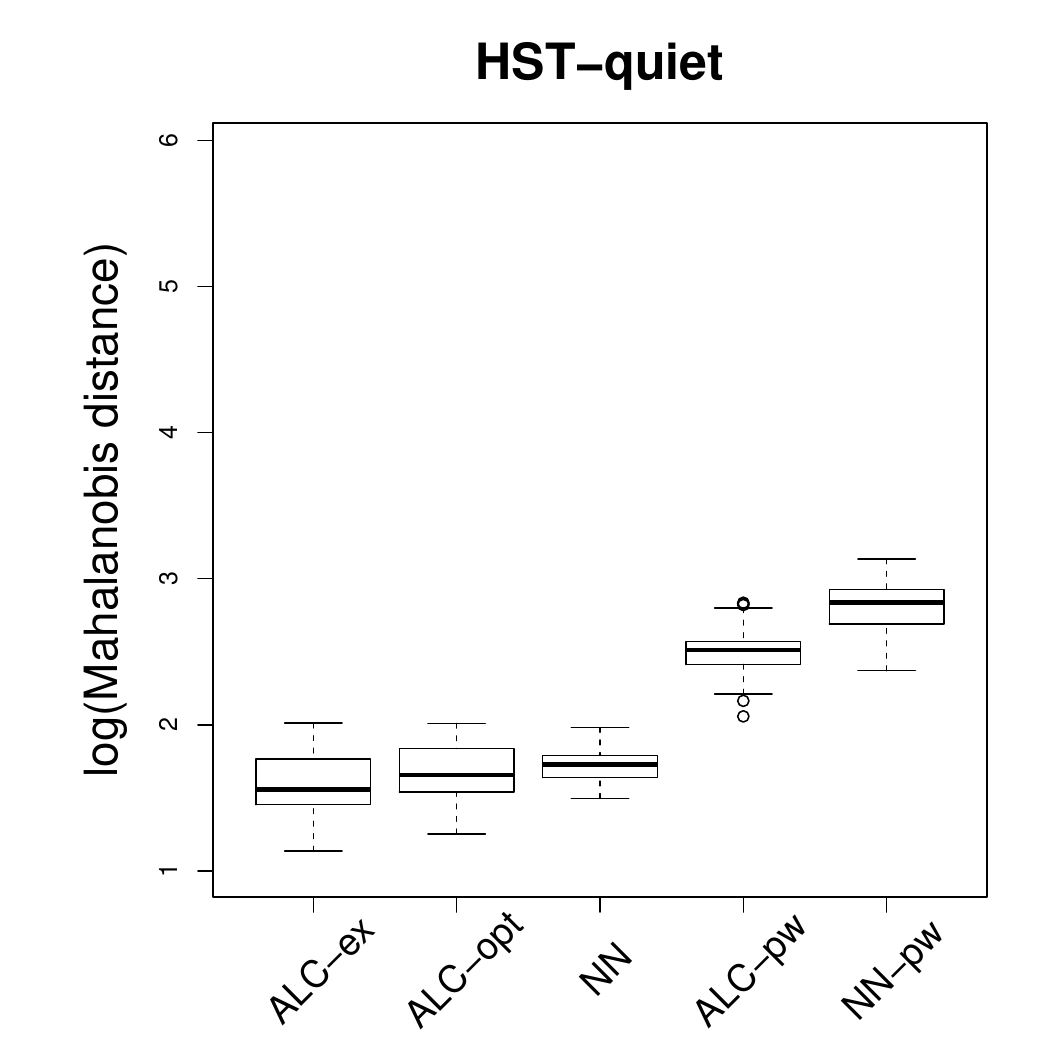}
\includegraphics[scale=0.24,trim=8 0 26 15,clip=TRUE]{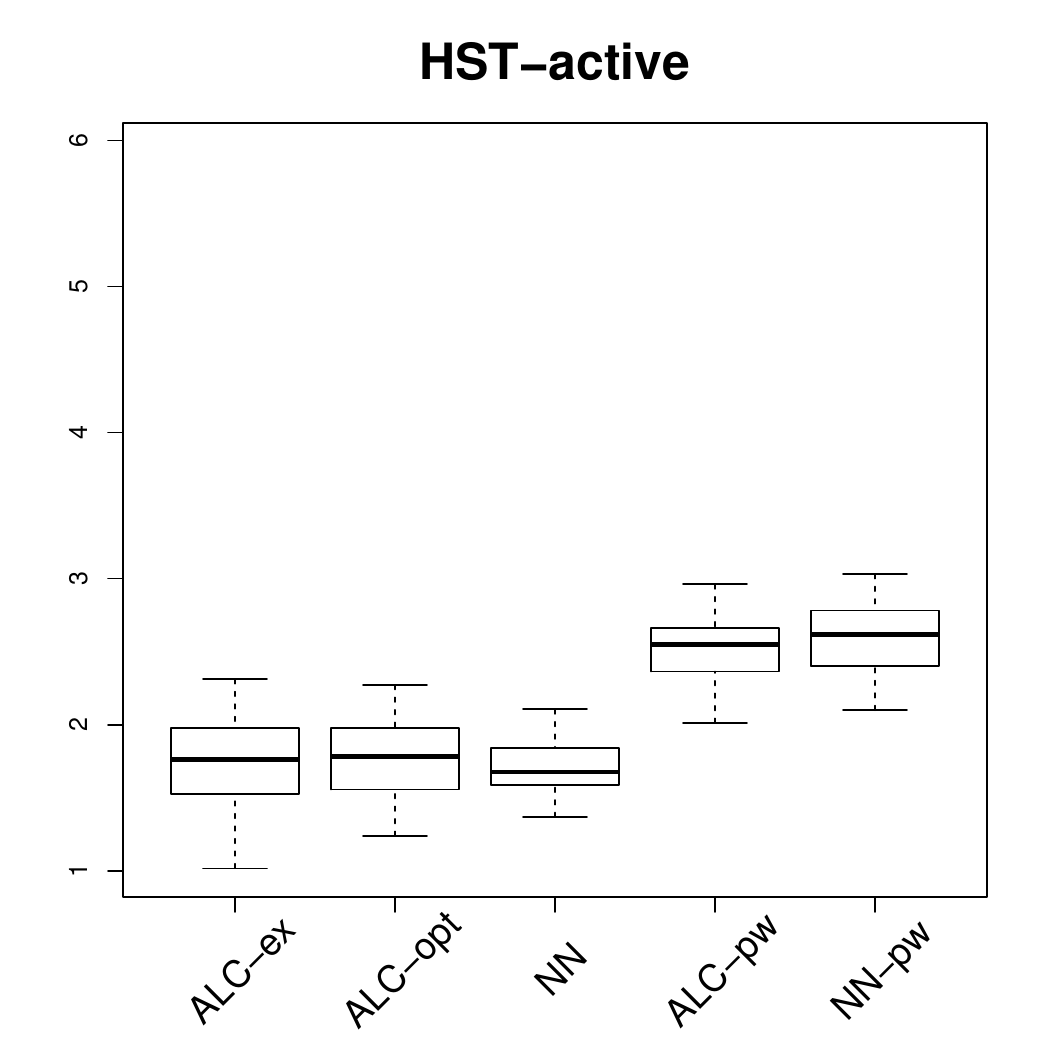}
\includegraphics[scale=0.24,trim=8 0 26 15,clip=TRUE]{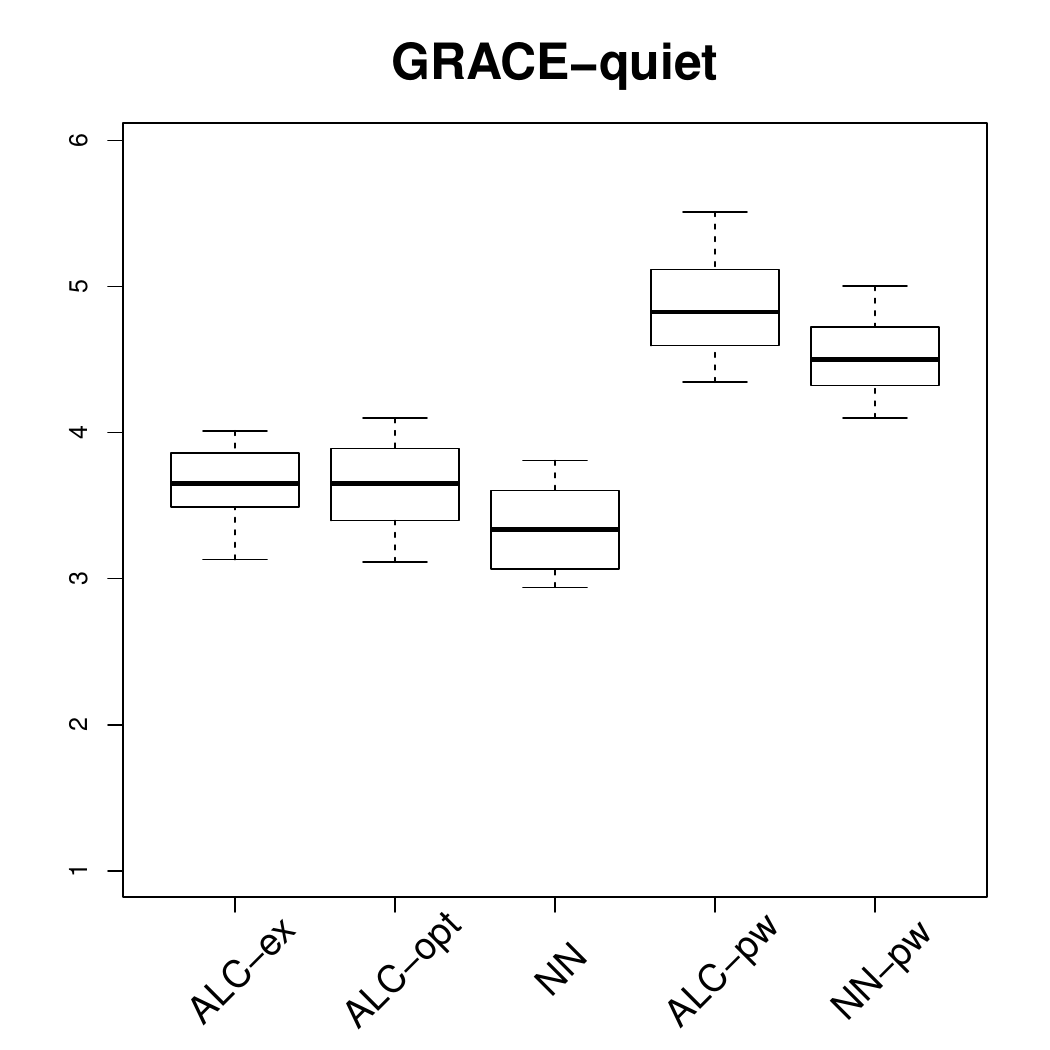}
\includegraphics[scale=0.24,trim=8 0 26 15,clip=TRUE]{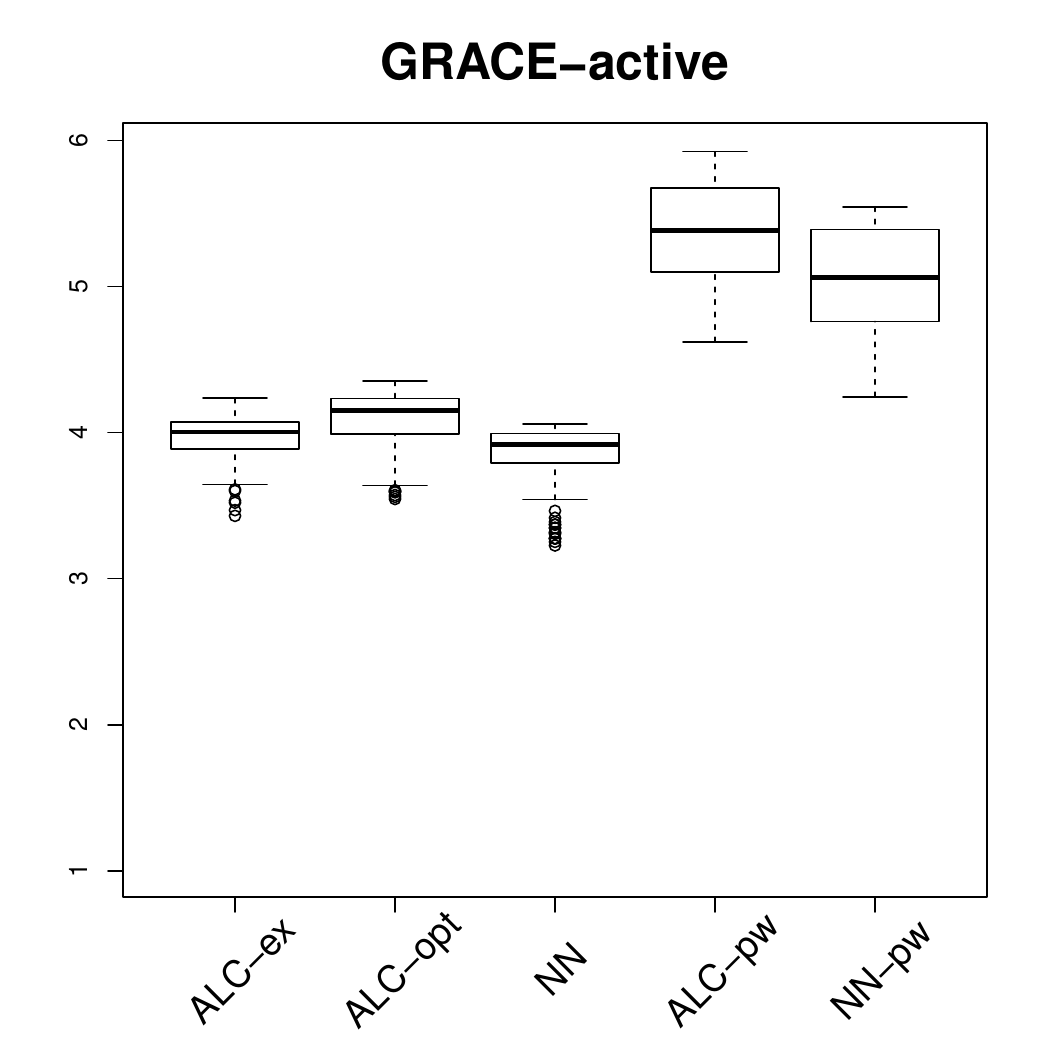}
\caption{Comparing log Mahalanobis distances among ALC-ex, ALC-opt, NN, ALC-pw, 
and NN-pw under quiet and active regimes.}
\label{f:UQ_traj}
\end{figure}
The results, measuring mean and covariance accuracy simultaneously via
Mahalanobis distance, are displayed in Figure \ref{f:UQ_traj} mimicking the
format of our synthetic comparison in Figure \ref{f:metric_compare}.  The RMSPE
boxplots for each comparator were strikingly similar, so we chose not to
display those in order to economize on space.  It is clear from the boxplots
that joint sampling dominates the pointwise alternatives, which we attribute
to their ability to more accurately capture the underlying covariance
structure along the trajectory.   Also notice that the distances for GRACE are
larger than the ones for HST, perhaps owing to the double-sized training set
for the latter.

\begin{table}[ht!]
\centering
\begin{tabular}{l|rrrrrr}
{\bfseries Trajectory} & {\bfseries ALC-ex} & {\bfseries ALC-opt} & {\bfseries NN} & {\bfseries ALC-pw} & {\bfseries NN-pw} & {\bfseries TPMC} \\ 
  \hline
HST-quiet & $\ 83.267$ & $\ 25.047$ & $\ 21.074$ & $\ 89.921$ & $\ 49.565$ & $\ 728$ \\ 
  HST-active &  $\ 82.014$ &  $\ 24.670$ &  $\ 20.815$ & $\ 112.020$ &  $\ 60.356$ & $\ 509$ \\ 
  GRACE-quiet & $\ 69.692$ & $\ 12.642$ &  $\ 9.649$ & $\ 83.952$ & $\ 37.460$ & $\ 12461$ \\ 
  GRACE-active & $\ 68.363$ & $\ 12.302$ &  $\ 9.366$ & $\ 81.356$ & $\ 39.488$ & $\ 13012$
\end{tabular}
\caption{Median computation time in seconds among joint (ALC-ex, ALC-opt, NN) and pointwise (ALC-pw, and NN-pw) comparators.}
\label{t:UQ_time}
\end{table}

Table \ref{t:UQ_time} shows the median computation time for each comparator,
including the actual TPMC evaluations. GRACE predictions are faster due to the
smaller training data size and lower input dimension.  Observe that an exhaustive ALC
search is much slower than the derivative-based analogue, ALC-opt, and not much
slower than the fastest method, NN.  Of course, all of the predictors are
orders of magnitude faster than running TPMC.

\section{Discussion}
\label{sec:discuss}

In this paper we are motivated by a large-scale computer model emulation
problem coming from aeronautics.  Demands for accuracy in that literature
require combining a large simulation effort with a high fidelity response
surface.  Unfortunately, those two are incompatible using current methods.
Gaussian processes buckle under the weight of cubic operations, precluding
training on large data.  Even recent approximations, such as the {\tt laGP}, which substantially increase the
size capability of GPs, come up short on this task.  We have proposed several
extensions to {\tt laGP} which, while being motivated by a particular problem,
we think have the potential of broad applicability.

Our first enhancement was more technological than methodological, extending
the local covariance structure to handle anisotropy while maintaining
parallelizability.  This required implementing a careful locking strategy with
{\tt OpenMP} pragmas in order to guarantee thread-safety when accessing legacy
libraries.  Our first methodological contribution involved a multiresolution
global/local variation that fits global lengthscales from data subsets.  An
initial, {\em ad hoc}, approach via random data subsets worked well, but left
something to be desired on a theoretical front.  We were lucky to come across
the work of \citet{liu:2014} and \citet{liu:hung:2015}, preferring more
structured BLHS for fast
consistent reproduction of global MLE of lengthscales. 

When prediction is the ultimate goal, the utility of such a result is not
readily apparent.  Moreover, the MLE of lengthscales, whether globally or locally,
need not lead to the best predictions \citep{zhang:2004}.  Nevertheless, we
showed empirically that if the data can be pre-scaled using those estimated
lengthscales, then subsequent local prediction, say via {\tt laGP}, could be
quite powerful indeed. This is borne out in several synthetic examples as well
as on two real-world satellite drag simulators, where we demonstrate that it is
possible (with GPs) to surpass a previously elusive 1\% accuracy benchmark.
Future work may investigate \citeauthor{zhang:2004}'s insight to target
elongated lengthscales, relative to the MLE, in both local and global
contexts.  Such analysis may help explain why a random subset outperforms BLHS
in the borehole experiment summarized in the left panel of Figure
\ref{f:synthetic_compare}.

A second methodological contribution comes in the form of an extension of the
notion of ``locale" in the {\tt laGP} framework, from one location to a set of
locations.  We facilitate that extension with an aggregate ALC criterion and,
more importantly, provide the closed form derivatives that make optimization
tractable. Besides offering a more realistic application of emulation in the
satellite drag application, namely, of predicting along likely trajectories in
LEO, the idea offers a significant technical ability not
previously enjoyed by local GP methods, that of a full predictive covariance
structure, rather than simply a pointwise one.  We envision several potential
extensions to this framework.  A natural variation may be to allow elements of
the local path (or set) to be treated differentially, say by weighting. Such
an extension may prove useful in modeling dynamic computer simulations via
local GPs \citep{Zhang:etal:2016}, replacing hard clusterings with weights,
say.  Another would be to augment the criterion defined in Eq.~(\ref{eq:vdxy}) to include
a full covariance for elements of path $W$, rather than a simple sum of
variances, via Mahalanobis distances or proper scores
\citep[Eq.~(25)][]{gneiting:raftery:2007}.  We believe that this could be
accommodated in the same framework, i.e., via analytic derivatives for
library-based optimization, however we doubt that the scheme would remain
computationally tractable as it would be cubic ($\mathcal{O}(|W|^3j^2)$) rather than linear in $|W|$, i.e., $\mathcal{O}(|W|j^2)$.

Finally, all methods reported in this paper have
been included as additions to the {\tt laGP} package for {\sf R} on CRAN
\citep{laGP,gramacy:jss:2016}, and made available as a public Bitbucket
mercurial repository:
\url{https://bitbucket.org/gramacylab/lagp}.  The TPMC simulation code,
described in more detail in Appendix \ref{sec:tpm}, and all code supporting
the empirical work in this paper, is available in a separate public
Bitbucket mercurial repository: \url{https://bitbucket.org/gramacylab/tpm}.
Our {\sf R} interface to the TPMC simulator may prove valuable in the future work
requiring real-world benchmarks, and supercomputer implementations without the
hassle of a bespoke interface.

\subsection*{Acknowledgments}

We are grateful to two referees for valuable comments which led to many
improvements in the paper. This work was completed in part with
resources provided by the University of Chicago Research Computing Center, and
by facilities for Advanced Research Computing at Virginia Tech.  TPMC and
IMPACT-PROP were developed under the IMPACT project at Los Alamos National
Laboratory.  The authors would like to gratefully acknowledge funding from the
National Science Foundation, awards DMS-1564438, DMS-1621722, DMS-1621746, and
DMS-1739097.

\appendix

\bigskip

\section{Test particle method and {\sf R} interface}
\label{sec:tpm}

In the TPMC simulations presented here, particles are initialized with the 
appropriate Maxwellian velocity distribution based on the speed of the 
spacecraft and the atmospheric temperature.  Particles are then randomly 
generated on the edges of the TPMC domain that encloses the spacecraft 
whose drag coefficient is being calculated.  Each particle is tracked (ray-traced) 
until it exits the domain (in which case its tracking is completed) or until it 
collides with the surface.  If a particle collides with the surface, the GSI model 
is used to appropriately re-orient the particle velocity vector and the particle is 
again tracked until it collides with the surface again or exits the domain.  One of 
the primary advantages of TPMC over analytic methods when computing drag 
coefficients is that it can handle multiple collisions with the surface whereas 
analytic methods are only valid for a single collision (because they assume 
that the incident particles have a Maxwellian velocity distribution which is no 
longer true for particles that have already reflected off the spacecraft surface).

The supplementary material contains a version of the original TPMC {\sf C}
code, developed at Los Alamos National Laboratory, which we have modified to
utilize {\tt OpenMP} for efficient symmetric multiprocessor parallelization
(i.e., utilization of multiple cores within a node).  A wrapper to that {\sf
C} library, written in {\sf R}, allows easy access to new simulations from
within {\sf R}. Everything is provided via a public Bitbucket mercurial
repository: \url{https://bitbucket.org/gramacylab/tpm.} The package
is not a CRAN package, but it is structured in a very similar way. After
cloning the repository, there are two steps
required to perform new TPMC simulations. First, to compile the {\sf C}
library, change into \verb!tpm/src! and execute

{\singlespacing
\begin{verbatim}
% R CMD SHLIB -o tpm.so *.c
\end{verbatim}}

\noindent Then, the second step involves sourcing wrapper files from within {\sf R}.

{\singlespacing
\begin{verbatim}
R> source("tpm/R/tpm.R")
\end{verbatim}}

Now we are ready to run our first TPMC simulations.  Below is code building a
data frame of eight uniformly random settings of the free parameters in Table
\ref{t:inputs}, repeated twice for sixteen settings total.

{\singlespacing
\begin{verbatim}
R> n <- 8
R> X <- data.frame(Umag=runif(n, 5500, 9500), Ts=runif(n, 100, 500),
+    Ta=runif(n, 200, 2000), theta=runif(n,-pi,pi), phi=runif(n,-pi/2,pi/2), 
+    alphan=runif(n), sigmat=runif(n))
R> X <- rbind(X,X)
\end{verbatim}}

Next we need to specify a mesh file.  The directory \verb!tpm/Mesh_Files!
contains all of the meshes used in this paper, and several others including
the ones for the International Space Station (ISS).  Below we specify the GRACE mesh.

{\singlespacing
\begin{verbatim}
R> mesh <- "tpm/Mesh_Files/GRACE_A0_B0_ascii_redone.stl"
\end{verbatim}}

\noindent Then specify the mixture vector over the chemical species, e.g.,
pure He.

{\singlespacing
\begin{verbatim}
R> moles <- c(0,0,0,0,1,0)
\end{verbatim}}

With those inputs, a blocked TPMC batch may be invoked as follows.

{\singlespacing
\begin{verbatim}
R> system.time(y <- tpm(X, moles=moles, stl=mesh, verb=0))

##   user     system  elapsed 
## 3204.680    1.136  265.502 
\end{verbatim}}

\noindent The elapsed time is 266 seconds (30 minutes), cumulative, for 
the sixteen evaluations.  That is an average of seventeen seconds per
evaluation. However, by inspecting the ``user'' time we can see that the
execution was parallelized.  The desktop this was performed on is an 8-core
64-bit Intel i7-6900K CPU at 3.20 GHz. The cores are hyperthreaded, meaning
that it can often perform sixteen numerical tasks in parallel rather than the
usual eight. However, the speedup is not 16-fold since some of the inputs
take longer than others.  Instead we observed about a 12-fold speedup,
amortized over all sixteen runs.  On a single core machine, the execution time
would have been about 53 minutes, or about 3.3 minutes per simulation.

The replication in the design can give us an indication of the level of noise
in these Monte Carlo simulations.

{\singlespacing
\begin{verbatim}
R> mean((y[1:n] - y[(n+1):length(y)])^2)

## [1] 4.120074e-05
\end{verbatim}}

\noindent So the noise is higher than machine precision, but low relative to the range
of {\tt y}-values.

{\singlespacing
\begin{verbatim}
R> range(y)

## [1] 1.812294 3.419442
\end{verbatim}}

The {\sf R} wrapper routines contain a {\tt tpm.parallel} command that allows
further cluster-style parallelization through the built-in {\sf R} library. It
works similarly to {\tt tpm}, taking a cluster object created by the {\tt
makeCluster} command, establishing socket or MPI connections to computer
nodes.  The {\tt tpm.parallel} command was used on University of Chicago's
Research Computer Center's {\tt midway} cluster to perform TPMC evaluations
used for training and testing of HST and GRACE satellite predictors in Section
\ref{sec:satemu}.  Our supplementary material includes the data files from those
runs, along with the smaller runs compiled by \citet{metha:etal:2014} using
the earlier stand-alone {\sf C} version of the TPMC library.

To independently verify the accuracy results reported by \citet{metha:etal:2014} 
we offer the following code re-creating their test for GRACE in pure He. First, read 
in the data.

{\singlespacing
\begin{verbatim}
R> train <- read.table("satdrag/GRACE/CD_GRACE_1000_He.dat")
R> test <- read.table("satdrag/GRACE/CD_GRACE_100_He.dat")
R> nms <- c("Umag", "Ts", "Ta", "alphan", "sigmat", "theta", "phi", "drag")
R> names(train) <- names(test) <- nms
R> print(r <- apply(rbind(train, test)[,-8], 2, range))

##          Umag       Ts        Ta       alphan       sigmat        theta
## [1,] 5501.933 100.0163  201.2232 0.0008822413 0.0007614135 1.270032e-05
## [2,] 9497.882 499.8410 1999.9990 0.9999078000 0.9997902000 6.978310e-02
##              phi
## [1,] -0.06978125
## [2,]  0.06971254
\end{verbatim}}

\noindent The output above allows us to compare ranges of inputs against
Tables \ref{t:inputs}--\ref{t:reduced}.  Next,  code those inputs in the unit 
cube as follows.  

{\singlespacing 
\begin{verbatim}
R> X <- train[,1:7]; XX <- test[,1:7]
R> for(j in 1:ncol(X)){
+      X[,j] <- X[,j] - r[1,j]; XX[,j] <- XX[,j] - r[1,j]; 
+      X[,j] <- X[,j]/(r[2,j]-r[1,j]); XX[,j] <- XX[,j]/(r[2,j]-r[1,j])
+  }
\end{verbatim}} 

Finally, we are ready to fit a GP.  The code below uses the (non-approximate)
separable covariance GP capability built into the {\tt laGP} package.

{\singlespacing
\begin{verbatim}
R> library(laGP)
R> fit.gp <- newGPsep(X, train[,8], 2, 1e-6, dK=TRUE)
R> mle <- mleGPsep(fit.gp)
R> p <- predGPsep(fit.gp, XX, lite=TRUE)
R> rmspe <- sqrt(mean((100*(p$mean - test[,8])/test[,8])^2))
R> rmspe

## [1] 0.7401575
\end{verbatim}}

\noindent The final calculation is RMSPE, and observe that this is below the 
desired 1\% benchmark.  Similar results are obtained with the other species, 
whose data files are also provided in the supplementary material.

\section{Compactly supported covariances v.~{\tt laGP}}
\label{sec:csc}

Here we follow an example provided by the CSC authors, \citet{kaufman:etal:2012}: 

\begin{center}
\url{https://www.stat.berkeley.edu/~cgk/rcode/assets/SparseEmExample.R}
\end{center}

\noindent Below is an implementation of the borehole function (Section
\ref{sec:bench}), followed by a random partition into training and testing
sets via a joint LHS.  This is almost verbatim from the code linked above, but 
we re-create it here in order to add our own {\tt laGP}-based comparisons.

{\singlespacing
\begin{verbatim}
R> borehole <- function(x){
+    rw <- x[1] * (0.15 - 0.05) + 0.05
+    r <-  x[2] * (50000 - 100) + 100
+    Tu <- x[3] * (115600 - 63070) + 63070
+    Hu <- x[4] * (1110 - 990) + 990
+    Tl <- x[5] * (116 - 63.1) + 63.1
+    Hl <- x[6] * (820 - 700) + 700
+    L <-  x[7] * (1680 - 1120) + 1120
+    Kw <- x[8] * (12045 - 9855) + 9855
+    m1 <- 2 * pi * Tu * (Hu - Hl)
+    m2 <- log(r / rw)
+    m3 <- 1 + 2*L*Tu/(m2*rw^2*Kw) + Tu/Tl
+    return(m1/m2/m3)
}

## Training-testing partition
R> n <- 4000
R> npred <- 500
R> dim <- 8
R> library(lhs)
R> x <- randomLHS(n+npred, dim)
R> y <- apply(x, 1, borehole)
R> ypred.0 <- y[-(1:n)]; y <- y[1:n]
R> xpred <- x[-(1:n),]; x <- x[1:n,]
\end{verbatim}}

\noindent Then comes the fitting functions, which are partitioned into routines 
that find the desired level of sparsity, which for this example is 99\%, followed
by sampling from the posterior at that level, and then gathering predictions
after burn-in.

{\singlespacing
\begin{verbatim}
R> mc <- find.tau(den=1-0.99, dim=ncol(x)) * ncol(x)
R> time1 <- system.time(
+    tau <- mcmc.sparse(y, x, mc=mc, degree=2, maxint=2, B=2000, verbose=F))
R> index <- seq(500+1, 2000, by=10)
R> time2 <- system.time(ypred.sparse <- 
+    pred.sparse(tau[index,], x, y, xpred, degree=2, maxint=2,verbose=F))
\end{verbatim}}

\noindent Here is a similar code for a 99.9\% sparse version.

{\singlespacing
\begin{verbatim}
R> mc <- find.tau(den=1-0.999, dim=ncol(x)) * ncol(x) 
R> time5 <- system.time(
+    tau <- mcmc.sparse(y, x, mc=mc, degree=2, maxint=2, B=2000, verbose=F))
R> index <- seq(500+1, 2000, by=10)
R> time6 <- system.time(ypred.sparse <- 
+    pred.sparse(tau[index,], x, y, xpred, degree=2, maxint=2, verbose=F))
\end{verbatim}}

\noindent For a fair comparison, \citet{kaufman:etal:2012} provide a non-sparse alternative.

{\singlespacing
\begin{verbatim}
R> time3 <- system.time(phi <- mcmc.nonsparse(y, x, B=200, verbose=F))
R> index <- seq(50+1, 200, by=10)
R> time4 <- system.time(ypred.nonsparse <- 
+    pred.nonsparse(phi[index,], x, y, xpred, 2, verbose=F))
\end{verbatim}}

\noindent Here is how these comparators fare in terms of computation time: about a factor 
of 3 speed-up by being 99\% sparse, and 10$\times$ better than that with 99.9\%
sparsity.

{\singlespacing
\begin{verbatim}
R> times <- c(sparse=as.numeric(time1[3]+time2[3]), 
+    dense=10*as.numeric(time3[3]+time4[3]), 
+    s999=as.numeric(time5[3]+time6[3]))
R> times

##   sparse   dense     s999 
##  3409.157 9020.870  258.196
\end{verbatim}}

\noindent The proper scores \citep[Eq.~(27)][]{gneiting:raftery:2007}, with larger being
better, follow a predictable pattern: the more dense the higher the accuracy.

{\singlespacing
\begin{verbatim}
R> s2s <- ypred.sparse$var
R> s2n <- ypred.nonsparse$var
R> s29 <- ypred.sparse$var
R> scores <- c(sparse=mean(- (ypred.sparse$mean - ypred.0)^2/s2s - log(s2s)),
+   dense=mean(-(ypred.nonsparse$mean - ypred.0)^2/s2n - log(s2n)),
+   s999=mean(-(ypred.sparse$mean - ypred.0)^2/s29 - log(s29)))
R> scores

##   sparse    dense     s999 
## -1.616263 -2.636446 -1.796645
\end{verbatim}}

Now what about {\tt laGP}?  The two commands below derive predictions based on
an isotropic and separable Gaussian correlation structure, respectively.

{\singlespacing
\begin{verbatim}
R> out <- aGP(x, y, xpred, d=list(max=20), omp.threads=8, verb=0)
R> outs <- aGPsep(x, y, xpred, d=list(max=20), omp.threads=8, verb=0)
\end{verbatim}}

\noindent In terms of time, {\tt laGP} is more than an order of magnitude
faster than even the sparsest CSC version.

{\singlespacing
\begin{verbatim}
R> times <- c(times, aGP=as.numeric(out$time), aGPs=as.numeric(outs$time)))
R> times

##  sparse     dense     s999       aGP      aGPs 
## 3409.157   9020.870  258.196    9.059    8.929
\end{verbatim}}

\noindent Not only are they faster, but also they are (much) more accurate,
especially the separable version.

{\singlespacing
\begin{verbatim}
R> s2 <- out$var; s2s <- outs$var
R> scores <- c(scores, aGP=mean(-(out1$mean - ypred.0)^2/s2 - log(s2)),
+    aGPs=mean(-(out1$mean - ypred.0)^2/s2s - log(s2s))))
R> scores

##   sparse       dense          s999         aGP         aGPs 
## -1.6162629   -2.6364460   -1.7966449   -0.6024819   0.1057021
\end{verbatim}}

Finally, as a demonstration of the potential of multiresolution modeling,
consider a {\em single} random subset with 1000 points fit with full-GP
subroutines in the {\tt laGP} package.

{\singlespacing
\begin{verbatim}
R> da <- darg(list(mle=TRUE, max=100), x)
R> subs <- sample(1:nrow(x), 1000, replace=FALSE)
R> gpsi <- newGPsep(x[subs,], y[subs], rep(da$start, dim), g=1e-3, dK=TRUE)
R> that <- mleGPsep(gpsi, tmin=d2$min, tmax=d2$max, ab=d2$ab, maxit=200)
R> psub <- predGPsep(gpsi, xpred, lite=TRUE)
R> deleteGPsep(gpsi)
\end{verbatim}}

\noindent Then we can scale the predictors by the square root of estimated lengthscales 
and fit another {\em isotropic} local approximate GP, initialized at a lengthscale of 1. 

{\singlespacing
\begin{verbatim}
R> scale <- sqrt(that$d)
R> xs <- x; xpreds <- xpred
R> for(j in 1:ncol(xs)){
+      xs[,j] <- xs[,j] / scale[j]
+      xpreds[,j] <- xpreds[,j] / scale[j]
+  }
R> outmr <- aGP(xs, y, xpreds, d=list(start=1, max=20), omp.threads=8, 
+  verb=0)
\end{verbatim}}

\noindent The new multiresolution predictor is the best of all, but notice that 
the subset GP is also very good.  Apparently the compactly supported
covariance structure, which generated our {\tt dense} results above, does not
offer enough smoothness for the borehole data.

{\singlespacing
\begin{verbatim}
R> s2sub <- out3$var; s2mr <- out4$var
R> scores <- c(scores, sub=mean(-(psub$mean-ypred.0)^2/s2sub - log(s2sub)),
+    aGPmr=mean(-(out3$mean - ypred.0)^2/s2mr - log(s2mr)))
R> scores

##   sparse      dense       s999        aGP       aGP2        aGPs
## -1.6162629  -2.6364460 -1.7966449 -0.6024819 -0.5693178  0.1057021
##     sub       aGPsm 
##  0.7415339  1.1909690
\end{verbatim}}

\noindent Further improvements (scores above 5) are possible with {\tt aGPsep}
with a smaller nugget value than the default, e.g.,
\verb!g=1/10000000!, but we leave those extensions to the curious reader.

\section{Derivative derivation (separable GP version)}
\label{sec:dd}

Here we derive the expression behind the components of the gradient of the
``joint" ALC defined in Eq.~(\ref{eq:dalc}) from Section \ref{sec:dalc}.  We 
allow any form of the correlation function where derivatives may be calculated 
relative to the spatial locations, $x$.  Following Eq.~(\ref{eq:dalc}), we take
the derivative with respect to each coordinate of $x_{j+1}$. For
$\ell=1, \dots, p$,
\begin{align*}
\frac{\partial}{\partial (x_{j+1})_\ell} &\left\{v_j(W) - v_{j+1}(W)\right\} 
= \frac{\partial}{\partial (x_{j+1})_\ell} \left\{\frac{1}{|W|}\sum_{w \in W} \left[v_j(w)-v_{j+1}(w)\right] \right\} \\
&=  \frac{1}{|W|}\sum_{w \in W} \frac{\partial}{\partial (x_{j+1})_\ell}\Big\{  k_j^\top(w)G_j(x_{j+1})v_j(x_{j+1})k_j(w) \\ 
& \quad\quad\quad\quad\quad\quad\quad\quad\quad\quad
 + 2k_j^\top(w)g_j(x_{j+1})K(x_{j+1}, w) + K(x_{j+1}, w)^2/ v_j(x_{j+1}) \Big\},
\end{align*}
where
\begin{align}
&\frac{\partial}{\partial (x_{j+1})_\ell}\left\{k_j^\top(w)G_j(x_{j+1})v_j(x_{j+1})k_j(w) + 2k_j^\top(w)g_j(x_{j+1})K(x_{j+1}, w) + \frac{K(x_{j+1}, w)^2}{v_j(x_{j+1})} \right\} \label{eq:dds1}\\
&= k_j^\top(w)\left\{ \left [\frac{\partial}{\partial (x_{j+1})_\ell} G_j(x_{j+1})\right]v_j(x_{j+1})+G_j(x_{j+1})\left[\frac{\partial}{\partial (x_{j+1})_\ell} v_j(x_{j+1})\right]\right\}k_j(w) \nonumber \\
&\quad +2k_j^\top(w)\left\{\left [\frac{\partial}{\partial (x_{j+1})_\ell} g_j(x_{j+1})\right]K(x_{j+1}, w)+g_j(x_{j+1})\left [\frac{\partial}{\partial (x_{j+1})_\ell} K(x_{j+1}, w)\right]\right\} \nonumber\\
&\quad +2K(x_{j+1}, w)\left[\frac{\partial}{\partial (x_{j+1})_\ell} K(x_{j+1}, w)\right] v_j^{-1}(x_{j+1})+K(x_{j+1}, w)^2\left[\frac{\partial}{\partial (x_{j+1})_\ell} v_j^{-1}(x_{j+1})\right]. \nonumber 
\end{align}
Particularly,
\begin{align}
\frac{\partial}{\partial (x_{j+1})_\ell} G_j(x_{j+1})&=2\left[\frac{\partial}{\partial (x_{j+1})_\ell} g_j(x_{j+1})\right]g^\top_j(x_{j+1}),\label{eq:dds11}\\
\frac{\partial}{\partial (x_{j+1})_\ell} g_j(x_{j+1})&=-K^{-1}_j\left\{\!\left[\frac{\partial}{\partial (x_{j+1})_\ell} k_j(x_{j+1})\right]v_j^{-1}(x_{j+1})+k_j(x_{j+1})\left[\frac{\partial}{\partial (x_{j+1})_\ell} v_j^{-1}(x_{j+1})\right]\!\right\}, \nonumber \\
\frac{\partial}{\partial (x_{j+1})_\ell} v_j(x_{j+1})&=2\left[\frac{\partial}{\partial (x_{j+1})_\ell} k^\top_j (x_{j+1}) K_j^{-1} k_j (x_{j+1})\right]. \label{eq:dds13}
\end{align}
By plugging Eqs.~(\ref{eq:dds11}--\ref{eq:dds13}), Eq.~(\ref{eq:dds1}) can be reorganized as follows:
\begin{equation}
\begin{aligned}
\mbox{Eq.~(\ref{eq:dds1})} = &-k^\top_j(w)K_J^{-1}\left[\dot{k}_\ell + k_j(x_{j+1})av_j^{-1}(x_{j+1})\right]g_j^\top(x_{j+1})k_j(w)\nonumber\\
& - k^\top_j(w)g_j(x_{j+1})\!\left[\dot{k}^\top_\ell K_J^{-1} + k_j^\top (x_{j+1})K_J^{-1}av_j^{-1}(x_{j+1})\right]\!k_j(w)\nonumber\\
&-k_j^\top (w) g_j(x_{j+1})g_j^\top(x_{j+1})ak_j(w) \nonumber \\
&+2k_j^\top(w) \left\{-K_J^{-1}\left[\dot{k}_\ell+ k_j(x_{j+1})av_j^{-1}(x_{j+1})\right]v_j^{-1}(x_{j+1})K(x_{j+1},w)+ g_j(x_{j+1})\dot{K}_\ell\right\}\nonumber \\
&+ K(x_{j+1},w)\left[2\dot{K}_\ell + K(x_{j+1},w)av_j^{-1}(x_{j+1})\right]v_j^{-1}(x_{j+1}),\nonumber \\
\end{aligned}
\end{equation}
where the macros $\dot{k}_\ell$ (a $j$-vector), $\dot{K}_\ell$ and $a$ (both scalars) are
\begin{align*}
&\dot{k}_\ell =\frac{\partial}{\partial (x_{j+1})_\ell} k_j (x_{j+1})= \left\{- \frac{2\left[(x_{j+1})_\ell - (x_j)_\ell\right]}{ \theta_\ell} \exp\left\{-\sum_{\ell=1}^p \frac{\left[(x_{j+1})_\ell-(x_j)_\ell\right]^2}{\theta_\ell}\right\}\right\}_{j\times 1}, \nonumber \\
&\text{where $x_j$ is any location within the current local design}.\\
&\dot{K}_\ell =\frac{\partial}{\partial (x_{j+1})_\ell} K(x_{j+1}, w)= -\frac{2\left[(x_{j+1})_\ell - w_\ell \right]}{\theta_\ell} \exp\left\{-\sum_{\ell=1}^p \frac{\left[(x_{j+1})_\ell-w_\ell\right]^2}{\theta_\ell} \right\},\\
&a = 2 \frac{\partial k_j^\top (x_{j+1})}{\partial (x_{j+1})_\ell} K_j^{-1} k_j (x_{j+1}). 
\end{align*}
Eq.~(\ref{eq:dds1}) can be further simplified to\\ 
\begin{equation}
-2k^\top_j(w)K^{-1}_J\left[ \frac{\partial k_j (x_{j+1}) }{\partial (x_{j+1})_\ell}  + \frac{ k_j (x_{j+1}) a}{v_j(x_{j+1})} \right]b-abc + 
2 c \frac{\partial K(x_{j+1}, w)}{\partial (x_{j+1})_\ell}  \nonumber
\end{equation}
where the macros $b$ and $c$ (both scalars) are
\begin{equation}
b =g_j^\top (x_{j+1})k_j(w) - K(x_{j+1},w)/v_j(x_{j+1}), \quad \mbox{and} \quad c =g_j^\top (x_{j+1})k_j(w) + K(x_{j+1},w)/v_j(x_{j+1}) \nonumber.  
\end{equation}
The final expression for the components of the gradient of the ``joint" ALC is as follows:
\begin{align}
\frac{\partial}{\partial (x_{j+1})_\ell} &
\left\{v_j(W) - v_{j+1}(W)\right\} \nonumber \\
=&\frac{1}{|W|}\sum_{w \in W} \left\{-2k^\top_j(w)K^{-1}_J\left[ \frac{\partial k_j (x_{j+1}) }{\partial (x_{j+1})_\ell}  + \frac{ k_j (x_{j+1}) a}{v_j(x_{j+1})} \right]b-abc + 
2 c \frac{\partial K(x_{j+1}, w)}{\partial (x_{j+1})_\ell}\right\}.\nonumber
\end{align}

\section{2d representative predictive paths}
\label{sec:rpp}

\begin{figure}[ht!]
\centering
\includegraphics[scale=0.42, trim=50 30 0 30]{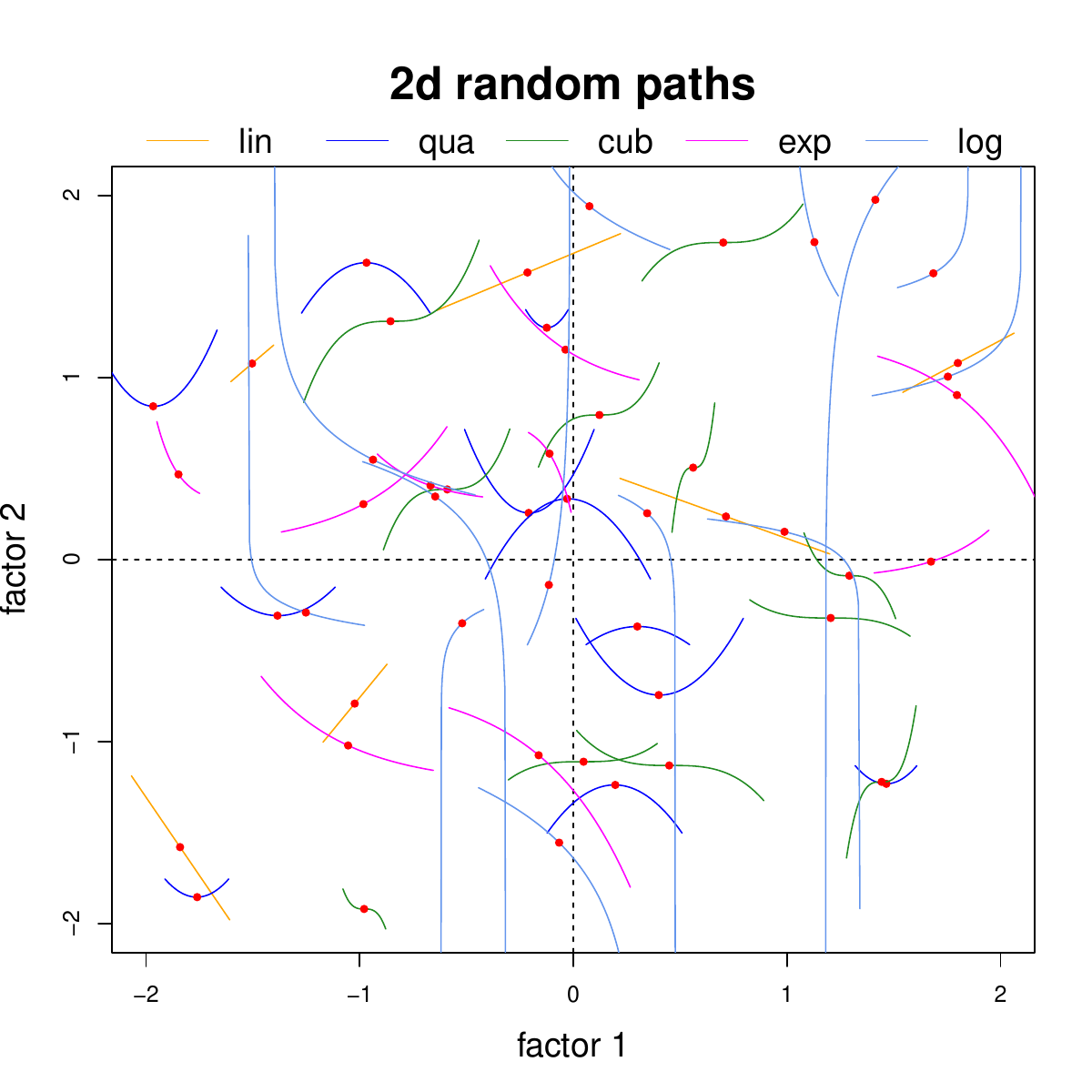}\vspace{-0.2cm}
\caption{A representative plot of 2d predictive paths}
\label{f:rpp}
\end{figure}

Figure \ref{f:rpp} shows a representative plot of 2d predictive paths
generated randomly as follows. 
First, a line type (linear, quadratic, cubic,
exponential, and natural logarithm) is chosen uniformly at random.
The line is then drawn, via a collection of discrete points, from the origin
via specifications (e.g., resolution and length) provided by the user.   The
discrete set of coordinates is then shifted and scaled, uniformly at random,
into a specified 2d rectangle ($[-2,2]^2$ in the figure) with the restriction
that at least half of the points comprising the line lie within the rectangle.  
A centering dot (mostly for visual effect) is calculated 
as the midway point on the discrete set of points making up the line.

\section{{\tt laGP} versus others}
\label{sec:rgasp}

A referee suggested we compare {\tt laGP} with some other methods, and
provided the beginnings of a script that applied {\tt RobustGaSP}
\citep{gu:etal:2018} and {\tt DiceKriging} \citep{roustant:etal:2012} on a
small ($N=400$) borehole example (Section \ref{sec:borehole}).  We augmented
that script to include {\tt laGP}'s full GP capability as a comparator.  The
full GP functionality in {\tt laGP} is bare-bones: simple MLE calculations via
(\ref{eq:gpk}) for unknown lengthscale hyperparameters.  This is extended to
global MLEs of lengthscale from large data sets via BLHS, subsequent local
sub-design and local MLE estimation as explained in the main body of our
manuscript, and detailed by \citet{gramacy:apley:2015}.  Once the local design
has been determined, inference for hyperparameters is fundamentally not
different than in the full GP case---just with less data.
\citeauthor{gu:etal:2018}'s careful approach to GP inference is impressive in
small-$N$ contexts, and \citeauthor{roustant:etal:2012}'s ``DICE''-family of
methods is extensive in terms of the breadth of modeling options. Yet, as we
summarize below, {\tt laGP} methods are competitive.  Of course, the local
approximation {\tt laGP} provides is unmatched when it comes to data size.
Neither of these packages can handle data an order of magnitude larger than
the $N=400$ in this small study, let alone the two-to-three magnitudes larger
which are the target of this manuscript.  We augmented the script provided by
the referee, which is very similar to the one summarized by Figure
\ref{f:synthetic_compare}, to include {\tt laGP}-based comparators, and
wrapped it in a {\tt for} loop to obtain thirty repetitions, and include it
along with our supplementary material.

\begin{figure}[ht!]
\centering
\includegraphics[scale=0.32,trim=0 0 0 50,clip=TRUE]{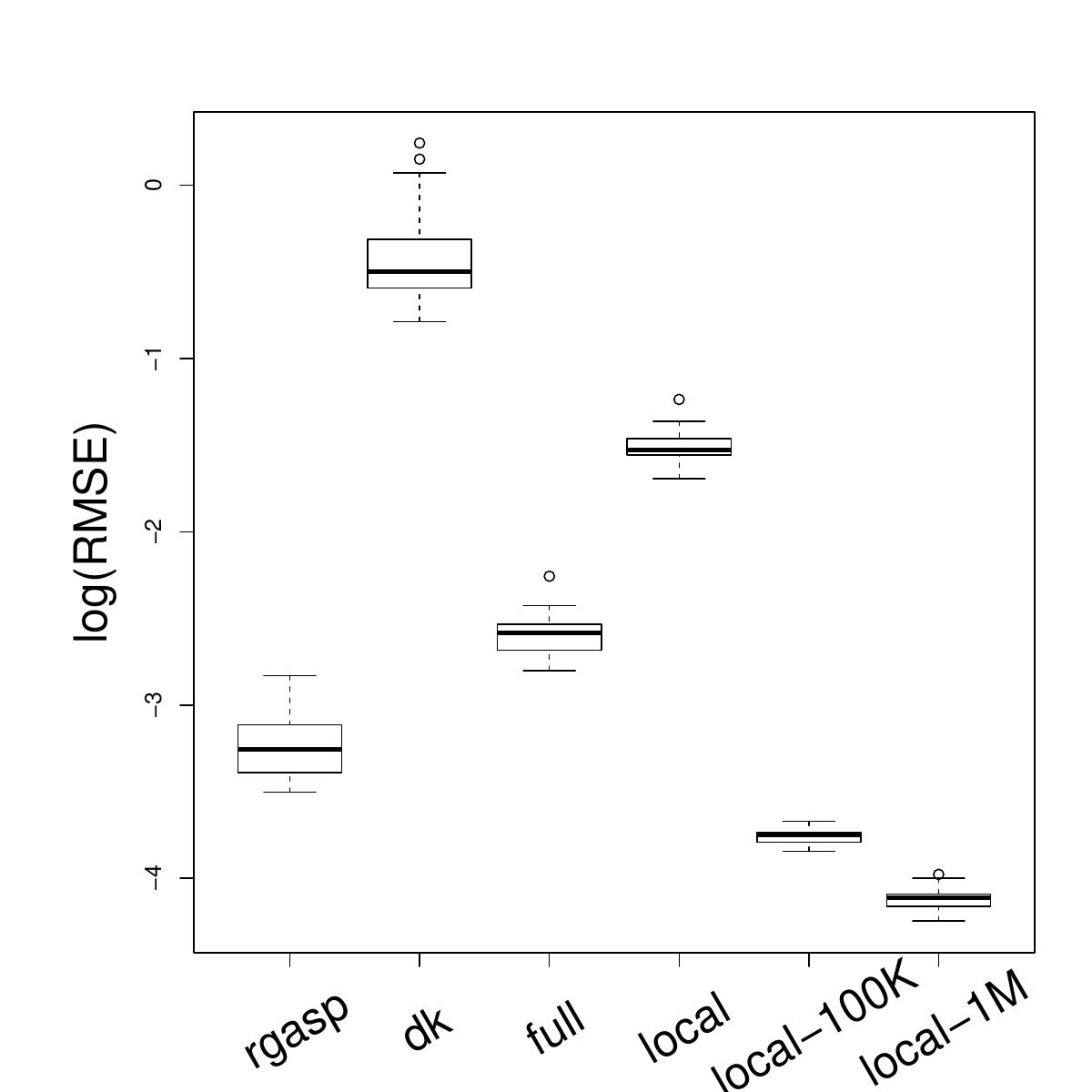} \hspace{0.5cm}
\includegraphics[scale=0.32,trim=0 0 0 50,clip=TRUE]{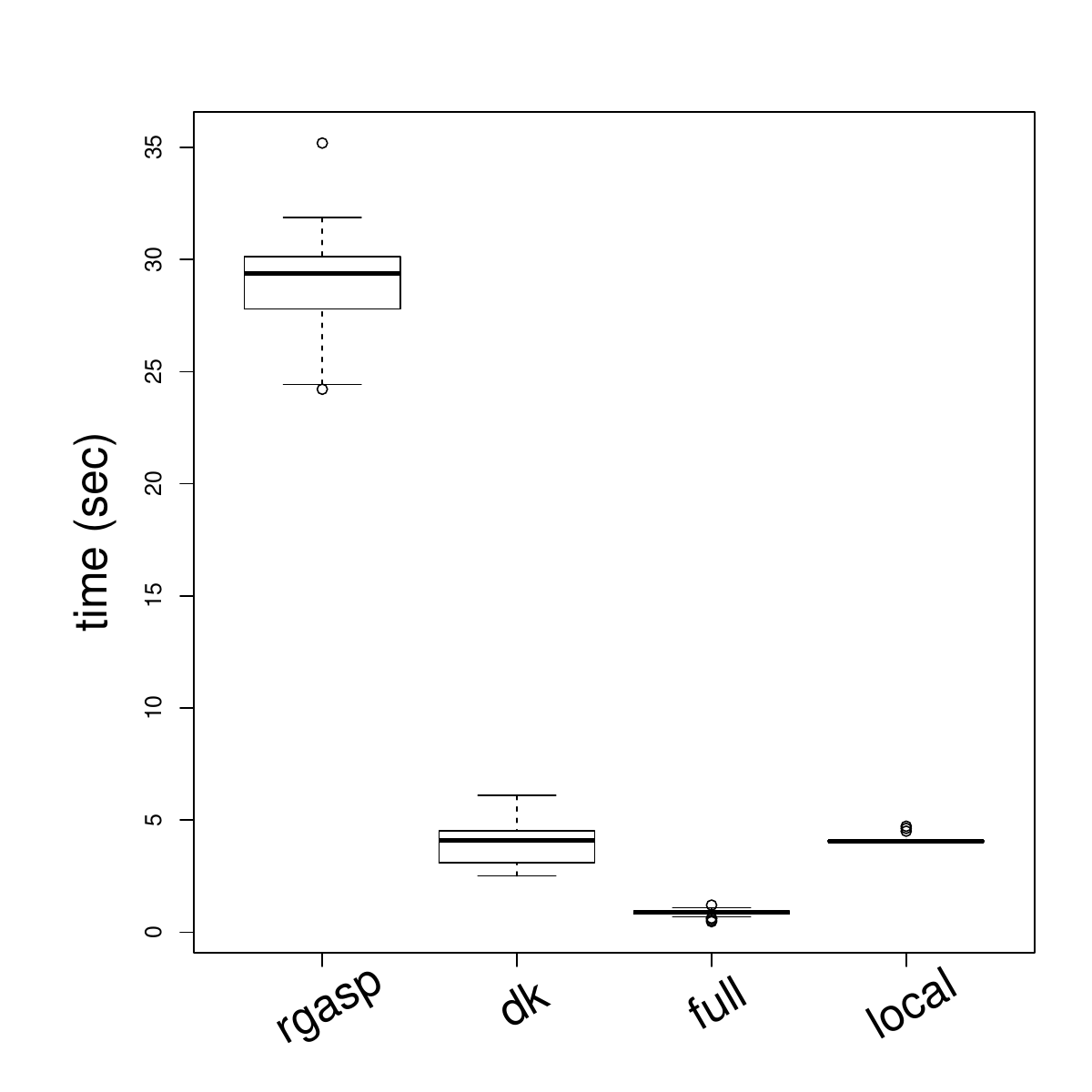}
\caption{Comparisons between {\tt RobustGaSP}, {\tt DiceKriging}, and {\tt
laGP}-based methods including full and local variations on an $N=400$-sized
borehole example. {\em Left}: empirical distribution of log RMSEs. {\em
Right}: empirical distribution of computational cost via wall-clock time (in
seconds). Comparators ``local-100K'' and ``local-1M'' correspond to the
``alc.sb'' comparators from Figure \ref{f:synthetic_compare} and are based on
the multifidelity framework of Section \ref{sec:mrlaGP}.}
\label{f:rgasp}
\end{figure}

The results of that experiment are provided in Figure \ref{f:rgasp}.  The left
panel shows out-of-sample RMSE in log space, and the right panel collects
execution time. The first four boxplots, from left to right in either panel,
correspond to full GPs fit to $N=400$ training sites.  {\tt RobustGaSP} is the
best here, but it is also thirty times slower than the second best, via the
full GP functionality in {\tt laGP}.  The local modeling features of {\tt
laGP} are competitive on this small example, but not better than full GP
modeling.  When the data are small enough for full GP modeling, local
approximations are neither necessary nor beneficial.  However, when the data
are bigger they are indeed quite necessary, and they represent an efficient way
to leverage the information available in such samples, beating the {\tt
RobustGaSP} comparator.  Note that the runtime for {\tt laGP} in these larger
$N$ settings is nearly identical to the smaller $N$ since both use a local
neighborhood size of $n=50$.  The only difference is an $N \log N$
pre-processing step to which is nearly instantaneous even with large $N$.

It is possible that the ideas behind {\tt RobustGaSP} could be ported over to
the local setting, however, it is not immediately obvious how to do that (from
an implementation perspective) at this time.  Clearly the requisite order of
magnitude larger runtime will limit the size of problems on which such an
approach can be applied.

\newpage
\bibliography{../laGP/laGP,../gpu/gpu,../rays/rays,../calib/calib,satdrag,rebuttal}
\bibliographystyle{jasa}

\end{document}